\documentstyle[12pt,fleqn,epsf,rotate,subfigure]{article}
  \newcommand{\ie}{i.e.$\!$}
  \newcommand{\heavi}{\Theta}
  \newcommand{\ee}{{\mathrm{e}}}
  \newcommand{\ii}{{\mathrm{i}}}
  \newcommand{\SMALLCAP}  [2]{\caption[#1]{{\small #2}}}
  \newcommand{\pippo}{^{^{^{^{^{^{^{^{}}}}}}}}}

\newcommand{\sectio}[1]{\section{#1}\setcounter{equation}{0}}
\renewcommand{\theequation}{\arabic{section}.\arabic{equation}} 
\def \vec#1{{\mathbf #1}}
\def \bbox#1{{\mathbf #1}}

\renewcommand{\d}{{d}}
\newcommand{\e}{{e}}
\newcommand{\p}{{\vec{p}}}
\newcommand{\q}{{\vec{q}}}
\newcommand{\r}{{\vec{r}}}
\newcommand{\G}{{\cal G}}

\def \omn#1{\vec{\Omega}_{#1}}
\def \fref#1{fig.\ref{#1}}
\def \cd{{\cal D}}
\newcommand{\k}{{\vec{k}}}
\newcommand{\K}{{\vec{K}}}

\newcommand{\sgn}{{\rm sign}}
\newcommand{\kf}{{k_{\rm F}}}

\def \virg{\;\;,}
\def \point{\;\;.}
\def \up{\uparrow}
\def \down{\downarrow}

\renewcommand{\S}{{\mathbf{S}}}

\def \omn#1{\vec{\Omega}_{#1}}
\def \fref#1{fig.\ref{#1}}
\begin{document}

\epsfverbosetrue

%\debugoff
\title{Fermi liquids and Luttinger liquids}

\author{H.J. Schulz\\
{\em Laboratoire de Physique des Solides, Universit\'e Paris--Sud}\\
{\em 91405 Orsay, France}\\
\phantom{ppppppppppp}\\
G. Cuniberti\\
{\em I. Institut f\"ur Theoretische Physik, Universit\"at Hamburg}\\
{\em Jungiusstra\ss e 9, 20355 Hamburg, Germany}\\
\phantom{ppppppppppp}\\
P. Pieri\\
{\em Dipartimento di Matematica e Fisica, and Sezione INFM}\\
{\em Universit\`a di Camerino, 62032 Camerino, Italy}
}

\date{}
\maketitle

\tableofcontents
\section{Introduction}
In these lecture notes, corresponding roughly to lectures given at the
summer school in Chia Laguna, Italy, in September 1997, an attempt is made
to present the physics of three--dimensional interacting fermion systems
(very roughly) and that of their one--dimensional counterparts, the
so--called Luttinger liquids (in some more detail). These subjects play a
crucial role in a number of currently highly active areas of research: high
temperature and organic superconductors, quantum phase transitions,
correlated fermion systems, quantum wires, the quantum Hall effect,
low--dimensional magnetism, and probably some others. Some understanding of
this physics thus certainly should be useful in a variety of areas, and it
is hoped that these notes will be helpful in this.

As the subject of these lectures was quite similar to those delivered at Les
Houches, some overlap in the notes\cite{schulz_houches} was
unavoidable. However, a number of improvements have been made, for example a
discussion of the ``Klein factors'' occurring in the bosonization of
one--dimensional fermions, and new material added, mainly concerning spin
chains and coupled Luttinger liquids. Some attempt has been made to keep
references up to date, but this certainly has not always been successful, so
we apologize in advance for any omissions (but then, these are lecture
notes, not a review article).

\sectio{Fermi Liquids}
Landau's Fermi liquid theory\cite{landau_1,landau_2,landau_3} is concerned with
the properties of a
many--fermion system at low temperatures (much lower than the Fermi energy)
in {\em the normal state}, i.e. in the absence or at least at temperatures
above any symmetry breaking phase transition (superconducting, magnetic, or
otherwise).  The ideal example for Landau's theory is liquid helium 3, above
its superfluid phase transition, however, the conceptual basis of Landau's
theory is equally applicable to a variety of other systems, in particular
electrons in metals. Quantitative applications are however more difficult
because of a variety of complications which appear in real systems, in 
particular the absence of
translational and rotational invariance and the presence of electron--phonon
interactions, which are not directly taken into account in Landau's
theory. Subsequently, I will first briefly discuss the case of a
noninteracting many--fermion system (the Fermi {\em gas}), and then turn to
Landau's theory of the interacting case (the {\em liquid}), first from a
phenomenological point of view, and then microscopically.  A much more
detailed and complete exposition of these subjects can be found in the
literature \cite{pines_nozieres,agd,nozieres,kadanoff_baym,baym_pethick}.

\subsection{The Fermi Gas}
In a noninteracting translationally invariant systems, the single-particle
eigenstates are plane waves
\begin{equation}
 |\k \rangle = \frac{1}{\sqrt\Omega} \e^{{\rm i} \k \cdot \vec{r}}
\end{equation}
with energy
\begin{equation}
\label{e:ek}
\varepsilon_{\k} = \frac{\k^2}{2m} \virg
\end{equation}
where $\Omega$ is the volume of the system, and we will always use units so
that $\hbar =1$.  The ground state of an $N$--particle system is the
well--known Fermi sea: all states up to the {\em Fermi wavevector} $\kf$ are
filled, all the other states are empty. For spin--1/2 fermions the relation
between particle number and $\kf$ is
\begin{equation}
N = \Omega \frac{\kf^3}{3\pi^2} \point
\end{equation}
The energy of the last occupied state is the so--called {\em Fermi
energy} $E_{\rm F} = \kf^2/(2m)$, and one easily verifies that
\begin{equation}
E_{\rm F} = \frac{\partial E_0(N)}{\partial N} = \mu(T=0)
\end{equation}
i.e. $E_{\rm F}$ is the zero--temperature limit of the chemical potential
($E_0(N)$ in the formula above is the ground state energy).

\begin{figure}[htb]
\centerline{
\subfigure[]{\epsfysize 6cm \rotate[l]{\epsffile{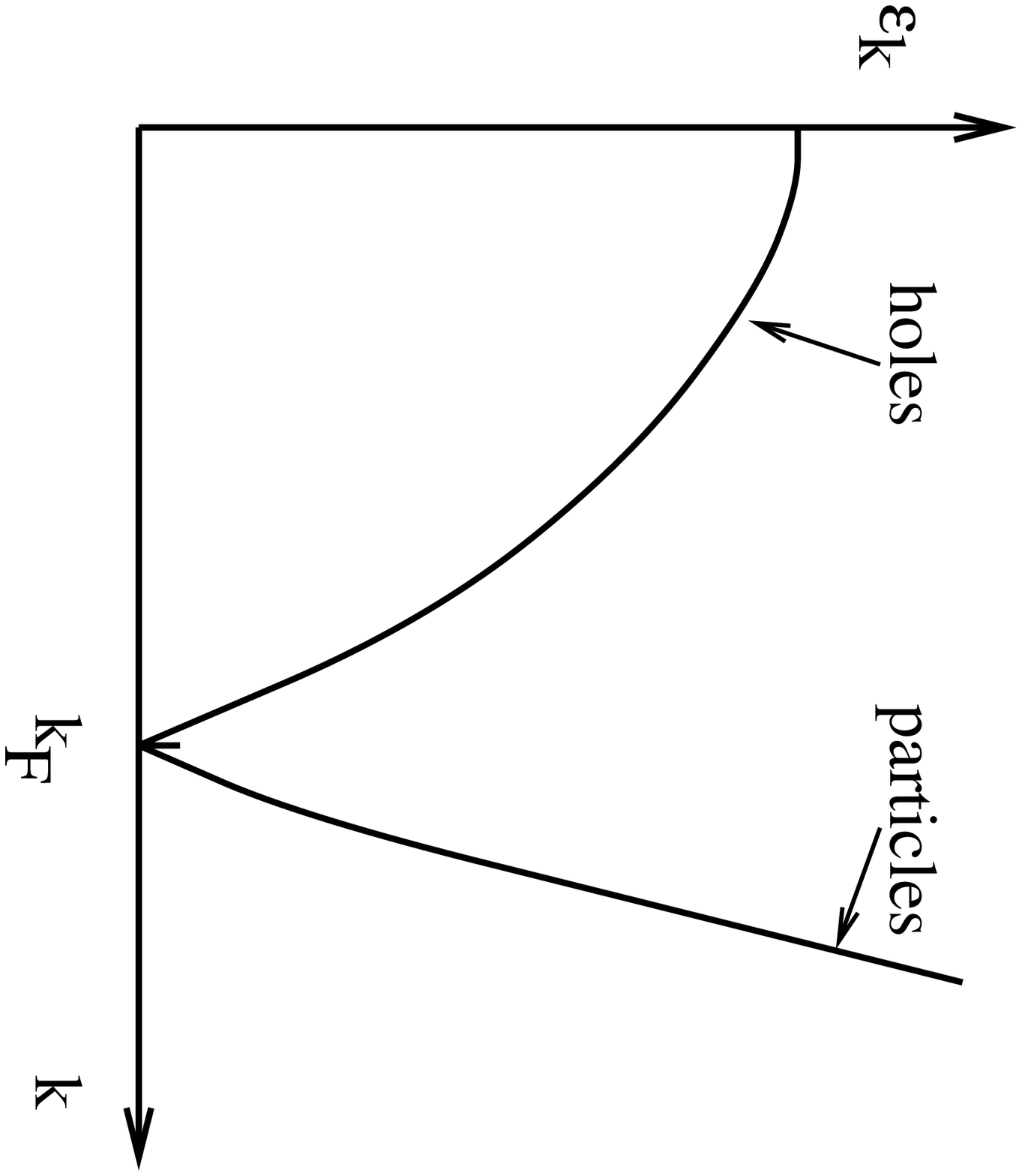}}}
\hspace*{1cm}
\subfigure[]{\epsfysize 6cm \rotate[l]{\epsffile{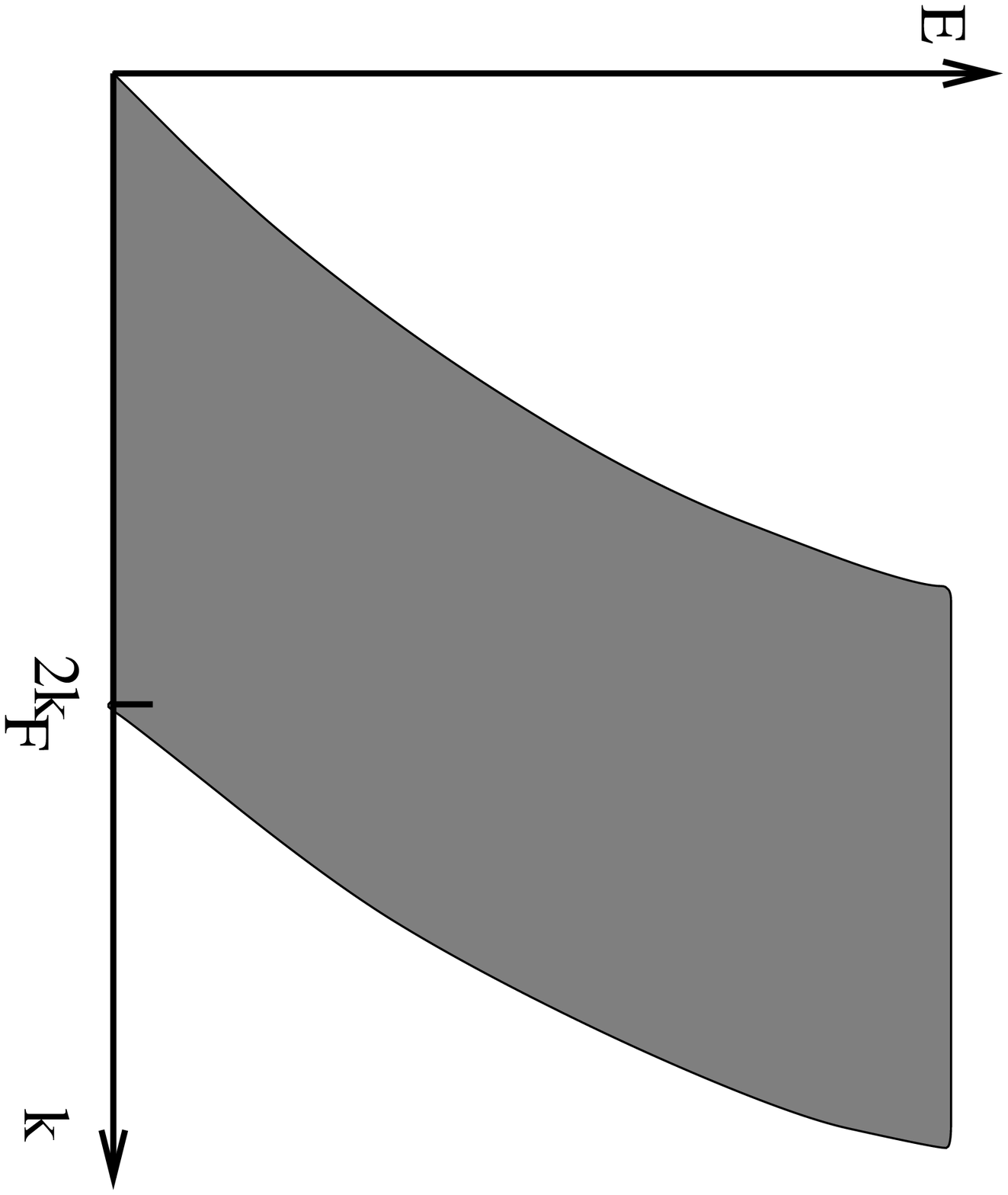}}}}
\SMALLCAP{e}{(a) the energy--momentum relation for the elementary particle ($k>\kf$)
and hole ($k<\kf$) excitations; and (b) the particle--hole continuum.}
\label{f2:1}
\end{figure}

It is usually convenient to define the Hamiltonian in a way so that the
absolute ground state has a well--defined fixed particle number. This is
achieved simply by including the chemical potential $\mu$ in the definition
of the Hamiltonian, i.e. by writing
\begin{equation}
\label{eq:h0}
H = \sum_{\k} \xi_{\k} n_{\k} \virg
\end{equation}
where $n_\k$ is the usual number operator, $\xi_\k = \varepsilon_\k - \mu$,
and the spin summation is not written explicitly (at finite temperature this
of course brings one to the usual grand canonical description where small
fluctuations of the particle number occur). With this definition
of the Hamiltonian, the elementary excitations of the Fermi gas are
\begin{itemize}
\item addition of a particle at wavevector $\k$ ($\delta n_\k=1$). This
requires
$|\k| > \kf$, and thus the energy of this excitation is 
$\varepsilon_\k=\epsilon_\k -
\mu > 0$.
\item destruction of a particle at wavevector $\k$ ($\delta n_\k=-1$),
i.e. creation of a hole. This requires
$|k| < \kf$, and thus the energy is $\varepsilon_k=\mu-\epsilon_\k
 > 0$.
\end{itemize}
The dispersion relation of the elementary particle and hole excitation
is shown in fig.\ref{f2:1}a.
These excitations change the total number of particles. Construction of
states at constant particle number is of course straightforward: one takes
one particle from some state $\k$, with $|k|<\kf$, and puts it into a state
$\k'$, with $|\k'|>\kf$. These {\em particle--hole excitations} are
parameterized by the {\em two} quantum numbers $\k$,$\k'$ and thus form a
continuum, as shown in fig.\ref{f2:1}b. The restriction on the allowed
values of $\k,\k'$ insures that all particle--hole states have positive
energy. Higher excited states, i.e.  states with many particles and many
holes, are straightforwardly constructed, the only restriction being imposed
by the Pauli principle.

Thermodynamic quantities are easily obtained and are all determined by
the density of states at the Fermi energy. For example, the specific
heat obeys the well known linear law $C(T) = \gamma T$, with
\begin{equation}
\label{e:gam}
\gamma = \frac{2\pi^2}{3} N(E_{\rm F}) k_{\rm B}^2
\end{equation}
and similarly the (Pauli) spin susceptibility $\chi$ and the
compressibility $\kappa$ are given by
\begin{eqnarray}
\chi & = & 2 N(E_{\rm F}) \mu_{\rm B}^2 \\
\label{e:kap}
\kappa & = & 2 N(E_{\rm F}) /\rho^2 \point
\end{eqnarray}
Here for the quadratic dispersion relation (\ref{e:ek}) the density of
states (per spin) at the Fermi energy is given by $N(E_{\rm F}) = m
\kf/(2\pi^2)$, but it
should be emphasized that eqs. (\ref{e:gam}) to (\ref{e:kap}) are valid
for an arbitrary density of states, in particular in solids where
bandstructure effects can change the electronic dispersion relation
quite drastically. Thus, for noninteracting electrons one expects the
so--called ``Wilson ratio''
\begin{equation}
R_W = \frac{\pi^2 k_{\rm B}^2}{3 \mu_{\rm B}^2} \frac{\chi}{\gamma}
\end{equation}
to be unity, independently of details of the bandstructure. Any
deviation from unity is necessarily an indication of some form of
interaction effect.

\subsection{Landau's theory of Fermi Liquids}
\subsubsection{Basic hypothesis}
Landau's theory is to a large extent based on the idea of a continuous
and one--to--one correspondence
between the eigenstates (ground state {\em and} excited states) of the
noninteracting and the interacting system. For this to be an acceptable
hypothesis it is crucial that the interactions do not lead to any form
of phase transition or symmetry--broken ground state. 

In particular one can consider a state
obtained by adding a particle (with momentum $|\vec{p}| > \kf$) to the
noninteracting ground state:
\begin{equation} \label{eq:p}
|\vec{p},N+1\rangle = a_{\vec{p}}^+ |0,N\rangle \point
\end{equation}
Here $a_{\vec{p}}^+$ is a fermion creation operator for momentum state
$\vec{p}$, and $|0,N\rangle$ is the $N$--particle ground state of the
noninteracting system. Now we add some form of particle--particle
interaction. In a translationally invariant system, interactions conserve
total momentum, and thus even after switching on the interaction the state
still has total momentum $\vec{p}$. However, the interaction of the added
particle with the filled Fermi sea, as well as the interaction of the
particles in the sea amongst themselves, will change the distribution of
particles in $\k$--space, and of course also modify the energy of our
state. The complex formed by the particle added at $\vec{p}$ and the
perturbed distribution of the other particles is called a {\em Landau
quasiparticle}. The Pauli principle implied $|\vec{p}| > \kf$ in the absence
of interactions, and by the continuity hypothesis the same restriction
remains valid in the interacting case. In particular, the value of $\kf$,
which imposes a lower limit on the allowed momentum of the quasiparticle, is
unchanged by the interactions.

Analogous considerations can be performed
for a state obtained by destruction of a particle
(e.g. creation of a hole):
\begin{equation} \label{eq:ph}
|\vec{p},N-1\rangle = a_{\vec{-p}} |0,N\rangle \point
\end{equation}
Note that due to the momentum $-\vec{p}$ the total momentum of this state is
indeed $\vec{p}$. 

The quasi--particle concept has a certain number of limitations, mainly due
to the fact that, as will be discussed below, the lifetime of a
quasi--particle is finite. However, for excitations close to $\kf$ one has
$1/\tau \propto (\varepsilon -E_{\rm F})^2$, i.e. the lifetime becomes much 
longer than the inverse excitation energy, and the quasi--particles 
therefore are reasonably well defined. In practice,
this means that Landau's theory is useful for phenomena at energy scales
much smaller than the Fermi energy, but inapplicable otherwise. In metals,
where $E_{\rm F} \approx 3\ldots 5 eV$, this restriction is not too serious when
one is concerned with thermodynamic or transport properties.
One should also note that the ground state energy itself has important contributions
from states well below $E_{\rm F}$, and therefore is not accessible to Landau's
theory. 

\subsubsection{Equilibrium properties}
In order to derive physical quantities from the picture of the
low--energy excitations, we need some information about the
energetics of the quasiparticles and of their interactions. To be
specific, starting from the ground state quasiparticle distribution
\begin{eqnarray}
\nonumber
n_0(\k) & = & 1 \; \mbox{if $|\k| < \kf$} \\
\label{eq:n0}
       & = & 0 \; \mbox{if $|\k| > \kf$}
\end{eqnarray}
one considers changes in quasiparticle occupation
number of the the form $n_0(k) \rightarrow n_0(k) + \delta n(k)$, i.e.
$\delta n(k) = 1$ represents an excited quasi--particle, $\delta n(k) =
-1$ an excited quasi--hole (with the notation $k=(\k,\sigma)$, and
$\sigma = \uparrow, \downarrow$ the spin index). The corresponding change
in energy is
\begin{equation}
\label{eq:de}
\delta E = \sum_k \varepsilon_\k^0 \delta n(k)
 + \frac{1}{2\Omega}  \sum_{kk'} f(k,k') \delta n(k) \delta n(k')
\virg
\end{equation}
where the first and second term represent  the energy of a single
quasi--particle and the interaction between quasiparticles,
respectively. To be more precise, we assume that the chemical potential
is included in the Hamiltonian, as in eq.(\ref{eq:h0}). Consequently,
$\varepsilon^0_\k$ vanishes on the Fermi surface, and, given that
we are mainly interested in phenomena in the vicinity of $\kf$, it is
sufficient to retain the lowest order term in an expansion around
$|\k| = \kf$. One thus writes
\begin{equation}
\label{eq:ek0}
\varepsilon_\k^0 = \frac{\kf}{m^*}(|\k| - \kf)
\virg
\end{equation}
thus defining the {\em effective mass} $m^*$ which is different from the
``bare" mass $m$ due to interaction effects that could in
principle be calculated from a microscopic theory of the system.

The energy of a quasi--particle added to the system is easily obtained from
eq.(\ref{eq:de}) by calculating the difference in $\delta E$ between a state
with $\delta n(k) = 1$ and a state with $\delta n(k) = 0$. One finds
\begin{equation} \label{eq:eqp}
\varepsilon_k = \varepsilon^0_\k + \frac{1}{\Omega} \sum_{k'} f(k,k') \delta
n(k') \virg
\end{equation}
i.e. the energy of an added quasi--particle is not just the ``bare''
quasiparticle energy $\varepsilon^0_\k$ but also depends, via the
interaction term, on the presence of the other quasi--particles.
Given that the non--interacting particles obey Fermi--Dirac statistics, the
quasi--particles do so too, and consequently, the occupation probability of a
quasi--particle state is given by
\begin{equation} \label{eq:nqp}
n(k) = \frac{1}{\e^{\beta \varepsilon_k}+1} \point
\end{equation}
Note that the full and not the bare quasi--particle energy enters this
expression. In principle, $n(k)$ thus has to be determined self-consistently
from eqs.(\ref{eq:eqp}) and (\ref{eq:nqp}).

For the subsequent calculations, it is convenient to transform the
quasiparticle interaction $f(k,k')$. First, spin symmetric and
antisymmetric $f$--functions are defined via
\begin{eqnarray}
\nonumber
f(\k\uparrow,\vec{k'}\uparrow) & = &
f^s(\k,\k') + f^a(\k,\k') \\
f(\k\uparrow,\vec{k'}\downarrow) & = &
f^s(\k,\k') - f^a(\k,\k')
\end{eqnarray}
Moreover, given the implicit restrictions of the theory, one is only
interested in processes where all involved particles are very close to the
Fermi surface. Under the assumption that the interaction functions are
slowly varying as a function of $\k$, one then can set $|\k| = |\k'| = \kf$.
Because of rotational symmetry, the $f$--functions then can only depend on
the angle between $\k$ and $\k'$, called $\theta$. One can then expand the
$f$--function in a Legendre series as
\begin{equation} \label{eq:fleg}
f^{a,s}(\k,\k') = \sum_{L=0}^\infty f_L^{a,s} P_L(\cos \theta) \virg
\cos \theta = \frac{\k \cdot \k'}{\kf^2} \virg
\end{equation}
where the $P_L$ are the Legendre polynomials. 
 Finally, one usually puts
these coefficients into dimensionless form by introducing
\begin{equation} \label{eq:FL}
F_L^{a,s} = \frac{\kf m^*}{\pi^2} f_L^{a,s} \point
\end{equation}

We are now in a position to calculate some equilibrium properties. The first
one will be the specific heat at constant volume
\begin{equation} \label{eq:cv}
C_\Omega = \frac{1}{\Omega} \frac{\partial U}{\partial T}
\end{equation}
where $U$ is the internal energy. The temperature--dependent part of $U$
comes from thermally excited quasi--particles, as determined by the
distribution (\ref{eq:nqp}). In principle, in this expression
$\varepsilon_k$ is itself temperature--dependent, because of the temperature
dependent second term in eq.(\ref{eq:eqp}). However, one can easily see
that this term only gives contributions of order $T^2$, and therefore can be
neglected in the low--temperature limit. Consequently, one can indeed
replace $\varepsilon_k$ by $\varepsilon^0_\k$, and then one only has to
replace the bare mass by $m^*$ in the result for a non--interacting system
to obtain
\begin{equation} \label{eq:cv2}
C_\Omega = \frac{m^* \kf}{3} k_{\rm B}^2 T \point
\end{equation}

The spin susceptibility (at $T=0$) is related to the second derivative of the
ground state energy with respect to the (spin) magnetization $M$:
\begin{equation} \label{eq:chi}
\chi = \left[ \Omega \frac{\partial^2 E_0}{\partial M^2} \right]^{-1} \point
\end{equation}
Spin magnetization is created by increasing the number of $\up$ spin particles
and decreasing the number of $\down$ spins ($M = \mu_{\rm B}(N_\up -N_\down)$),
i.e. by changing the Fermi wavevectors for up and down spins: $\kf
\rightarrow \kf + \delta \kf$ for $\sigma = \up$ and $\kf
\rightarrow \kf - \delta \kf$ for $\sigma = \down$. 

By calculating with eq.(\ref{eq:de}) the corresponding change of the 
ground state energy, we obtain from eq.(\ref{eq:chi}): 
\begin{equation} \label{eq:chi1}
\chi = \frac{1}{1+F_0^a} \frac{\mu_{\rm B}^2 \kf m^*}{\pi^2} \virg
\end{equation}
Note that here, and contrary to the specific heat, interactions enter not
only via $m^*$ but also explicitly via the coefficient $F_0^a$, which is 
the only coefficient that appears here because the distortion of the Fermi 
distribution function is antisymmetric in the spin index and has 
spherical symmetry ($L=0$).
The Wilson ratio is then 
\begin{equation} \label{eq:rw1}
R_W = \frac{1}{1 + F_0^a} \point
\end{equation}

Following a similar reasoning, one can calculate the compressibility 
$\kappa$ of a Fermi liquid: 
\begin{equation} \label{eq:ka}
\kappa = -\frac{1}{\Omega} \frac{\partial \Omega}{\partial P} =
\left[ \Omega \frac{\partial^2 E_0}{\partial \Omega^2} \right]^{-1}
=\frac{m^* \kf}{\pi^2 \rho^2 (1+F^s_0)}\point
\end{equation}

It is also interesting that in a translationally invariant system as we have
considered here, the effective mass is not independent of the interaction
coefficients. One can show indeed, by exploiting the Galilean invariance 
of the system, that
\begin{equation} \label{eq:mm}
\frac{m^*}{m} = 1 + F_1^s/3 \point
\end{equation}

\subsubsection{Nonequilibrium properties}
\label{sec:noneq}
As far as equilibrium properties are concerned, Landau's theory is
phenomenological and makes some important qualitative predictions, the most
prominent being that even in the presence of interactions the
low--temperature specific heat remains linear in temperature and that the
spin susceptibility tends to a constant as $T\rightarrow 0$. However,
Landau's theory has little quantitative predictive power because the crucial
Landau parameters have actually to be determined from experiment. The
situation is different for non--equilibrium situations, where the existence
of new phenomena, in particular collective modes, is predicted.  These modes
are another kind of elementary excitations which, contrary to
quasiparticles, involve a coherent motion of the whole system.  We shall not
enter into the details of the treatment of non--equilibrium properties of
Fermi liquids (see refs.\cite{pines_nozieres,kadanoff_baym,baym_pethick})
and just briefly sketch the general conceptual framework and some of the
more important results.  To describe non--equilibrium situations, one makes
two basic assumptions:
\begin{itemize}
\item Deviations from equilibrium are described by a {\em Boltzmann
equation} for a space-- and time-- dependent {\em quasiparticle
distribution function} $n(\k,\r,t)$, which describes the density of
quasi--particles of momentum and spin $(\k,\sigma)$ at point $\r$ and
time $t$. At equilibrium, $n$ is of course given by eq.(\ref{eq:n0}).
The fact that in the distribution function one specifies simultaneously
momentum and position of course imposes certain restrictions, due to
the quantum--mechanical nature of the underlying problem. 
More precisely, spatial and temporal variations of the distribution function
have to be slow compared to the typical wavelength and frequency of the 
quasiparticles. We have then the conditions $v_{\rm F} |\q|, |\omega| 
< E_{\rm F}$, where $\q$ and $|\omega|$ set the scale of the spatial and
temporal variations of $n(\r,t)$.
\item Because of the $\r$--dependent $n$, the quasiparticle energy is
itself, via eq.(\ref{eq:eqp}), $\r$--dependent. One then assumes  the following 
quasi--classical equations of motion
\begin{eqnarray}
\nonumber
\dot{\r} &=& \vec{\nabla}_\k \varepsilon_\k(\r) \\
\label{eq:mot}
\dot{\k} &=& -\vec{\nabla}_\r \varepsilon_\k(\r) \point
\end{eqnarray}
Note in particular that a space--dependent distribution function gives
rise, via the $f(k,k')$ function, to a force acting on a quasiparticle.
\end{itemize}

By linearizing the Boltzmann equation and studying the collisionless 
regime, where the collision term in the Boltzmann equation can be 
neglected, one finds collective mode solutions which correspond to 
oscillations of the Fermi surface. The most important one is the 
longitudinal symmetric mode which, like ordinary sound, involves
fluctuations of the particle density. This kind of sound appears, 
however, in a regime where ordinary sound cannot exist (the existence
of collisions is indeed crucial for the propagation of ordinary sound
waves) and is a purely quantum effect. Since collisions can always be 
neglected at very low temperatures, this new kind of sound has been called by 
Landau {\em zero sound}.
The collision term of the Boltzmann equation is on the contrary essential to 
calculate the quasiparticle lifetime $\tau$. One can find indeed, for
a quasiparticle of energy $\varepsilon_\p$
\begin{equation} \label{eq:tau}
\tau^{-1} \propto m^{*3}
\frac{(\pi T)^2 + \varepsilon_\p^2}{1+\\e^{-\beta \varepsilon_\p}}
\point
\end{equation}
The most important result here is the divergence of the lifetime for low
energies and temperatures as $\tau \propto \max(\varepsilon_\p,T)^{-2}$, so
that the product $\varepsilon_\p \tau$ in fact diverges as the Fermi surface
is approached. This shows that the quasiparticle becomes a well--defined
(nearly--) eigenstate at low excitation energies, i.e. in the region where
Landau's theory is applicable. On the other hand, at higher energies the
quasiparticle becomes less and less well--defined. One may note that
initially we had assumed that a quasiparticle is an {\em exact eigenstate}
of the interacting system, which was obtained from a noninteracting
eigenstate by switching on the interaction, and therefore should have
infinite lifetime. We now arrive at the conclusion that the lifetime is not
strictly infinite, but only very long at low energies. In the following
section we will try to clarify this from a microscopical point of view.

\subsection{Microscopic basis of Landau's theory}
At our current knowledge, it does not seem generally possible to {\em
derive} Landau's theory starting from some microscopic Hamiltonian, apart
possibly in perturbation theory for small interactions. It is however
possible to formulate the basic hypotheses in terms of microscopic
quantities, in particular one-- and two--particle Green functions. This
will be outlined below.

\subsubsection{Quasiparticles}
As far as single particle properties are concerned it is sufficient to
consider the one--particle Green function
\begin{equation} \label{eq:gf1}
{\cal G}(k,\tau) = - \langle T_\tau a_k(\tau) a^+_k(0) \rangle \virg
\end{equation}
where $\tau$ is the usual (Matsubara) imaginary time. In this quantity,
interaction effects appear via self--energy corrections $\Sigma$ in the Fourier
transformed function
\begin{equation} \label{eq:gf2}
{\cal G}(k,\omega) = \frac{1}{{\rm i} \omega - \varepsilon_k^{00}
-\Sigma(k,\omega)} \point
\end{equation}
Here $\varepsilon_k^{00}$ is the bare particle energy, without any
effective mass effects.
Excitation energies of the system then are given by the poles of ${\cal G}
(k,\omega)$. In these terms, Landau's assumption about the existence of
quasiparticles is equivalent to assuming that $\Sigma (k,\omega)$ is
sufficiently regular close to the Fermi surface as to allow an expansion for
small parameters. Regularity in $(\k,\omega)$ space implies that in real
space the self--energy has no contributions that decay slowly in time and/or
space. Given that the self--energy can be calculated in terms of the
effective interaction between particles this is certainly a reasonable
assumption when the particle--particle interaction is short--range (though
there is no formal prove of this). For Coulomb interactions, screening has
to be invoked to make the effective interaction short ranged.

\begin{figure}[htb]
\begin{center}
\mbox{\epsfysize 8cm \rotate[l]{\epsffile{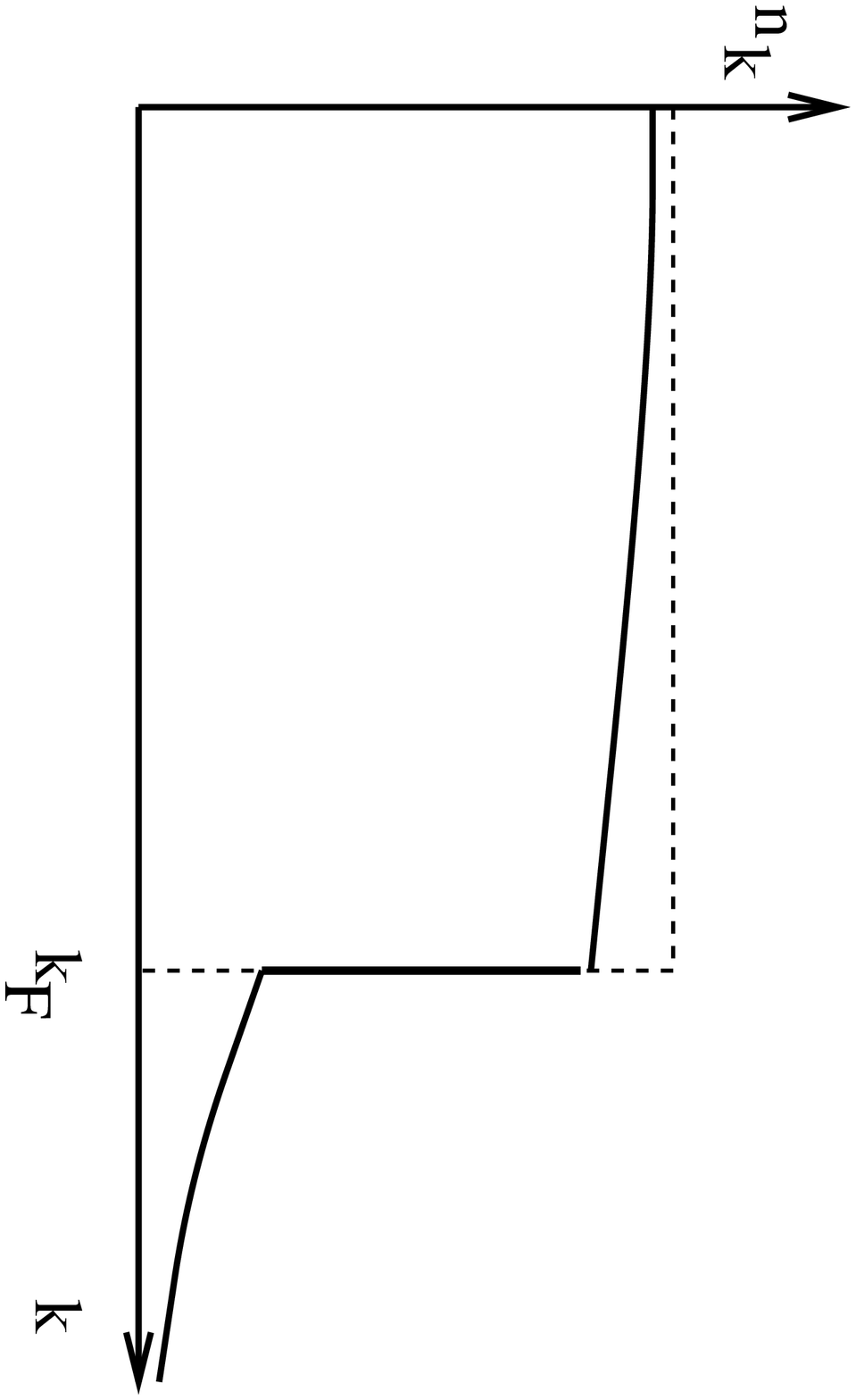}}}
\end{center}
\SMALLCAP{q}{The momentum distribution function $n_\k = \langle a^+_\k a_\k
\rangle$ in the interacting
(full line) and noninteracting (dashed line) cases.}
\label{f2:4}
\end{figure}
One can further notice that $\Sigma(\kf,0)$ just renormalizes the chemical
potential. Given that we want to work at fixed particle number we can absorb
this term in the effective $\mu$. Expanding then to first order around the
Fermi surface, the Green function takes the form
\begin{equation} \label{eq:gf3}
{\cal G} (k,\omega) = \frac{z}{{\rm i}\omega - \varepsilon_\k^0} \point
\end{equation}
where $\varepsilon_\k^0$ has the form (\ref{eq:ek0}) of the phenomenological
approach, with
\begin{equation} \label{eq:meff}
m^* = m \left(1-\frac{\partial \Sigma}{\partial \omega}\right)
\left( 1 + \frac{m}{\kf} \frac{\partial \Sigma}{\partial k}\right)^{-1}
\virg
\end{equation}
and the {\em quasiparticle renormalization factor} is
\begin{equation} \label{eq:z}
z = \left(1-\frac{\partial \Sigma}{\partial \omega}\right)^{-1} \point
\end{equation}
All derivatives are to be taken at the Fermi surface and at $\omega=0$. One
should notice that a sum rule imposes that the frequency--integrated
spectral density
\begin{equation}\label{eq:spec}
A(k,\omega) = -\frac{1}{\pi} {\rm Im}\, {\cal G}(k,i\omega \rightarrow \omega +
i
\delta)
\end{equation}
equals unity. Consequently, in order to fulfill the sum rule, if $z<1$ there
 has to be a contribution in addition to the {\em quasiparticle pole} in
 eq.(\ref{eq:gf3}). This is the so--called ``incoherent background'' from
 single and multiple particle--hole pair excitations which can extend to
 rather high energies but becomes small close to the Fermi surface.

The form (\ref{eq:gf3}) gives rise to a jump in the momentum distribution
function at $\kf$ of height $z$, instead of unity in the noninteracting case
(fig.\ref{f2:4}).  In addition to the jump, the incoherent background gives
rise to a contribution which is continuous through $\kf$.

A finite quasiparticle lifetime arises if the expansion of
$\Sigma(k,\omega)$ is carried to second order. Then eq.(\ref{eq:gf3})
generalizes to
\begin{equation}\label{eq:gf4}
{\cal G} (k,\omega) = \frac{z}{{\rm i}\omega - \varepsilon_\k^0+ {\rm
i} \sgn(\omega) \tau(\omega)^{-1}}
\virg
\end{equation}
where $\tau(\omega)$ is typically given by an expression like
eq.(\ref{eq:tau}). 

\subsubsection{Quasiparticle interaction}
\begin{figure}[hb]
\centering \mbox{
\subfigure[]{\epsfysize 6cm \rotate[r]{\epsffile{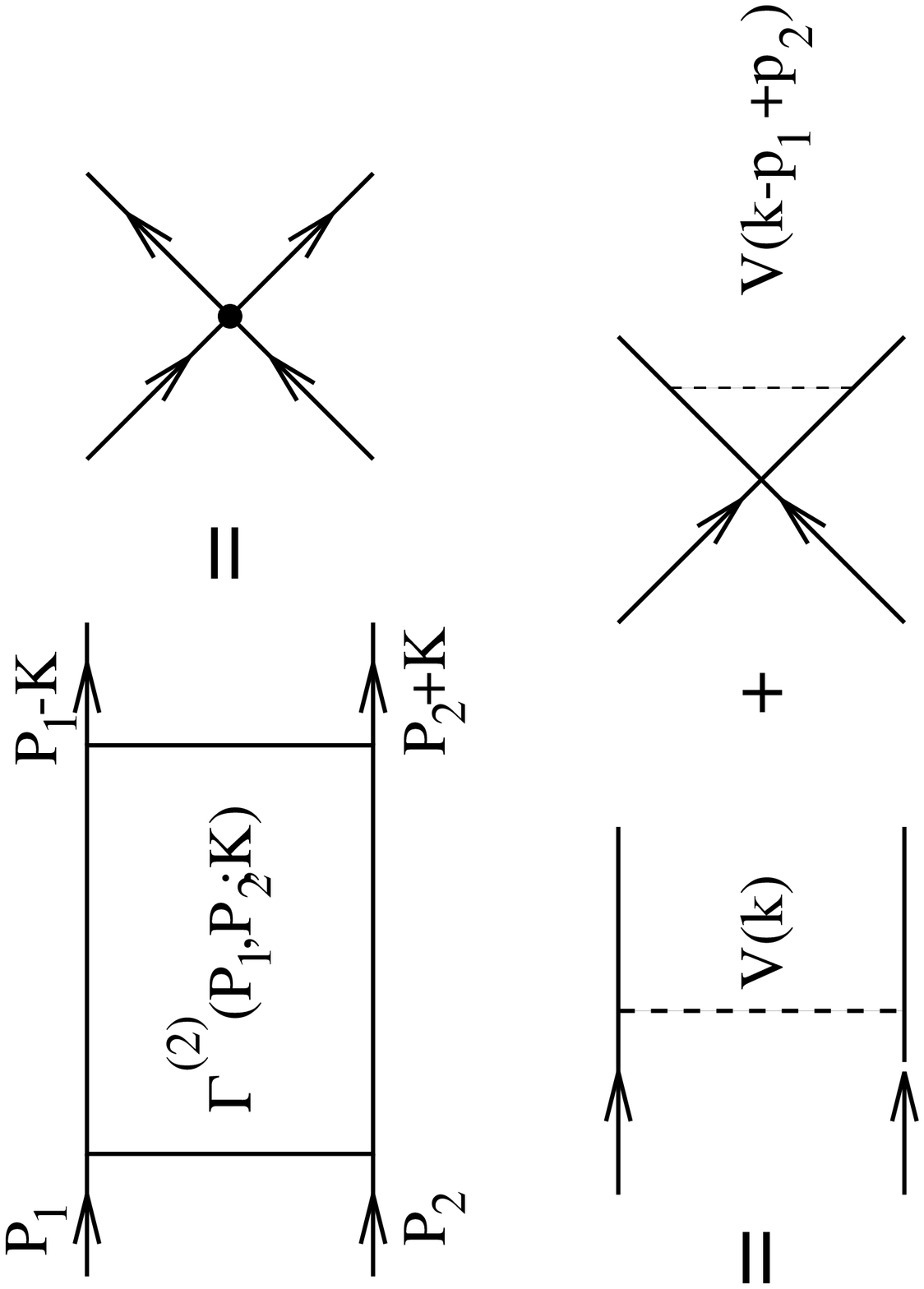}}}
\hspace*{1cm}
\subfigure[]{\epsfysize 6cm \rotate[l]{\epsffile{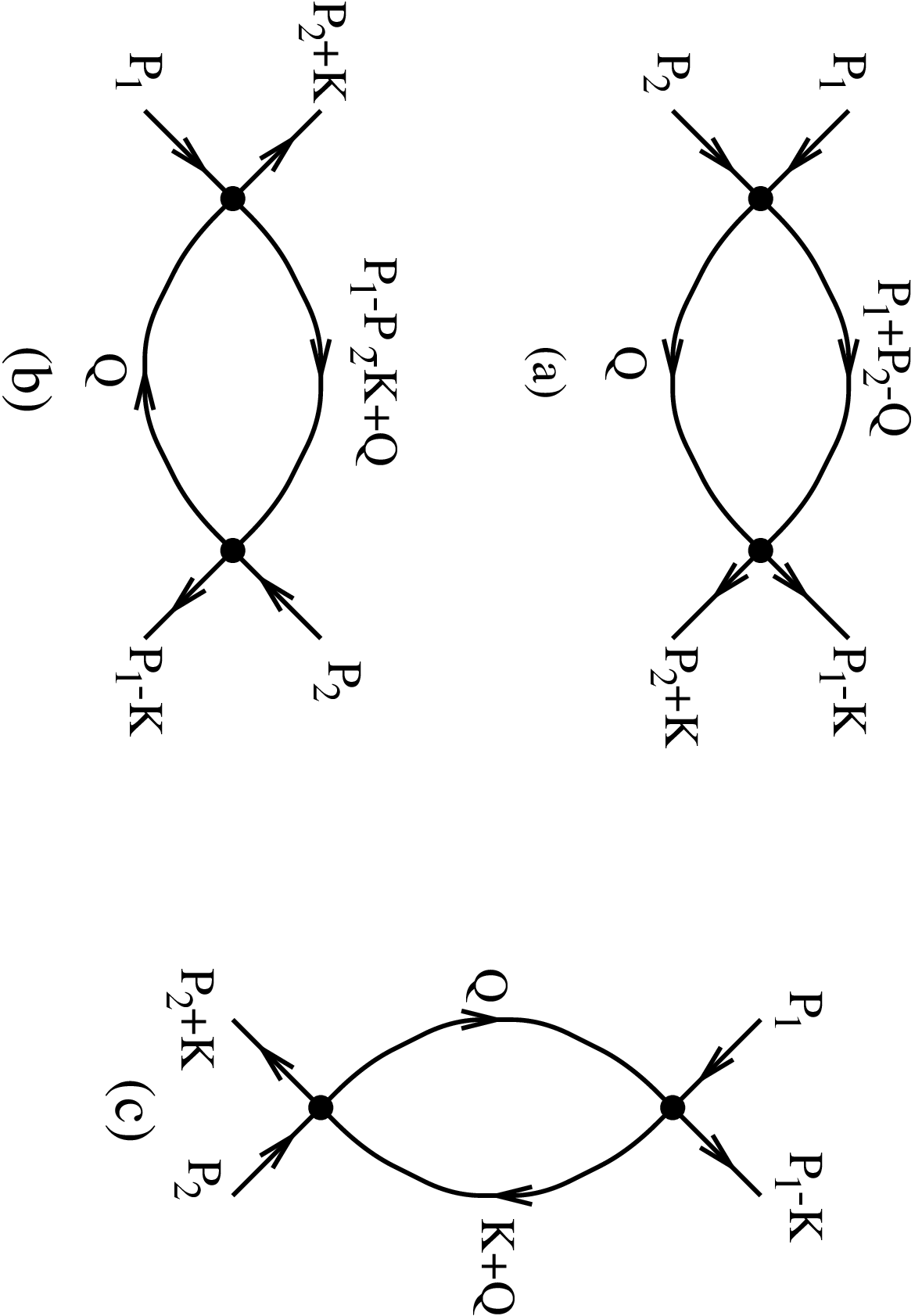}}}}
\SMALLCAP{t}{The lowest (a) and second order (b) contributions to the two--particle vertex
$\Gamma^{(2)}$. Note that the external lines do not represent actual Green
functions but only indicate the external ``connections''}
\label{f2:5}
\end{figure}

The quasiparticle interaction parameters $f(k,k')$ are expected to be
connected to the the two--particle vertex function.  This function, which we
will denote $\Gamma^{(2)}(P_1,P_2;K)$ describes the scattering of two
particles from initial state $P_1,P_2$ to the final state $P_1-K,P_2+K$, and
the notation is $P_i = (\p_i,\omega_i,\sigma_i)$. The contribution of first
and second order in the interaction potential $V(\k)$ are shown in
fig.\ref{f2:5}.
Let us now study the case of small transfer $K$, but arbitrary $P_{1,2}$,
only restricted to be close to the Fermi surface. One then notices that
diagram \ref{f2:5}b (part c) gives rise to singularities, because the poles
of the two intervening Green functions coalesce. On the other hand
diagrams \ref{f2:5}b (part a) and \ref{f2:5}b (part b) remain nonsingular
for small $K$.  This motivates one to introduce a {\em two--particle
irreducible function} $\tilde{\Gamma}^{(2)}$ which is the sum of all
contributions which do not contain a single product ${\cal G}(Q) {\cal
G}(K+Q)$. This function then is nonsingular for small $K$, and consequently
the total vertex function is determined by the integral equation
\begin{equation} \label{eq:gam1}
\Gamma(P_1,P_2;K) = \tilde{\Gamma}(P_1,P_2) +
\frac{1}{(2\pi)^4} \int \d\omega \d^3 q \tilde{\Gamma}(P_1,Q) \G (Q) \G(K+Q)
\Gamma(Q,P_2,K) \point 
\end{equation}
For simplicity, the spin summation is omitted here. The singular
contribution now comes from small $K$ and $Q$ in the vicinity of the Fermi
surface. In this area the $Q$--dependence of the $\Gamma$'s in
eq.(\ref{eq:gam1}) is non--singular and can be neglected. The energy and
radial momentum integral over $Q$ can then be done, leading to
\begin{equation} \label{eq:gam2}
\Gamma(P_1,P_2;K) = \tilde{\Gamma}(P_1,P_2) +
\frac{z^2 \kf^2}{(2\pi)^3} \int \d^2\Omega_\q  \tilde{\Gamma}(P_1,Q) 
\frac{\hat{\q} \cdot \k}{\omega - v_{\rm F} \hat{\q} \cdot \k}
\Gamma(Q,P_2,K) \virg
\end{equation}
where $\hat{\q}$ is a vector on the Fermi surface, and $\d^2\Omega_\q$ is
the corresponding angular integration.
Here only the quasiparticle pole in $\G$ has been taken into account. The
contribution from the incoherent parts can in principle be absorbed into the
definition of $\tilde{\Gamma}$.

The expression (\ref{eq:gam2}) is clearly singular because it has radically
different behavior according to whether one first sends $\k$ or $\omega$ to
zero. In the first case, the limit of $\Gamma$ can be related to the Landau 
$f$--function, while the second one is relevant for the calculation of 
the transition probabilities determining the lifetime of quasiparticles. 
Here we will consider only the former case.
Sending $\k$ to zero first one finds straightforwardly 
\begin{equation} \label{eq:gamo}
\lim_{\omega\rightarrow 0}(\lim_{\k\rightarrow 0} \Gamma) \equiv
\Gamma^\omega(P_1,P_2)  = \tilde{\Gamma}(P_1,P_2) \point
\end{equation}
Closer inspection then shows that in this case the poles of the two
Green functions in eq.(\ref{eq:gam1}) are always on the same side of the
real axis and consequently the singular term in eq.(\ref{eq:gam1}) vanishes.
To make the identification between $\Gamma^\omega(P_1,P_2)$ and the 
Landau $f$--function  we notice that the density
response function at energy--momentum $K$, whose poles give the collective
(zero--sound) modes, contains the interactions via $\Gamma(P_1,P_2,K)$. 
In particular, the existence of a pole in the response function implies a pole
in $\Gamma$. The comparison between the equations for this pole and those 
which can be obtained through the Boltzmann equation within Landau's theory
allows then the identification
\begin{equation} \label{eq:gamf}
f(k,k') = z^2 \Gamma^\omega(k,k') \point
\end{equation}

\subsection{Summary}
The basic assumption of Landau's theory is the existence of low--energy
quasiparticles with a very long lifetime, and their description in terms of
a rather simple energy functional, eq.(\ref{eq:de}). From this a number of
results for thermodynamic properties is obtained. At this level, the theory
is of little quantitative power because the Landau parameters are not
determined. Qualitatively, however, the predictions are important: the
low--temperature thermodynamic properties of an interacting fermion system
are very similar to those of a noninteracting system, the interactions only
lead to quantitative renormalizations.  Actual quantitative predictions are
obtained when one extends the theory to nonequilibrium properties, using the
Boltzmann equation.\cite{pines_nozieres} A new phenomenon predicted (and
actually observed in $\rm^3He$ \cite{abel_zero_sound}) is the existence of
collective excitations, called ``zero sound''. This approach also allows the
calculation of the quasiparticle lifetime and its divergence as the Fermi
energy is approached, as well as the treatment of a number of transport
phenomena.

As already mentioned, the ideal system for the application of Landau's
theory is $\rm ^3He$, which has both short--range interaction and is
isotropic. The application to electrons in metals is more problematic.
First, the interactions are long--ranged (Coulombic). This can however be
accommodated by properly including screening effects. More difficulties, at
least at the quantitative level, arise because metals are naturally
anisotropic. This problem is not of fundamental nature: even when the Fermi
surface is highly anisotropic, an expansion like eq.(\ref{eq:de}) can still
be written down and thus interaction parameters can be defined. However, a
simple Legendre expansion like eq.(\ref{eq:fleg}) is not in general
possible and the description of the quasiparticle interaction in terms of
a few parameters becomes impossible. An exception case, with a very nearly
spherical Fermi surface, are the alkali metals, where a determination of
Landau parameters can indeed be attempted.\cite{pines_nozieres} It should be
noticed that the difficulties with the Landau description of metals are not
of conceptual nature and in particular do not invalidate the quasiparticle
concept but are rather limitations on the usefulness of the theory for
quantitative purposes.

Landau's theory can be interpreted in terms of microscopic quantities like
Green functions (the {\em quasiparticle pole}) and interaction vertices, as
discussed above. It should however be emphasized that these arguments do
provide a microscopic interpretation of Landau's picture, rather than
proving its correctness.  Similar remarks apply to the calculated diverging
quasiparticle lifetime: this at best show that Landau's picture is
internally consistent. Considerable progress towards a deeper formal
understanding of Fermi liquid theory has been made in recent
years.\cite{feldman92,feldman93}

\sectio{Renormalization group for interacting fermions}
In this chapter, we will consider properties of interacting fermions in the
framework of renormalization group theory. This will serve two purposes:
first, the treatment of one--dimensional interacting fermions, which will be
considered in considerable detail in the following chapters, gives rise to
divergences which can only be handled by this approach. Results obtained in
this way will be an essential ingredient in the subsequent discussion of
``Luttinger liquids''. More generally, the renormalization group method will
clarify the status of both Landau's Fermi liquid theory and the Luttinger
liquid picture as renormalization group fixed points, thus establishing a
link with a number of other phenomena in condensed matter physics. We will
formulate the problem in terms of fermion functional integrals, as done by
Bourbonnais in the one--dimensional case \cite{bourbon_couplage} and more
recently for two and three dimensions by Shankar \cite{shankar_revue}. For
the most part, I will closely follow Shankar's notation.

Before considering the interacting fermion problem in detail, let us briefly
recall the general idea behind the renormalization group, as formulated by
Kadanoff and Wilson: one is interested in the statistical mechanics of a
system described by some Hamiltonian $H$. Equilibrium properties then are
determined by the partition function
\begin{equation} \label{eq:z0}
Z = \sum_{\mbox{configurations}} \e^{-\beta H} =
\sum_{\mbox{configurations}} \e^{-S} \virg
\end{equation}
where the second equality defines the {\em action} $S=\beta H$. Typically,
the action
contains degrees of freedom at wavevectors up to some {\em cutoff}
$\Lambda$, which is of the order of the dimensions of the Brillouin
zone. One wishes to obtain an ``effective action'' containing only the
physically most interesting degrees of freedom. In standard phase transition
problems this is the vicinity of {\em the point} $\k=0$, however, for the
fermion problem at hand the {\em surface} $|\k| = \kf$ is relevant, and the
cutoff has to be defined with respect to this surface. In order to
achieve this one proceeds as follows:
\begin{enumerate}
\item Starting from a cutoff--dependent action $S(\Lambda)$ one eliminates
all degrees of freedom between $\Lambda$ and $\Lambda/s$, where $s$ is a
factor larger than unity. This gives rise to a new action $S'(\Lambda' =
\Lambda/s)$.
\item One performs a ``scale change'' $\k \rightarrow s \k$. This brings the
cutoff back to its original value and a new action $S'(\Lambda)$ is 
obtained. 
Because of the degrees of freedom integrated out, coupling
constants (or functions) are changed.
\item One chooses a value of $s$ infinitesimally close to unity: $s =
1+\varepsilon$, and performs the first two steps iteratively. This then
gives rise to differential equations for the couplings, which (in favorable
circumstances) can be integrated until all non--interesting degrees of
freedom have been eliminated.
\end{enumerate}

\subsection{One dimension}
\label{sec:1d}
The one--dimensional case, which has interesting physical applications,
will here be mainly used to clarify the procedure. Let us first consider a
noninteracting problem, e.g. a one--dimensional tight--binding model defined
by
\begin{equation} \label{eq:hti}
H = \sum_k \xi_k a^\dagger_k a^{\phantom{\dagger}}_k \virg \quad \xi_k = -2 t \cos k -\mu \virg
\end{equation}
where $t$ is the nearest--neighbor hopping integral. We will consider the
metallic case, i.e. the chemical potential is somewhere in the middle of the
band. Concentrating on low--energy properties, only states close to the
``Fermi points'' $\pm\kf$ are important, and one can then linearize the
dispersion relation to obtain
\begin{equation} \label{eq:hlin}
H= \sum_{k,r=\pm} v_F(rk-\kf) a^\dagger_{kr} a^{\phantom{\dagger}}_{kr} \virg
\end{equation}
where $v_F= 2 t \sin \kf$ is the Fermi velocity, and the index $r$
differentiates between right-- and left--going particles, i.e. particles
close to $\kf$ and $-\kf$. To simplify subsequent notation, we (i) choose
energy units so that $v_F=1$, (ii) translate $k$-space so that zero energy
is at $k=0$, and (iii) replace the $k$--sum by an integral. Then
\begin{equation} \label{eq:hl2}
H= \sum_{r=\pm} \int_{-\Lambda}^\Lambda \frac{\d k}{2\pi} rk a^\dagger_r(k) a^{\phantom{\dagger}}_r(k)
\point
\end{equation}

For the subsequent renormalization group treatment we have to use a
functional integral formulation of the problem in terms of Grassmann
variables (a detailed explanation of this formalism is given by Negele and
Orland \cite{negele_orland}). The partition function becomes
\begin{equation} \label{eq:zl}
Z(\Lambda) = \int {\cal D}\phi \e^{-S(\Lambda)} \virg
\end{equation}
where ${\cal D}\phi$ indicates functional integration over a set of Grassmann
variables. The action is
\begin{equation} \label{eq:sl1}
S(\Lambda) = \int_0^\beta \d\tau\left\{\sum_{r=\pm} \int_{-\Lambda}^\Lambda \frac{\d
k}{2\pi} \phi_r^*(k,\tau)\partial_\tau\phi^{\phantom{*}}_r(k,\tau) + H(\phi^*,\phi)
\right\} \virg
\end{equation}
where the zero--temperature limit $\beta\rightarrow \infty$ has to be taken,
and $H(\phi^*,\phi)$ indicates the Hamiltonian, with each $a^\dagger$ replaced by
a $\phi^*$, and each $a$ replaced by a $\phi$. Fourier transforming with
respect to the imaginary time variable
\begin{equation} \label{eq:sl2}
\phi_r(k,\tau) = T \sum_{\omega_n} \phi_r(k,\omega_n) \e^{-{\rm
i}\omega_n\tau} \quad (\omega_n = 2\pi(n+ 1/2)T)
\end{equation}
and passing to the limit $T\rightarrow 0$ one obtains the noninteracting action
\begin{equation} \label{eq:sl3}
S_0(\Lambda) = \sum_{r=\pm} \int_{-\infty}^\infty \frac{\d\omega}{2\pi}
\int_{-\Lambda}^\Lambda \frac{\d k}{2\pi} \phi_r^*(k,\omega)[-{\rm
i}\omega+rk] \phi^{\phantom{*}}_r(k,\omega) \point
\end{equation}
We notice that this is diagonal in $k$ and $\omega$ which will greatly
simplify the subsequent treatment. Because of the units chosen, $\omega$ has
units of (length)$^{-1}$ (which we will abbreviate as $L^{-1}$), and then
$\phi_r(k,\omega)$ has units $L^{3/2}$.

We now integrate out degrees of freedom. More precisely, we will integrate
over the strip $ \Lambda/s < |k| < \Lambda$, $-\infty < \omega <\infty$. The
integration over {\em all} $\omega$ keeps the action local in time. One then
has
\begin{equation} \label{eq:zz}
Z(\Lambda) = Z(\Lambda,\Lambda/s) Z(\Lambda/s) \virg
\end{equation}
where $Z(\Lambda,\Lambda/s)$ contains the contributions from the integrated
degrees of freedom, and $Z(\Lambda/s)$ has the same form of
eq.(\ref{eq:zl}). The new action is then $S'_0(\Lambda/s)=S_0(\Lambda/s)$. 
Introducing the scale change
\begin{equation} \label{eq:sca}
k' = ks \virg \quad \omega' = \omega s \virg \quad \phi' = \phi s^{-3/2}
\end{equation}
one easily finds that $S'_0(\Lambda) = S_0(\Lambda)$. The action
does not change therefore under scale change (or {\em renormalization}): we are at a
{\em fixed point}. One should notice that the scale change of $k$ implies
that $k'$ is quantized in in units of $\Delta k' = 2 \pi s/L$, i.e. eliminating
degrees of freedom actually implies that we are considering a shorter
system, with correspondingly less degrees of freedom. This means that even
though the action is unchanged the new $Z(\Lambda)$ is the partition
function of a shorter system. To derive this in detail, one has to take into
account the change in the functional integration measure due to
the scale change on $\phi$.

Before turning to the problem of interactions, it is instructive to consider
a quadratic but diagonal perturbation of the form
\begin{equation} \label{eq:s2}
\delta S_2 = \sum_{r=\pm} \int_{-\infty}^\infty \frac{\d\omega}{2\pi}
\int_{-\Lambda}^\Lambda \frac{\d k}{2\pi} \mu(k,\omega)
\phi_r^*(k,\omega)\phi^{\phantom{*}}_r(k,\omega) \point
\end{equation}
We assume that $\mu(k,\omega)$ can be expanded in a power series
\begin{equation} \label{eq:pow}
\mu(k,\omega) = \mu_{00} + \mu_{10}k+\mu_{01}{\rm i}\omega + \ldots
\end{equation}
Under the scale change (\ref{eq:sca}) one then has
\begin{equation} \label{eq:msca}
\mu_{nm} \rightarrow s^{1-n-m} \mu_{nm} \point
\end{equation}
There now are three cases:
\begin{enumerate}
\item a parameter $\mu_{nm}$ grows with increasing $s$. Such a parameter is
called {\em relevant}. This is the case for $\mu_{00}$.
\item a parameter remains unchanged ($\mu_{10}$,$\mu_{01}$). Such a
parameter is {\em marginal}.
\item Finally, all other parameter decrease with increasing $s$. These are
called {\em irrelevant}.
\end{enumerate}
Generally, one expects relevant parameters, which grow after elimination of
high--energy degrees of freedom, to strongly modify the physics of the
model. In the present case, the relevant parameter is simply a change in
chemical potential, which doesn't change the physics much (the same is true
for the marginal parameters). One can easily see that another relevant
perturbation is a term coupling right-- and left--going particles of the
form $m(\phi^*_1 \phi^{\phantom{*}}_2 +\phi^*_2 \phi^{\phantom{*}}_1)$. This
term in fact does lead to a basic change: it leads to the appearance of a
gap in the spectrum.

Let us now introduce fermion--fermion interactions. The general form of
the interaction term in the action is
\begin{equation}\label{eq:s4}
 S_I = \int_{k\omega} u(1234) \phi^*(1) \phi^*(2) \phi(3) \phi(4)
\point
\end{equation}
Here $\phi(3)$ is an abbreviation  for $\phi_{r_3}(k_3,\omega_3)$, and
similarly for the other factors, while $u$ is an interaction function to be
specified. The integration measure is
\begin{equation}\label{eq:mea}
\int_{k\omega} = \left( \prod_{i=1}^4 \int_{-\infty}^\infty
\frac{\d\omega_i}{2\pi} \int_{-\Lambda}^\Lambda \frac{\d k_i}{2\pi}
\right) \delta(k_1+k_2-k_3-k_4)
\delta(\omega_1+\omega_2-\omega_3-\omega_4) \point
\end{equation}
We now note that the dimension of the integration measure is $L^{-6}$,
and the dimension of the product of fields is $L^6$. This in particular
means that if we perform a series expansion of $u$ in analogy to
eq.(\ref{eq:pow}) the constant term will be $s$--independent, i.e.
marginal, and all other terms are irrelevant. In the following we will
thus only consider the case of a constant ($k$-- and
$\omega$--independent) $u$.

These considerations are actually only the first step in the analysis:
in fact it is quite clear that (unlike in the noninteracting case above)
integrating out degrees of freedom will not in general leave the
remaining action invariant. To investigate this effect, we use a more
precise form of the interaction term:
\begin{equation}
S_I =\int_{k\omega} \sum_{ss'} \left\{ g_1\phi^*_{s+}(1)  \phi^*_{s'-}(2)
\phi^{\phantom{*}}_{s'+}(3)  \phi^{\phantom{*}}_{s-}(4) 
+ g_2 \phi^*_{s+}(1)  \phi^*_{s'-}(2)
\phi^{\phantom{*}}_{s'-}(3)  \phi^{\phantom{*}}_{s+}(4)\right\} \point
\label{eq:g1g2}
\end{equation}
Here we have reintroduced spin, and the two coupling constants
$g_1$ and $g_2$ denote, in the original language of eq.(\ref{eq:hlin}),
backward
($(\kf,-\kf)\rightarrow(-\kf,\kf)$) and forward
($(\kf,-\kf)\rightarrow(\kf,-\kf)$) scattering. Note that in the absence
of spin the two processes are actually identical.

\begin{figure}[htb]
\begin{center}
\mbox{\epsfysize 5cm \epsffile{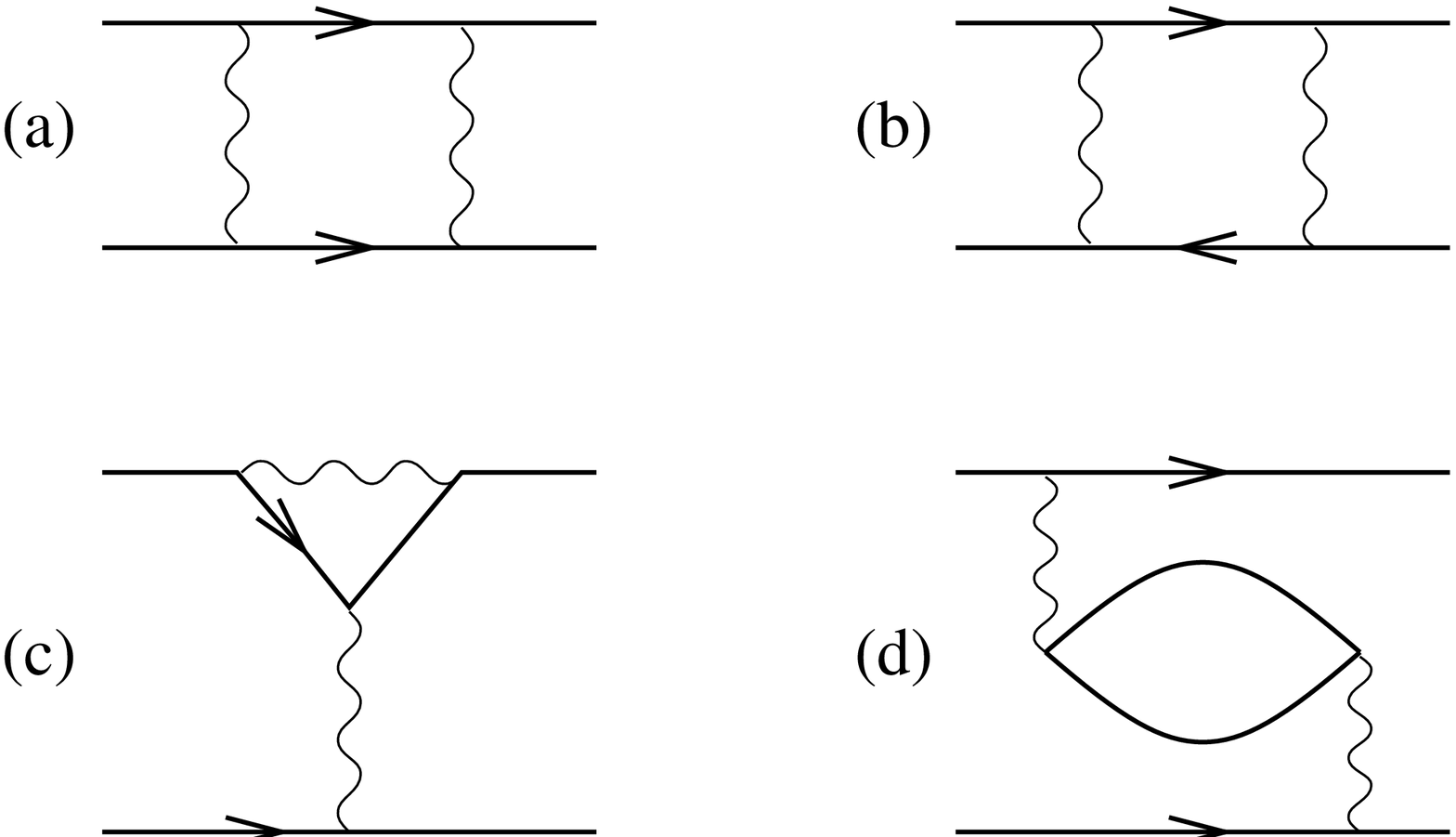}}
\end{center}
\SMALLCAP{t}{The diagrams intervening in the renormalization of the coupling
constants $g_1$ and $g_2$. Note that in (b) the direction of one arrow is
reversed with respect to (a), i.e. this is a particle--hole diagram.}
\label{f3:1}
\end{figure}

Now, the Kadanoff--Wilson type mode elimination can be performed via
\begin{equation}\label{eq:esp}
\e^{-S'} = \int {\cal D} \bar{\phi} \e^{-S} \virg
\end{equation}
where $ {\cal D} \bar{\phi}$ denotes integration only over degrees of
freedom in the strip $\Lambda/s < |k| < \Lambda$. Dividing the field
$\phi$ into $\bar{\phi}$ (to be eliminated) and $\phi'$ (to be kept),
one easily sees that the noninteracting action can be written as $S_0 =
S_0(\phi') + S_0(\bar{\phi})$. For the interaction part, things are a
bit more involved:
\begin{equation}\label{eq:sI}
S_I = \sum_{i=0}^4 S_{I,i} = S_{I,0} + \bar{S}_I \point
\end{equation}
Here $S_{I,i}$ contains $i$ factors $\bar{\phi}$. We then obtain
\begin{equation} \label{eq:spp}
\e^{-S'} = \e^{-S_0(\phi') -S_{I,0}} \int {\cal D} \bar{\phi}
\e^{-S_0(\bar{\phi}) - \bar{S}_I} \point
\end{equation}
Because $\bar{S}_I$ contains up to four factors $\bar{\phi}$, the
integration is not straightforward, and has to be done via a perturbative
expansion, giving 
\begin{equation} \label{eq:sbar}
\int {\cal D} \bar{\phi}
\e^{-S_0(\bar{\phi}) - \bar{S}_I} = Z_0(\Lambda,\Lambda/s) 
\exp\left[-\sum_{i=1}^\infty \frac{1}{n!} \langle \bar{S}_I^n
\rangle_{\bar{0},con} \right] \virg
\end{equation}
where the notation $\langle \ldots \rangle_{\bar{0},con}$ indicates
averaging over $\bar{\phi}$ and only the connected diagrams are to be
counted. It can be easily seen, moreover, that because of the $U(1)$
invariance of the original action associated to the particle number
conservation, terms which involve an odd number of $\phi'$ or $\bar{\phi}$
fields are identically zero. The first order cumulants give corrections to
the energy and the chemical potential and are thus of minor importance. The
important contributions come from the second order term $\langle
S_{I,2}^2\rangle_{\bar{0},con}$ which after averaging leads to terms of the
form $\phi'^* \phi'^* \phi' \phi'$, i.e. to corrections of the interaction
constants $g_{1,2}$. The calculation is best done diagrammatically, and the
four intervening diagram are shown in fig.\ref{f3:1}.

One can easily see that not all of these diagrams contribute corrections to
$g_1$ or $g_2$. Specifically, one has
\begin{eqnarray}
\nonumber
\delta g_1 & \propto &g_1 g_2 [(a) + (c)] + 2 g_1^2 (d) \\ 
\label{eq:dg}
\delta g_2 & \propto &(g_1^2+ g_2^2) (a)  +  g_2^2 (b)  
\end{eqnarray}
where the factor 2 for diagram $(d)$ comes from the spin summation over the
closed loop. Because the only marginal term is the constant in $u(1234)$,
one can set all external energies and momenta to zero. The integration over
the internal lines in diagram $(a)$ then gives
\begin{eqnarray}
\nonumber
(a) & = & \int_s \frac{\d k}{2\pi}\int \frac{\d \omega}{2\pi} \frac{1}{{\rm
i}\omega-k} \frac{1}{-{\rm i}\omega-k} \\
& = &\int_{\Lambda/s}^\Lambda \frac{\d
k}{2\pi} \frac{1}{k} = \frac{1}{2 \pi} \d \ell \virg
\label{eq:inta}
\end{eqnarray}
where $s = 1 +\d \ell$, and similarly the particle--hole diagrams $(b)$ to
$(d)$ give a contribution $- d\ell/(2 \pi)$. Performing this procedure
recursively, using at each step the renormalized couplings of the previous
step, one obtains the renormalization group equations
\begin{equation} \label{eq:rg}
\frac{d g_1}{d \ell} = - \frac{1}{\pi} g_1^2(\ell) \virg \quad \frac{d
g_2}{d \ell} = - \frac{1}{2\pi} g_1^2(\ell) \virg
\end{equation}
where $s= \e^\ell$. These equations describe the effective coupling
constants to be used after degrees of freedom between $\Lambda$ and
$\Lambda \e^\ell$ have been integrated out. As initial conditions one of
course uses the bare coupling constants appearing in eq.(\ref{eq:g1g2}).
Equations (\ref{eq:rg}) are easily solved. The
combination $g_1-2g_2$ is $\ell$--independent, and one has further
\begin{equation} \label{eq:gl}
g_1(\ell) = \frac{g_1}{1+g_1 \ell} \point
\end{equation}
There then are two cases: 
\begin{enumerate} 
\item{Initially, $g_1 \ge 0$. One then renormalizes to the {\em
fixed line} $g_1^* = g_1(\ell \rightarrow \infty) = 0$, $g_2^* = g_2 -g_1/2$,
i.e. one of the couplings has actually vanished from the problem, but there
is still the free parameter $g_2^*$. A case like this, where perturbative
corrections lead to irrelevancy, is called ``marginally irrelevant''.}
\item{ Initially, $g_1 < 0$. Then $g_1$ diverges at some finite value of
$\ell$. We should however notice that, well before the divergence, 
we have left the weak--coupling regime where the perturbative calculation
leading to the eq.(\ref{eq:rg}) is valid. We should thus not overinterpret
the divergence and just remember the renormalization towards strong
coupling. This type of behavior is called ``marginally relevant''.}
\end{enumerate}
We will discuss the physics of both cases in the next section.

Two remarks are in order here: first, had we done a straightforward
order--by--order perturbative calculation, integrals like eq.(\ref{eq:inta})
would have been logarithmically divergent, both for particle--particle and
particle--hole diagrams. This would have lead to inextricably complicated
problem already at the next order. Secondly, for a spinless problem, the
factor $2$ in the equation for $g_1(\ell)$ is replaced by unity. Moreover,
in this case only the combination $g_1-g_2$ is physically meaningful. This
combination then remains unrenormalized.

\subsection{Two and three dimensions}
\begin{figure}[htb]
\centerline{\epsfysize 6cm \epsffile{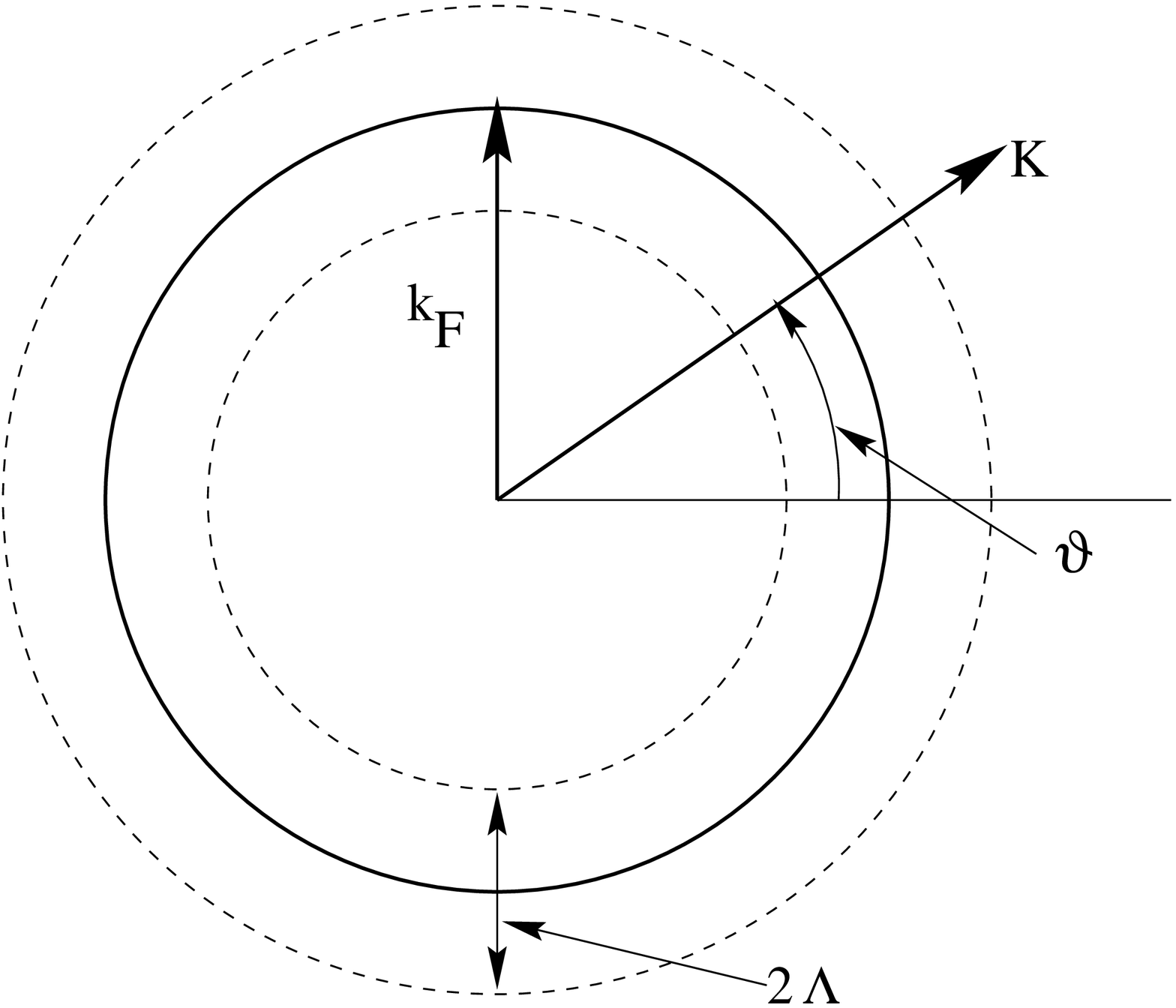}}
\SMALLCAP{t}{Fermi surface geometry in two dimensions.}
\label{f3:2}
\end{figure}

We will now follow a similar logic as above to consider two and more
dimensions. Most arguments will be made for the two--dimensional case, but
the generalization to three dimensions is straightforward. The argument is
again perturbative, and we thus start with free fermions with energy
\begin{equation} \label{eq:xi2d}
\xi_\K = \frac{\K^2}{2m} - \mu = v_F k + O(k^2) \quad (v_F= \kf/m) \point
\end{equation}
We use upper case momenta $\K$ to denote momenta measured from zero, and
lower case to denote momenta measured from the Fermi surface: $k = |\K| -\kf$. 
The Fermi surface geometry now is that of a circle as shown in fig.\ref{f3:2}.
One notices in particular that states are now labeled by two quantum numbers
which one can take as radial ($k$) and angular ($\theta$). Note that the
cutoff is applied around the low--energy excitations at $|\K| - \kf$, not
around $\K=0$. The
noninteracting action then takes the form
\begin{equation} \label{eq:s02d}
S_0 = \kf \int_{-\infty}^\infty \frac{\d\omega}{2\pi} \int_0^{2\pi}
\frac{\d\theta}{2\pi} \int_{-\Lambda}^\Lambda\frac{\d k}{2\pi}
[\phi^*(k\theta\omega) (-i \omega - k) \phi(k\theta\omega)]  \point
\end{equation}
One notices that this is just a (continuous) collection of one--dimensional
action functional, parameterized by the variable $\theta$. The prefactor
$\kf$ comes from the two--dimensional integration measure $\d^2 K = (\kf +k)
\d k \d\theta$, where the extra factor $k$ has been neglected because it is
irrelevant, as discussed in the previous section.

The general form of the interaction term is the same as in the
one--dimensional case
\begin{equation} \label{eq:si2d}
 S_I = \int_{\K\omega} u(1234) \phi^*(1) \phi^*(2) \phi(3) \phi(4) \virg
\end{equation}
however, the integration measure is quite different because of
two--dimensional $\K$--space. Performing the integration over $\K_4$ and
$\omega_4$ in the two--dimensional analogue of eq.(\ref{eq:mea}), the
measure becomes 
\begin{equation} \label{eq:mea2d}
\int_{\K\omega} = \left( \frac{\kf}{2\pi} \right)^3
\left( \prod_{i=1}^3 \int_{-\infty}^\infty \frac{\d\omega_i}{2\pi} \int_0^{2\pi}
\frac{\d\theta_i}{2\pi} \int_{-\Lambda}^\Lambda\frac{\d k_i}{2\pi} \right)
\Theta(\Lambda-|k_4|)
\end{equation}
Here $\K_4 = \K_1 + \K_2 - \K_3$.
Now the step function poses a problem because one easily convinces oneself
that even when $\K_{1,2,3}$ are on the Fermi surface, in general $\K_4$ can
be far away from it. This is quite different from the one--dimensional case,
where everything could be (after a trivial transformation) brought back into
the vicinity of $k=0$.

To see the implications of this point, it is convenient to replace the sharp
cutoff in eq.(\ref{eq:mea2d}) by a soft cutoff, imposed by an exponential:
\begin{equation} \label{eq:soft}
\Theta(\Lambda-|k_4|) \rightarrow \exp(-|k_4|/\Lambda) \point
\end{equation}
Introducing now unit vectors $\omn{i}$ in the direction of $\K_i$ via $\K_i
= (\kf+k_i) \omn{i}$ one obtains
\begin{equation} \label{eq:k4}
k_4 = |\kf(\omn{1}+\omn{2}-\omn{3}) + k_1 \omn{1}+\ldots| - \kf
\approx \kf (|\vec{\Delta}|-1) 
\quad \vec{\Delta} = \omn{1}+\omn{2}-\omn{3} \point
\end{equation}
Now, integrating out variables leaves us with $\Lambda \rightarrow
\Lambda/s$ in eq.(\ref{eq:mea2d}) everywhere, including the exponential
cutoff factor for $k_4$. After the scale change (\ref{eq:sca}) the same form
of the action as before is recovered, with
\begin{equation} \label{eq:up}
u'(k_i',\omega_i',\theta_i') = \e^{-(s-1) (\kf/\Lambda) 
||\vec{\Delta}|-1|}
u(k_i/s,\omega_i/s,\theta_i) \point
\end{equation}
We notice first that nothing has happened to the angular variable, as
expected as it parameterizes the Fermi surface which is not
affected. Secondly, as in the one--dimensional case, the $k$ and $\omega$
dependence of $u$ is scaled out, i.e. {\em only the values $u(0,0,\theta_i)$
on the Fermi surface are of potential interest (i.e. marginal)}. Thirdly, the
exponential prefactor in eq.(\ref{eq:up}) suppresses couplings for which
$|\vec{\Delta}| \neq 1$. This is the most important difference with the
one--dimensional case.

\begin{figure}[htb]
\centering \mbox{
\subfigure[]{\epsfysize 5.5cm \epsffile{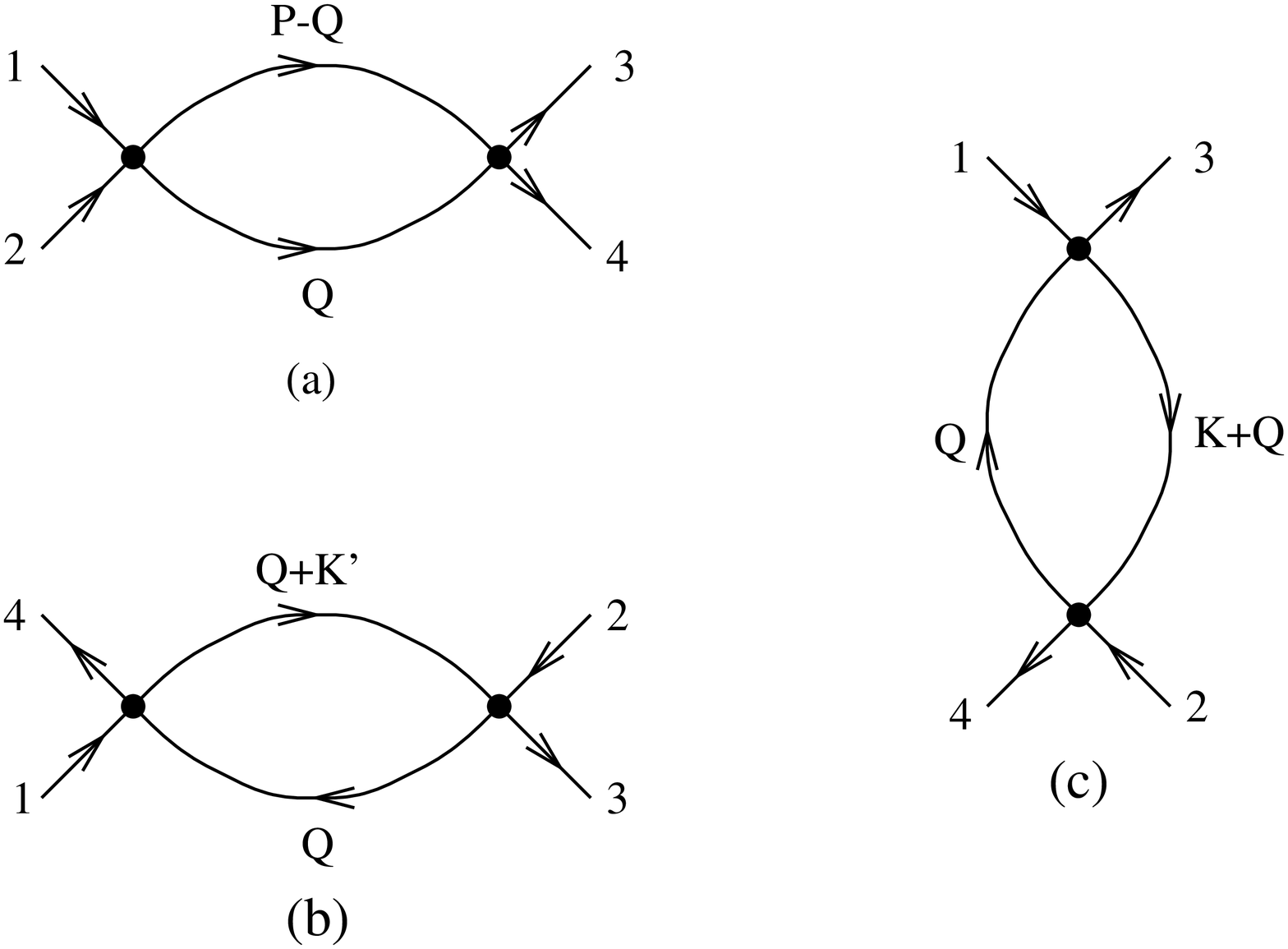}}
\hspace*{0.5cm}
\subfigure[]{\epsfysize 4.5cm \epsffile{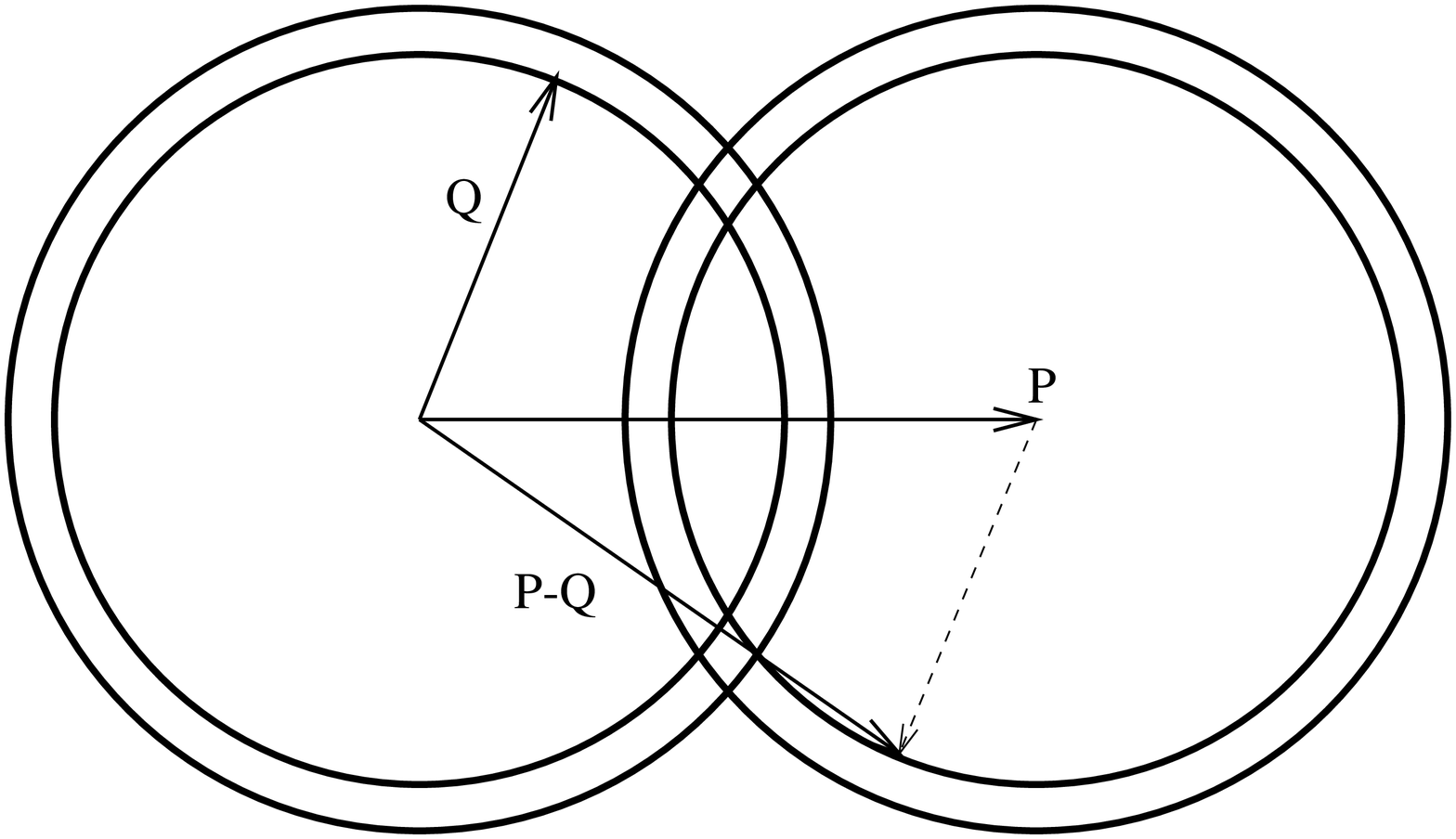}}}
\SMALLCAP{}{(a) second order diagrams renormalizing the coupling function. Here
$\vec{P}=\K_1+\K_2$,$\K_3=\K_1-\K$, and $\K_4=\K_1-\K'$. $\vec{Q}$ is the loop
integration variable. Note that these diagrams are actually identical to
those occurring in Landau's theory (fig.\ref{f2:5}b). (b) phase space for diagram (a). The rings are the degrees of freedom
to be integrated out between $\Lambda$ and $\Lambda/s$. Note that {\em only}
if $\vec{P}=0$ are $\vec{Q}$ and $\vec{P}-\vec{Q}$ simultaneously in the
area to be integrated, giving a contribution of order $\d\ell$.}
\label{f3:3}
\end{figure}

A first type of solution to $|\vec{\Delta}| = 1$ is 
\begin{eqnarray}
\nonumber
\omn{1} = \omn{3} & \Rightarrow & \omn{2} = \omn{4} \virg \mbox{or} \\
\label{eq:pos}
\omn{1} = \omn{4} & \Rightarrow & \omn{2} = \omn{3} \point
\end{eqnarray}
These two cases only differ by an exchange of the two outgoing particles,
and consequently there is a minus sign in the respective matrix
element. Both processes depend only on the angle $\theta_{12}$ between
$\omn{1}$ and $\omn{2}$, and we will write
\begin{equation} \label{eq:uf}
u(0,0,\theta_1,\theta_2,\theta_1,\theta_2) =
-u(0,0,\theta_1,\theta_2,\theta_2,\theta_1) = F(\theta_1-\theta_2) \point
\end{equation}
We can now consider the perturbative contributions to the renormalization of
$F$. To lowest nontrivial (second) order the relevant diagrams are those of
Fermi liquid theory and are reproduced in fig.\ref{f3:3}a.

Consider diagram (a). To obtain a contribution to the renormalization of
$F$, both $\vec{Q}$ and $\vec{P}-\vec{Q}$ have to lie in the annuli to be
integrated out. As can be seen from fig.\ref{f3:3}b, this will give a
contribution of order $\d\ell^2$ and therefore does not contribute to a
renormalization of $F$.  The same is true (if we consider the first case in
eq.(\ref{eq:pos})) for diagram (b). Finally, for diagram (c), because $\K$
is small, the poles of both intervening Green functions are on the same
side of the real axis, and here then the frequency integration gives a zero
result. For the second process in eq.(\ref{eq:pos}) the same considerations
apply, with the roles of diagrams (b) and (c) interchanged. The conclusion
then is that $F$ is not renormalized and remains marginal:
\begin{equation} \label{eq:df}
\frac{\d F}{\d \ell} = 0 \point
\end{equation}

The third possibility is to have $\omn{1} = - \omn{2}$,
$\omn{3}=-\omn{4}$. Then the angle between $\omn{1}$ and $\omn{3}$ can be
used to parameterize $u$:
\begin{equation} \label{eq:uv}
u(0,0,\theta_1,-\theta_1,\theta_3,-\theta_3) = V(\theta_1-\theta_3) \point
\end{equation}
In this case $\vec{P}=0$, and therefore in diagram (a) if $\vec{Q}$ is to be
eliminated, so is $-\vec{Q}$. Consequently, one has a contribution of order
$\d\ell$. For the other two diagrams, one finds again negligible
contributions of order $\d \ell^2$. Thus, one obtains
\begin{equation} \label{eq:rgfv}
\frac{\d V(\theta_1-\theta_3)}{\d \ell} = -\frac{1}{8\pi^2} \int_0^{2\pi}
\frac{\d\theta}{2\pi} V(\theta_1-\theta) V(\theta-\theta_3) \point
\end{equation}
This is a renormalization equation for a function, rather than for a
constant, i.e. one here has an example of a ``functional renormalization
group''. Nevertheless, a Fourier transform
\begin{equation} \label{eq:ftv}
V_\lambda = \int_0^{2\pi} \frac{\d\theta}{2\pi} \e^{i\lambda\theta} V(\theta)
\end{equation}
brings this into a more standard form:
\begin{equation} \label{eq:rgv0}
\frac{\d V_\lambda}{\d \ell} = - \frac{V_\lambda^2}{4\pi} \point
\end{equation}
This has the straightforward solution
\begin{equation} \label{eq:rgv}
V_\lambda = \frac{V_\lambda}{1+V_\lambda\ell/(4\pi)} \point
\end{equation}

From eqs.(\ref{eq:df}) and eq.(\ref{eq:rgv}) there are now two
possibilities:
\begin{enumerate}
\item At least one of the $V_\lambda$ is negative. Then one has a divergence
of $V_\lambda(\ell)$ at some finite energy scale. Given that this equation
only receives contributions from BCS--like particle--particle diagrams, the
interpretation of this as a superconducting pairing instability is
straightforward. The index $\lambda$ determines the relative angular
momentum of the particles involved.
\item All $V_\lambda > 0$. Then one has the fixed point $V_\lambda = 0$,
$F(\theta_1-\theta_2)$ arbitrary. What is the underlying physics of this
fixed point? One notices that here $\theta_3 = \theta_1$,
$\theta_4=\theta_2$, i.e. the marginal term in the action is
$\phi^*_{\theta_1}\phi^*_{\theta_2}\phi^{\phantom{*}}_{\theta_1}\phi^{\phantom{*}}_{\theta_2}$. In the
operator language, this translates into  
\begin{equation} \label{eq:hfp}
H_{int} \approx \int \d\theta_1 \d\theta_2 n_{\theta_1} n_{\theta_2} \point
\end{equation}
We now can recognize this as an operator version of Landau's energy
functional, eq.(\ref{eq:de}). The fixed point theory is thus identified as
{\em Landau's Fermi liquid theory}.
\end{enumerate}

The generalization of the above to three dimensions is rather
straightforward. In addition to the forward scattering amplitudes $F$,
scattering where there is an angle $\phi_{12;34}$ spanned by the planes
$(\omn{1},\omn{2})$ and $(\omn{3},\omn{4})$ is also marginal. For
$\phi_{12;34}\neq 0$ these processes are the ones contributing to the
quasiparticle lifetime, as discussed in sec.\ref{sec:noneq}, however they do
not affect equilibrium properties. The (zero temperature) fixed point
properties thus still only depend on amplitudes for $\phi_{12;34}=0$,
i.e. the Landau $f$--function.

\sectio{Bosonization and the Luttinger Liquid}
\label{wcsec} The Fermi liquid picture described in the preceding two
sections is believed to be relevant for most three--dimensional itinerant
electron systems, ranging from simple metals like sodium to heavy--electron
materials. The best understood example of non--Fermi liquid properties is
that of interacting fermions in one dimension. This subject will be
discussed in the remainder of these lecture notes. We have already started
this discussion in section~\ref{sec:1d}, where we used a perturbative
renormalization group to find the existence of one marginal coupling, the
combination $g_1-2g_2$. This approach, pioneered by S\'olyom and
collaborators in the early 70's \cite{solyom_revue_1d}, can be extended to
stronger coupling by going to second or even third order
\cite{rezayi_3order} in perturbation theory. A principal limitation remains
however the reliance on perturbation theory, which excludes the treatment of
strong--coupling problems. An alternative method, which allows one to
handle, to a certain extent, strong--interaction problems as well, is
provided by the bosonization approach, which will be discussed now and which
forms the basis of the so--called Luttinger liquid description. It should be
pointed out, however, that entirely equivalent results can be obtained by
many--body techniques, at least for the already highly nontrivial case of
pure forward scattering \cite{dzyalo_larkin,evertz_luttinger}.

\subsection{Spinless model: representation of excitations}
\label{sec:bosonspl}
\begin{figure}[htb]
\centering{\mbox{\epsfxsize 5cm{\epsffile{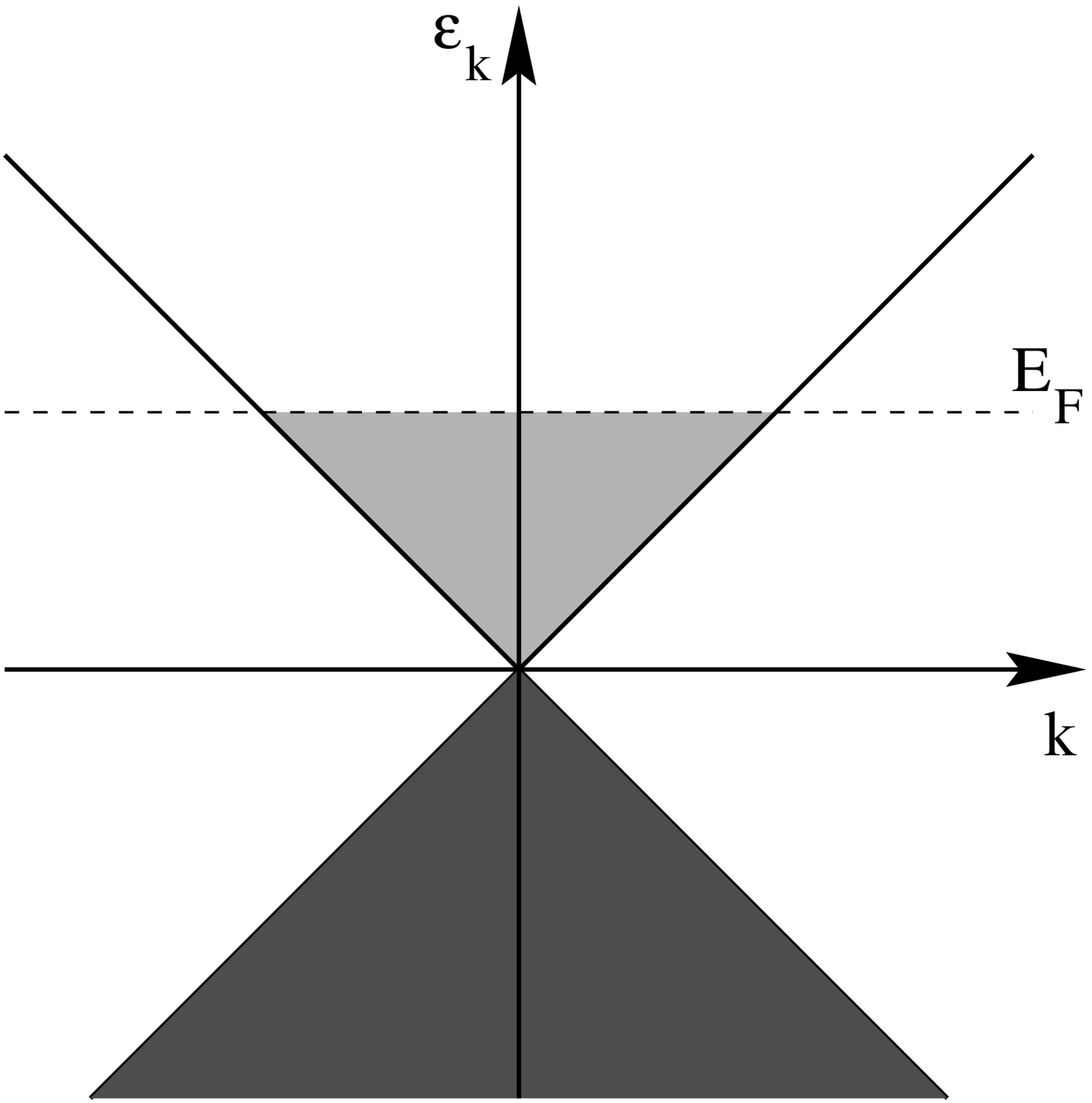}}}}
\SMALLCAP{t}{Single--particle energy spectrum of the Luttinger model. Occupied
states are shown in grey, the dark grey area represents the states added to
make the model solvable.}
\label{f4:1}
\end{figure}

The bosonization procedure can be formulated precisely, in the
form of operator identities, for fermions with a linear energy--momentum
relation, as discussed in section \ref{sec:1d}. To clarify notation, we
will use $a_+$--($a_-$--)operators for right--(left--)moving fermions. The
linearized noninteracting Hamiltonian, eq.(\ref{eq:hlin}) then becomes
\begin{equation}
\label{h0}
H_0 = v_{{\mathrm{F}}} \sum_{k} \left( 
\pippo
(k-k_{{\mathrm{F}}}) a_{+,k}^\dagger a_{+,k} +
 (-k-k_{{\mathrm{F}}}) a_{-,k}^\dagger a^{\phantom{\dagger}}_{-,k} \right)
\virg
\end{equation}
and the density of states is $N(E_{{\mathrm{F}}}) = 1/(\pi
v_{{\mathrm{F}}})$.  In the {\em Luttinger model}
\cite{luttinger_model,mattis_lieb_bos,schulz98}, one generalizes this
kinetic energy
by letting the momentum cutoff $\Lambda$ tend to infinity.  There then are
two branches of particles, ``right movers'' and ``left movers'', both with
unconstrained momentum and energy, as shown in figure~\ref{f4:1}.
At least for weak interaction, this addition of extra states far from the
Fermi energy is not expected to change the physics much. However, this
modification makes the model exactly solvable even in the presence of
nontrivial and possibly strong interactions.  Moreover, and most
importantly, many of the features of this model carry over even to strongly
interacting fermions on a lattice.

We now introduce the Fourier components of the particle density operator for
right and left movers:
\begin{equation}
\rho_\pm(q) = \sum_k a^\dagger_{\pm,k+q} a_{\pm,k} ,
%\rho_-(q) = \sum_k a^\dagger_{-,k+q} a^{\phantom{\dagger}}_{-,k} \;\; ,
\end{equation}
The noninteracting Hamiltonian (and a more general —model including
interactions, see below) can be written in terms of these operators in a
rather simple form and then be solved exactly. This
is based on the following facts:
\begin{enumerate}
\item{the density fluctuation operators $\rho_{\alpha}$, with $\alpha =\pm$,
obey Bose type commutation relations:
\begin{eqnarray}
\label{eq:b1}
\left[ \rho_{\alpha}(-q),\rho_{\alpha'}(q') \right]  = 
\delta_{\alpha \alpha'} \delta_{qq'} \frac{\alpha qL}{2\pi}.
\end{eqnarray}
The relation (\ref{eq:b1})  for $q\neq q'$ or $\alpha\neq\alpha'$
can be derived by straightforward operator algebra. The slightly
delicate part is eq.(\ref{eq:b1}) for $q=q'$. One easily finds
\begin{equation}\label{eq:b3}
[\rho_+(-q),\rho_+(q)] = \sum_k (\hat{n}_{k-q} - \hat{n}_k) \virg
\end{equation}
where $\hat{n}_k$ is an occupation number {\em operator}. In a usual system
with a finite interval of states between $-\kf$ and $\kf$ occupied, the
summation index of one of the $\hat{n}$ operators could be shifted, giving a
zero answer in eq.(\ref{eq:b3}). In the present situation, with an infinity
of states occupied below $\kf$, this is not so.  Consider for example the
ground state and $q>0$. Then each term in eq.(\ref{eq:b3}) with
$\kf<k<\kf+q$ contributes unity to the sum, all other terms vanish, thus
establishing the result (\ref{eq:b1}). More generally, consider a state with
all levels below a certain value $k_0$ ($<\kf$) occupied, but an arbitrary
number of particle hole pairs excited otherwise. One then has, assuming
again $q>0$,
\begin{eqnarray}
\nonumber
\sum_k (\hat{n}_{k-q} - \hat{n}_k) & = & \left(\sum_{k\geq k_0}
  + \sum_{k<k_0} \right) (\hat{n}_{k-q} - \hat{n}_k) \\
\nonumber
& = & \sum_{k\geq k_0} (\hat{n}_{k-q} - \hat{n}_k) \\
\nonumber
& = & \sum_{k\geq k_0-q}\hat{n}_k-\sum_{k\geq k_0}  \hat{n}_k \\
\label{eq:b5}
& = & \sum_{k_0-q\leq k< k_0} \hat{n}_k = \frac{L q}{2\pi}  \point
\end{eqnarray}
The result is independent of $k_0$, and one thus can take the limit
$k_0\rightarrow-\infty$. Together with an entirely parallel argument for
$\rho_-$, this then proves eq.(\ref{eq:b1}).  Moreover, for $q>0$ both
$\rho_+(-q)$ and $\rho_-(q)$ annihilate the noninteracting groundstate.  One
can easily recover canonical Bose commutation relations by introducing
normalized operators, e.g. $\!$ $\tilde{\rho}_+(q) = \sqrt{2\pi/(qL)}
\rho_+(q)$ would be a canonical creation operator, but we won't use this
type of operators in the following.}
\item The noninteracting Hamiltonian obeys a simple commutation relation
with the density operators. For example
\begin{equation}\label{eq:b4}
%[H_0,\rho_+(q)] = v_{{\mathrm{F}}} q \rho_+(q) \virg
\left[ H_0,\rho_{\alpha}(q) \right] = v_{{\mathrm{F}}} \alpha q 
\rho_{\alpha}(q) \virg
\end{equation}
\ie \ states created by $\rho_+(q)$ are eigenstates of $H_0$, with energy
$v_{{\mathrm{F}}} q$.
Consequently, the kinetic part of the Hamiltonian can be re--written as
a term bilinear in boson operators, \ie \ quartic in fermion operators:
\begin{eqnarray}
\label{h0b}
H_0 = \frac{2 \pi v_{{\mathrm{F}}}}{L} \sum_{q>0,\alpha=\pm}
 \rho_{\alpha}(q) \rho_\alpha (-q) \point
\end{eqnarray}
This equivalence may be made more apparent noting that $\rho_+(q)$
creates
particle--hole pairs that all have total momentum $q$. Their energy is
$\varepsilon_{k+q} - \varepsilon_k$, which, because of the linearity of
the spectrum, equals $v_{{\mathrm{F}}} q$, {\em independently of} $k$.
Thus, states
created by  $\rho_+(q)$ are linear combinations of individual
electron--hole excitations all with the same energy, and therefore are
also eigenstates of (\ref{h0}).
\item The above point shows that the spectra of the bosonic and
fermionic representations of $H_0$ are the same. To show complete
equivalence, one also has to show that the degeneracies of all the
levels are identical. This can be achieved calculating the partition
function in the two representations and demonstrating that they are
equal. This then shows that the states created by repeated application
of
$\rho_{\pm}$ on the ground state form a complete set of basis states
\cite{haldane_bosonisation,heidenreich_bosonisation}.
\end{enumerate}

We now introduce interactions between the fermions. As long as only
forward scattering of the type
$(k_{{\mathrm{F}}};-k_{{\mathrm{F}}}) \rightarrow
(k_{{\mathrm{F}}};-k_{{\mathrm{F}}})$ or 
$(k_{{\mathrm{F}}};k_{{\mathrm{F}}}) \rightarrow  
(k_{{\mathrm{F}}};k_{{\mathrm{F}}})$ is introduced, the
model remains exactly solvable.
The interaction Hamiltonian describing these processes takes the form
\begin{equation}
\label{hint}
H_{\mathrm{int}}  =  \frac{1}{2L} \sum_{q,\alpha=\pm} \left(
\pippo
g_2(q) \rho_\alpha(q) \rho_{-\alpha}(-q) +
g_4(q) \rho_\alpha(q) \rho_{\alpha}(-q)\right)
\point
\end{equation}
Here, $g_2(q)$ and $g_4(q)$ are the Fourier transforms of a real space
interaction potential, and in a realistic case one would of course have
$g_2(q) = g_4(q)=g(q)$, but it is useful to allow for differences between
$g_2$ and $g_4$.  For Coulomb interactions one expects $g_2, g_4 > 0$.  In
principle, the long--range part of the Coulomb repulsion leads to a singular
$q$--dependence. Such singularities in the $g_i$ can be handled rather
straightforwardly and can lead to interesting physical effects as will be
discussed below. Here I shall limit myself to nonsingular $g_2, g_4$.
Electron--phonon interactions can lead to effectively attractive
interactions between electrons, and therefore in the following I will not
make any restrictive assumptions about the sign of the constants. One should
however notice that a proper treatment of the phonon dynamics and of the
resulting retardation effects requires more care \cite{voit_phonon}.

Putting together (\ref{h0b}) and (\ref{hint}), the complete interacting
Hamiltonian, the {\em Tomonaga--Luttinger model}, then becomes a bilinear
form in boson operators that is easily diagonalized by a Bogolyubov
transformation. A first consequence is the expression for the excitation
spectrum
\begin{equation}
\label{omq}
\omega(q) = |q| 
\sqrt{
\left (v_{{\mathrm{F}}} + \frac{g_4(q)}{2\pi} \right)^2 - 
\left(\frac{g_2(q)}{2 \pi} \right)^2}
\point
\end{equation}
The diagonal boson operators are linear combinations of the original
$\rho$ operators, and consequently, these elementary excitations are
collective density oscillations, their energy being determined both by
the kinetic energy term and the interactions.

We note here that in order for the Bogolyubov transformation to be a
well--defined unitary transformation, $g_2(q)$ has to decrease at large $q$
at least as $|q|^{-1/2}$.\cite{haldane_bosonisation} On the other hand, the
large--$q$ behavior of $g_2$ is unimportant for the low--energy properties
of the model. We therefore in the following will almost always use a
$q$--independent $g_2$ and $g_4$. An approximate and frequently used way to
cure the divergences arising due to this procedure is to keep a parameter
$\alpha$ in subsequent formulae as a finite short--distance cutoff, of the
order of a lattice spacing.  One can then also include the ``backward
scattering'' $(k_{{\mathrm{F}}};-k_{{\mathrm{F}}})
\rightarrow (-k_{{\mathrm{F}}};k_{{\mathrm{F}}})$, because for
spinless electron this is just the exchange analogue of forward scattering
and does not constitute a new type of interaction. It is worthwhile
emphasizing here that the solution is valid for arbitrarily strong
interactions, no perturbative expansion is needed!

Up to this point, the construction does not allow for a direct
calculation of correlation functions like the one--particle Green
function or more generally any function involving individual creation or
destruction operators. This type of correlation function becomes
tractable by representing single particle operators in terms of the
boson operators. To this end, we introduce the field operators
\begin{eqnarray}
\phi(x) & = & -\frac{ \ii\pi}{L}\sum_{q \ne 0} \frac1{q}
\ee ^{-\alpha |q| /2 - \ii qx} \left( 
\pippo
\rho_+(q) + \rho_-(q) \right) - N \frac{\pi x}{L}
\virg \label{phi0}
\\
\Pi(x) & = & \frac{1}{L}\sum_{q \ne 0}
\ee ^{-\alpha |q| /2 - \ii qx} \left( 
\pippo
\rho_+(q) - \rho_-(q) \right) + \frac{J}{L} 
\label{Pi0}
\\
&& N = N_++N_- \virg \quad J=N_+-N_- \point
\label{eq:NJ}
\end{eqnarray}
Here $N_\pm$ are the numbers of particles added to the ground state on the
right-- and left--moving branch \ie \ 
\begin{eqnarray}
N_\pm = \rho_\pm (q=0) = \sum_k \left( \hat{n}_{\pm, k} - 
\langle \hat{n}_{\pm, k} \rangle_0 \right) . 
\end{eqnarray}
Because addition of a particle changes both
$N$ and $J$, one has the ``selection rule'' $(-1)^N=(-1)^J$;
$\alpha$ is a cutoff parameter which
(at least in principle, see the discussion above) has to be set to zero in
the end of any calculation. 
The fields $\phi$ and $\Pi$ then obey canonical boson
commutation relations:
\begin{equation}
[\phi(x),\Pi(y)] = \ii \delta (x-y) \virg
\end{equation}
and $\phi$ is related to the local particle density via
\begin{equation}
\label{eq:den}
\partial_x \phi = - \pi (\rho(x)-\rho_0) \virg
\end{equation}
where $\rho_0$ is the average particle density in the ground state.
More precisely, in a lattice model this would represent the slowly varying
components ($q\approx 0$) of the density, whereas components with $q\approx
2\kf$ would correspond to crossproducts between $\psi_\pm$.

The expression for the single fermion operators then is
\begin{equation} \label{singlepsi0}
\psi_\pm(x) = \lim_{\alpha \rightarrow 0} \frac1{\sqrt{2\pi
\alpha}} U_\pm \exp \left[
\pm \ii k_{{\mathrm{F}}} x \mp \ii \phi(x) + \ii \theta(x) \right]
\virg
\end{equation}
where the upper and lower sign refer to right-- and left--moving electrons
respectively, and
\begin{eqnarray}
\nonumber
\theta(x) &= & \pi \int^x \Pi (x^{\prime}) \, d x^{\prime} \\
& = & \frac{\ii\pi}{L}\sum_{q \ne 0} \frac{1}{q}
\ee ^{-\alpha |q| /2 - \ii qx} [\rho_+(q) - \rho_-(q)] + J 
\frac{\pi x}{L} \point
\end{eqnarray}
The $U$--operators (sometimes referred to as ``Klein factors''; they are
non--Hermitian: $U_\alpha^\dagger\neq U_\alpha$) decrease the total particle
number on one of the branches by unity and are necessary because the boson
fields all conserve the total particle number. These operators also insure
proper anticommutation between right- and left--going operators: one easily
checks that the ``chiral components'' $\phi_+=\theta-\phi$ and
$\phi_-=\theta+\phi$ commute with each other, one therefore needs the
anticommutation relations $[U_\alpha,U_\beta]_+=0$, etc. In the
thermodynamic limit $L\rightarrow\infty$, the fact that the $U$'s change the
particle number is of minor importance because this represents a shift of
$k_F$ by a quantity of order $1/L$, and one then can replace the
$U_\alpha$'s by Majorana (Hermitian) fermion operators obeying
$[\eta_\alpha,\eta_\beta]_+=0$,\cite{banks_sun} as discussed in more detail
in the appendix. In single--chain problems these effects play a minor role,
however in the many chain systems to be discussed at the end of these notes,
proper account of anticommutation is crucial.

A detailed derivation of the important eq.(\ref{singlepsi0}) as an operator
identity is given in the literature
\cite{haldane_bosonisation,heidenreich_bosonisation}.
However, a simple plausibility argument can be given: creating a particle at
site $x$ requires introducing a kink of height $\pi$ in $\phi$, \ie \ $\phi$
has to be shifted by $\pi$ at points on the left of $x$.  Displacement
operators are exponentials of momentum operators, and therefore a first
guess would be $\psi(x) \approx \exp (\ii \pi \int_{-\infty}^x \Pi(x')
dx')$.  However, this operator commutes with itself, instead of satisfying
canonical {\em anti}commutation relations.  Anticommutation is achieved by
multiplying with an operator, acting at site $x$, that changes sign each
time a particle passes through $x$.  Such an operator is $\exp (\pm \ii
\phi(x))$. The product of these two factors then produces
(\ref{singlepsi0}).

The full Hamiltonian can also be simply expressed in terms of $\phi$ and
$\Pi$. In the long--wavelength limit, neglecting the momentum dependence
 of the $g_i$, one can express the total Hamiltonian 
in the field phase formalism
\begin{equation}
H = H_0 + H_{\mathrm{int}}
= \int dx \left( 
\pippo
\frac{\pi u K}{2} \Pi(x)^2
       + \frac{u}{2\pi K} (\partial_x \phi )^2 \right) \point
\label{hbos0}
\end{equation}
This is obviously just the Hamiltonian of an elastic string, with the
eigenmodes corresponding to the collective density fluctuations of the
fermion liquid. It is important to notice that these collective modes are
the only (low--energy) excited states, and that in particular {\em there are
no well--defined single particle excitations}, nor are there the incoherent
particle--hole pair excitations typical of a Fermi gas. The parameters in
(\ref{hbos0}) are given by
\begin{equation} \label{uk0}
u = \sqrt{\left( v_{{\mathrm{F}}}+\frac{g_4}{2\pi} \right)^2 - 
\left( \frac{g_2}{2\pi} \right)^2} \virg \;
K = \sqrt{ \frac{2\pi v_{{\mathrm{F}}} +g_4 - g_2}{2\pi v_{{\mathrm{F}}} 
+g_4 + g_2}}
\point
\end{equation}
The energies of the eigenstates are $\omega(q) = u |q|$, in agreement with
eq. (\ref{omq}). From the continuity equation, the expression
(\ref{eq:den}) for the local particle density and the equation of motion of
$\phi$ the (number) current density is
\begin{equation} \label{eq:jx}
j(x) = uK \Pi(x) \point
\end{equation}
Note in particular that for $g_2= g_4$ one has 
$u K = v_{{\mathrm{F}}}$, \ie \ the
expression for the current density is interaction--independent. The relation
$u K = v_{{\mathrm{F}}}$ holds in particular for systems with full (Galilean)
translational invariance. On the other hand, in the continuum limit of
lattice systems this relation is in general not true.

The most remarkable result here is the ``collectivization'' of the
dynamics: there are no quasiparticle--like excitations. In fact there is
a rather simple physical picture explaining this: imagine accelerating
one particle a little bit in one direction. Very soon it will hit its
neighbor and transmit its momentum to it, and the neighbor will in turn
transmit its momentum to a further neighbor, and so on. Quite quickly,
the initial localized motion will have spread coherently through the
whole system. This picture can be formalized noting that in one
dimension the difference between a solid and a fluid is not
well--defined: whereas is higher dimensions solids and fluids are
differentiated by the presence or absence of long--wavelength transverse
modes, no transverse modes can exist in a system with movement along
only one direction. The long--wavelength modes thus can equally well be
considered as the phonons of a one--dimensional
crystal \cite{haldane_bosons,emery_magog}. Note that on the contrary in
dimensions larger than one the neighbors of any given particle can be
pushed aside, giving rise to a backflow that allows the particle to move
trough the system more or less freely.

The exclusive existence of collective excitations, at least at low energies,
is one of the typical properties of the {\em Luttinger liquid}. Rather than
discussing the physics of the spinless case in detail, we will turn now to
the more interesting case of fermions with spin.

\subsection{Model with spin; the concept of the Luttinger Liquid}
\label{spinhalfsec}
In the case of spin--1/2 fermions, all the fermion operators acquire an
additional spin index $s$. Following the same logic as above, the kinetic
energy then takes the form
\begin{eqnarray}
\nonumber
H_0 & = & v_{{\mathrm{F}}} \sum_{k,s} \left( \pippo 
(k-k_{{\mathrm{F}}}) a_{+,k,s}^\dagger a_{+,k,s} +
 (-k-k_{{\mathrm{F}}}) a_{-,k,s}^\dagger a^{\phantom{\dagger}}_{-,k,s} 
\right)\\
\label{h0s}
&=& \frac{2 \pi v_{{\mathrm{F}}}}{L} \sum_{q>0,\alpha=\pm,s}
\rho_{\alpha,s}(q) \rho_{\alpha,s}(-q)
%\left( \rho_{+,s}(q) \rho_{+,s}(-q) + \rho_{-,s}(-q) \rho_{-,s}(q) \right)
\virg
\end{eqnarray}
where density operators for spin projections $s=\uparrow,\downarrow$ have
been introduced:
\begin{equation}
\rho_{\pm,s}(q) = \sum_k a^\dagger_{\pm,k+q,s} a_{\pm,k,s} .
%\rho_{-,s}(q) = \sum_k b^\dagger_{k+q,s} b^{\phantom{\dagger}}_{k,s} \;\; \point
\end{equation}
There are now two types of interaction. First, the ``backward scattering''
$(k_{{\mathrm{F}}},s;-k_{{\mathrm{F}}},t)$ 
$ \rightarrow $ $ (-k_{{\mathrm{F}}},s;k_{{\mathrm{F}}},t)$ which for 
$s \neq t$ cannot
be re--written as an effective forward scattering (contrary to the spinless
case). The corresponding Hamiltonian is
\begin{equation}
H_{\mathrm{int},1}  =  \frac{1}{L} \sum_{k,p,q,s,t}
g_1 a_{+,k,s}^\dagger a_{-,p,t}^\dagger 
a^{\phantom{\dagger}}_{+,p+2k_{{\mathrm{F}}}+q,t} 
a^{\phantom{\dagger}}_{-,k-2k_{{\mathrm{F}}}-q,s}
\point
\end{equation}
And, of course, there is also the forward scattering, of a form similar to
the spinless case
\begin{eqnarray}
\label{hint2}
H_{\mathrm{int},2}  & = & \frac{1}{2L} \sum_{q,\alpha,s,t} 
\left( ^{^{^{^{^{^{}}}}}}
g_2(q) \rho_{\alpha,s}(q) \rho_{-\alpha,t}(-q) + 
g_4(q) \rho_{\alpha,s}(q)  \rho_{\alpha,t}(-q) \right)
\point
\end{eqnarray}

To go to the bosonic description, one introduces $\phi$ and $\Pi$ fields for
the two spin projections separately, and then transforms to charge and spin
bosons via $\phi_{\rho,\sigma}= (\phi_\uparrow \pm \phi_\downarrow)/\sqrt2$,
$\Pi_{\rho,\sigma}= (\Pi_\uparrow \pm \Pi_\downarrow)/\sqrt2$.  The
operators $\phi_\nu$ and $\Pi_\nu$ obey Bose--like commutation relations:
\begin{eqnarray}
[\phi_\nu(x),\Pi_{\mu}(y)] = \ii\delta_{\nu\mu}\delta (x-y) \virg 
\end{eqnarray}
and single fermion operators can be written in a form analogous to
(\ref{singlepsi0}):
\begin{equation} \label{singlepsi}
\psi_{\pm,s}(x) = \lim_{\alpha \rightarrow 0} \frac{\eta_{\pm,s}}{\sqrt{2\pi
\alpha}} \exp \left(
{\pm \ii k_{{\mathrm{F}}} x}
-\ii( \pm (\phi_\rho + s \phi_\sigma) + (\theta_\rho+s\theta_\sigma)
\pippo
)
/\sqrt{2} \right)
\virg
\end{equation}
where $\theta_\nu(x) = \pi \int^x \Pi_\nu(x') dx'$.

The full Hamiltonian $H = H_0 + H_{int,1} + H_{int,2}$ then takes the form
\begin{equation} \label{hbos}
H = H_\rho + H_\sigma + \frac{2g_1}{(2\pi \alpha )^2}
\int dx \cos (\sqrt8 \phi_\sigma ) \point
\end{equation}
Here $\alpha$ is a short--distance cutoff,
and for $\nu = \rho , \sigma $
\begin{equation} \label{hnu}
 H_\nu = \int dx \left(
 \pippo
 \frac{\pi u_\nu K_\nu}{2} \Pi_\nu^2
       + \frac{u_\nu}{2\pi K_\nu} (\partial_x \phi_\nu )^2 \right) \virg
\end{equation}
with
\begin{equation}
\label{uks}
u_\nu = \sqrt{\left( v_{{\mathrm{F}}}+\frac{g_{4,\nu}}{\pi} 
\right)^2 - \left( \frac{g_\nu}{2\pi}\right)^2} \virg \;
K_\nu = \sqrt{ \frac{2\pi v_{{\mathrm{F}}} +2g_{4,\nu}+ g_\nu}{2\pi 
v_{{\mathrm{F}}} +2g_{4,\nu}-g_\nu}} \virg
\end{equation}
and $g_\rho = g_1 -2g_2$, $g_\sigma = g_1$, $g_{4,\rho} = g_4$,
$g_{4,\sigma} = 0$. The choice of sign (which is the conventional one) for
the cosine--term in eq.(\ref{hbos}) corresponds to a particular ``gauge
choice'', as discussed in the appendix. 

For a noninteracting system one thus has $u_\nu = v_{{\mathrm{F}}}$ (charge
and spin velocities equal!)  and $K_\nu=1$. For $g_1=0$, (\ref{hbos})
describes independent long-wavelength oscillations of the charge and spin
density, with linear dispersion relation $\omega_\nu(k) = u_\nu |k|$, and
the system is conducting. As in the spinless case, there are no
single--particle or single particle--hole pair excited states.  This model
(no backscattering), usually called the Tomonaga--Luttinger model, is the
one to which the bosonization method was originally applied
\cite{luttinger_model,mattis_lieb_bos,tomonaga_model}.

For $g_1 \neq 0$ the cosine term has to be treated perturbatively.
We have already obtained the corresponding renormalization group equations
in the previous section (eq.(\ref{eq:rg})). In particular,
 for repulsive interactions ($g_1 >
0$), $g_1$ is renormalized to zero in the long-wavelength limit, and
at the fixed point one has $K_\sigma^* = 1$. The three remaining
parameters in (\ref{hbos}) then completely determine the long-distance
and low--energy properties of the system.

It should be emphasized that (\ref{hbos}) has been derived here for fermions
with linear energy--momentum relation. For more general (e.g. $\!$ lattice)
models, there are additional operators arising from band curvature and the
absence of high--energy single--particle states \cite{haldane_bosonisation}.
One can however show that all these effects are, at least for not very
strong interaction, irrelevant in the renormalization group sense, \ie \ they
do not affect the low--energy physics. Thus, {\em(\ref{hbos}) is still the
correct effective Hamiltonian for low--energy excitations}.  The lattice
effects however intervene to give rise to ``higher harmonics'' in the
expression for the single--fermion operators, \ie \ there are low energy
contributions at wavenumbers $q \approx (2m+1) k_{{\mathrm{F}}}$ for 
arbitrary integer $m$ \cite{haldane_bosons}.

The Hamiltonian (\ref{hbos}) also provides an explanation for the physics of
the case of negative $g_1$, where the renormalization group scales to strong
coupling (eq.(\ref{eq:rg})). In fact, if $|g_1|$ is large in
(\ref{hbos}), it is quite clear that the elementary excitations of
$\phi_\sigma$ will be small oscillations around one of the minima of the
cosine term, or possibly soliton--like objects where $\phi_\sigma$ goes from
one of the minima to the other. Both types of excitations have a gap, i.e.
for $g_1 < 0$ one has a {\em gap in the spin excitation spectrum}, whereas
the charge excitations remain massless. This can actually investigated in
more detail in an exactly solvable case \cite{luther_exact}. We will
subsequently concentrate on the case $g_1>0$, so that there is no spin gap,
however investigations of spectral functions as described below have also
been recently performed for the case with a spin gap.\cite{voit98}

\subsubsection{Spin--charge separation}
One of the more spectacular consequences of the Hamiltonian (\ref{hbos})
is the complete separation of the dynamics of the spin and charge
degrees of freedom. For example, in general one has $ u_\sigma \neq
u_\rho$, \ie \ the charge and spin oscillations propagate with
different velocities. Only in a noninteracting system or if some
accidental degeneracy occurs one does 
have $u_\sigma = u_\rho = v_{{\mathrm{F}}}$. 
To make the meaning of this fact more
transparent, let us create an extra particle in the ground state, at
$t=0$ and spatial coordinate $x_0$. The charge and spin densities then
are easily found, using
 $\rho(x) = -(\sqrt2/\pi) \partial_x \phi_\rho$
 (note that $\rho(x)$ is the deviation of the
density from its average value) and
 $\sigma_z(x) = -(\sqrt2/\pi) \partial_x \phi_\sigma$
:
\begin{eqnarray}
\nonumber
\langle 0 | \psi_+(x_0) \rho(x) \psi^\dagger_+(x_0) | 0 \rangle & = &
\delta(x-x_0) \virg \\
\langle 0 | \psi_+(x_0) \sigma_z(x) \psi^\dagger_+(x_0) | 0 \rangle & =
& \delta(x-x_0) \point
\end{eqnarray}
Now, consider the time development of the charge and spin
distributions. The time--dependence of the charge and spin density
operators is easily obtained from  (\ref{hbos}) (using the fixed point
value $g_1 = 0$), and one obtains
\begin{eqnarray}
\nonumber
\langle 0 | \psi_+(x_0) \rho(x,t) \psi^\dagger_+(x_0) | 0 \rangle & = &
\delta(x-x_0-u_\rho t) \virg \\
\langle 0 | \psi_+(x_0) \sigma_z(x,t) \psi^\dagger_+(x_0) | 0 \rangle &
= & \delta(x-x_0-u_\sigma t) \point
\end{eqnarray}
%\marginpar{left--moving part?}
Because in general $u_\sigma \neq u_\rho$, after some time charge and
spin will be localized at completely different points in space, i.e.
{\em charge and spin have separated completely}. 
%An interpretation of this surprising phenomenon in terms of the Hubbard model will be given in section~(\ref{hubsec}).

Here a linear energy--momentum relation has been
assumed for the electrons, and consequently the shape of the charge
and spin distributions is time--independent. If the energy--momentum
relation has some curvature (as is necessarily the case in lattice
systems) the distributions will widen with time. However this widening
is proportional to $\sqrt{t}$, and therefore much smaller than the
distance
between charge and spin. Thus, the qualitative picture of spin-charge
separation is unchanged.

\subsubsection{Physical properties}
The simple form of the Hamiltonian (\ref{hbos}) at the fixed point
$g_1^*=0$ makes the calculation of physical properties rather
straightforward. The specific heat now is determined both by the charge and
spin modes, and consequently the specific heat coefficient $\gamma$ is given
by
\begin{equation}       \label{gamma}
\gamma/\gamma_0 = \frac12 (v_{{\mathrm{F}}}/u_\rho + v_{{\mathrm{F}}} / 
u_\sigma) \point
\end{equation}
Here $\gamma_0$ is the specific heat coefficient of noninteracting
electrons of Fermi velocity $v_{{\mathrm{F}}}$.

The spin susceptibility and the compressibility are equally easy to
obtain. Note that in
(\ref{hbos}) the coefficient $u_\sigma / K_\sigma$ determines the energy
necessary to create a nonzero spin polarization, and, as in the spinless
case, $u_\rho
/ K_\rho$ fixes the energy needed to change the particle density. Given
the fixed point value $K_\sigma^*=1$, one finds
\begin{equation}  \label{chi}
\chi / \chi_0 = v_{{\mathrm{F}}}/u_\sigma \virg \quad \kappa / 
\kappa_0 = v_{{\mathrm{F}}} K_\rho
/ u_\rho \virg
\end{equation}
where $\chi_0$ and $\kappa_0$ are the susceptibility and
compressibility of the noninteracting
case. From eqs.(\ref{gamma}) and (\ref{chi}) the Wilson ratio is
\begin{equation}   \label{rw}
R_{\mathrm{W}} = \frac{\chi }{ \gamma} \frac{\gamma_0}{\chi_0} =
\frac{2 u_\rho}{u_\rho +u_\sigma } \point
\end{equation}

The quantity $\Pi_\rho(x)$ is proportional to the current density.
As before, the Hamiltonian commutes with the total current, one thus has
\begin{equation}  \label{sig0}
\sigma (\omega) = 2 K_\rho u_\rho \delta (\omega ) +
\sigma_{\mathrm{reg}}(\omega) \virg
\end{equation}
\ie \ the product $K_\rho u_\rho $ determines the weight of the {\sc dc} peak in
the conductivity. If the total current commutes with the Hamiltonian
$\sigma_{\mathrm{reg}}$ vanishes, however more generally this part of the
conductivity varies as $\omega^3$ at low frequencies \cite{giamarchi_millis}.

The above properties, linear specific heat, finite spin susceptibility,
and {\sc dc} conductivity are those of an ordinary Fermi liquid, the
coefficients $u_\rho, u_\sigma$, and $K_\rho$ determining
renormalizations with respect to noninteracting quantities. However, the
present system is {\em not a Fermi liquid}. This is in fact already
obvious from the preceding discussion on charge--spin separation,
and can be made more precise considering
the single--particle Green function. Using the
representation (\ref{singlepsi}) of fermion operators one  finds
(at the fixed point $g_1 = 0$)
\begin{eqnarray}
\nonumber
G^{{\mathrm{R}}}(x,t) & = & -\ii \heavi (t) \left\langle
\pippo
\left[ \psi_{+,s}(x,t),\psi_{+,s}^\dagger(0,0) \right]_+ \right\rangle \\
\label{eq:gs}
& = & -\frac{\heavi (t)}{\pi} \ee ^{\ii k_{{\mathrm{F}}}x} {\rm Re}
\left( \frac{1}{\sqrt{(u_\rho t -x) (u_\sigma t -x)}}
  \left[\frac{\alpha^2}{(\alpha+\ii u_\rho t)^2 + x^2} \right]^{\delta/2}
   \right),
\end{eqnarray}
where the presence of the Heaviside function ensures the retarded nature of
$G^{{\mathrm{R}}}$ and $[,]_+$ denotes the anticommutation of fermion
operators.  Note that this expression factorizes into a spin and a charge
contribution which propagate with different velocities. Fourier
transforming (\ref{eq:gs}) gives the momentum distribution function in the
vicinity of $k_{{\mathrm{F}}}$:
\begin{equation}
n_k \approx n_{k_{{\mathrm{F}}}} -  {\rm const.} \times
 {\rm sign}(k-k_{{\mathrm{F}}}) |k-k_{{\mathrm{F}}}|^\delta \;\; \virg
\end{equation}
and for the single-particle density of states (\ie \ the
momentum--integrated spectral density) one finds:
\begin{equation}
\label{spec}
N(\omega ) \approx |\omega |^\delta \point
\end{equation}
In both cases $\delta=(K_\rho + 1/K_\rho -2)/4$.  Note that for any $K_\rho
\neq 1$, \ie \ {\em for any nonvanishing interaction}, the momentum
distribution function and the density of states have power--law
singularities at the Fermi level, with a vanishing single particle density
of states at $E_{{\mathrm{F}}}$. This behavior is obviously quite different
from a standard Fermi liquid which would have a finite density of states and
a step--like singularity in $n_k$. The absence of a step at
$k_{{\mathrm{F}}}$ in the momentum distribution function implies the {\em
absence of a quasiparticle pole} in the one--particle Green function. In
fact, a direct calculation of the spectral function $A(k,\omega)$ from
(\ref{eq:gs}) \cite{voit_spectral,meden_spectral} shows that the usual
quasiparticle pole is replaced by a continuum, with a lower threshold at
$\min_\nu(u_\nu(k-k_{{\mathrm{F}}}))$ and branch cut singularities at
$\omega = u_\rho p$ and $\omega = u_\sigma p$:
\begin{eqnarray}
A(k,\omega) &\approx (\omega - u_\sigma (k-k_{{\mathrm{F}}}))^{\delta-1/2} 
%\virg
& \left|\omega - u_\rho ^{^{^{}}}(k-k_{{\mathrm{F}}})\right|^{(\delta-1)/2} 
\quad (u_\rho > u_\sigma)
\virg \\
A(k,\omega) &\approx (\omega - u_\rho (k-k_{{\mathrm{F}}}))^{(\delta-1)/2} 
%\virg
& \left|\omega - u_\sigma ^{^{^{}}}(k-k_{{\mathrm{F}}})\right|^{\delta-1/2} \quad (u_\rho < u_\sigma)
\point
\end{eqnarray}

\begin{figure}[htb]
\begin{center}
\centerline{ \epsfxsize 6.5cm \epsffile{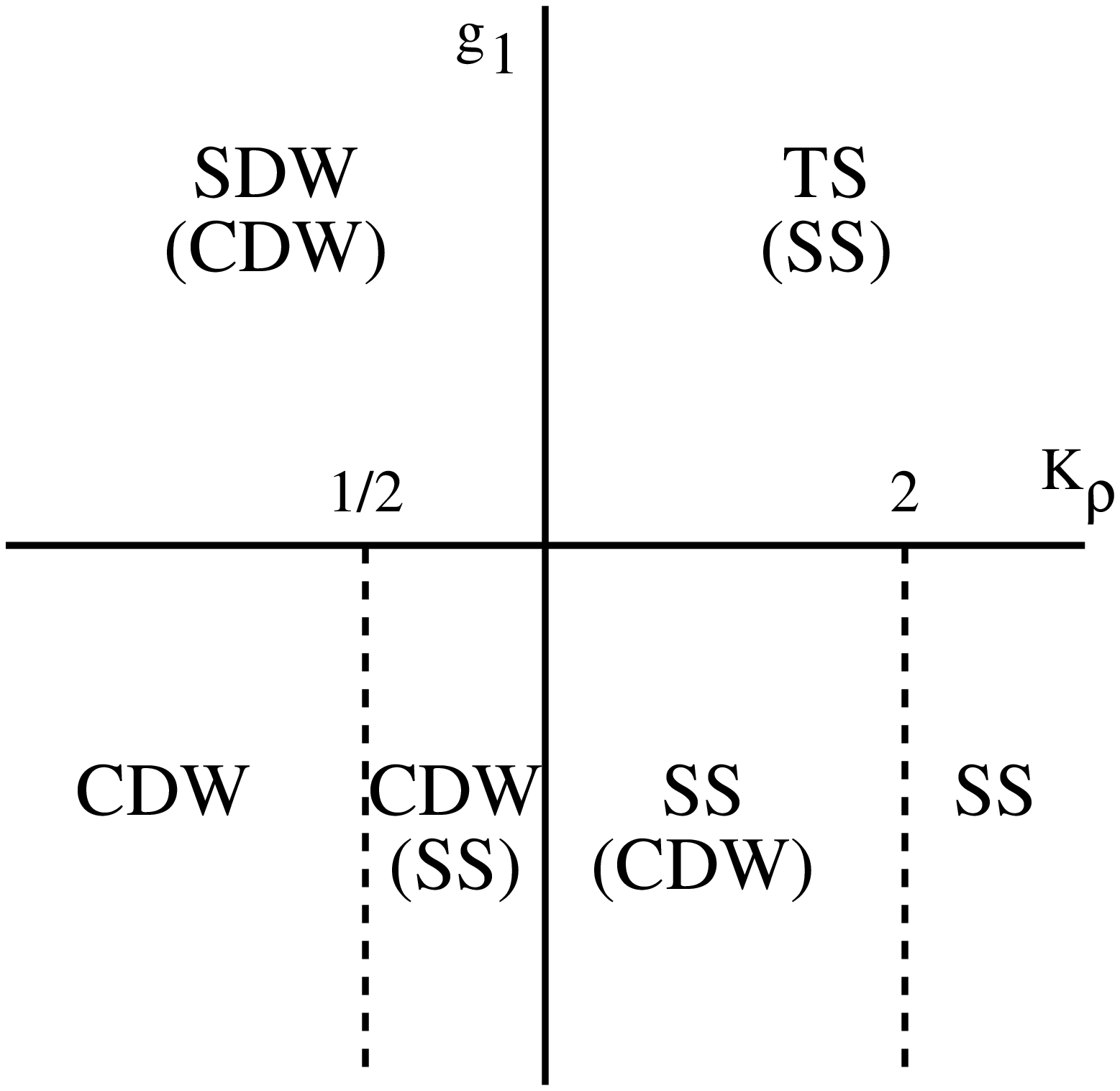}}
\end{center}
\SMALLCAP{}{Phase diagram for interacting spin--$1/2$ fermions.}
\label{f4:4}
\end{figure}

The coefficient $K_\rho$ also determines the long-distance decay of all
other correlation functions of the system: Using the representation
(\ref{singlepsi}) the charge and spin density operators at
$2k_{{\mathrm{F}}}$ are
\begin{eqnarray}
\label{ocdw}
O_{\mathrm{CDW}}(x) & = & \sum_s \psi_{-,s}^\dagger(x) \psi_{+,s}(x)
\nonumber\\ &=&
\lim_{\alpha \rightarrow 0}
\frac{\ee ^{2\ii k_{{\mathrm{F}}}x}}{\pi\alpha} \ee ^{- \ii \sqrt2 \phi_\rho(x)}
\cos [\sqrt2 \phi_\sigma(x)] \virg \\
O_{{\mathrm{SDW}}_x}(x) & = & \sum_{s} \psi_{-,s}^\dagger (x)
\psi_{+,-s} (x)\nonumber\\
&=& \lim_{\alpha \rightarrow 0}
\frac{\ee ^{2\ii k_{{\mathrm{F}}}x}}{\pi\alpha} \ee ^{- \ii \sqrt2 \phi_\rho(x)}
\cos [\sqrt2 \theta_\sigma(x)] \point
\end{eqnarray}
Similar relations are also found for other operators. It is important to
note here that all these operators decompose into a product of one factor
depending on the charge variable only by another factor depending only on
the spin field. Using the Hamiltonian (\ref{hbos}) at the fixed point $g_1^*
= 0$ one finds for example for the charge and spin correlation
functions\footnote{The time- and temperature dependence is also easily
obtained, see ref.\cite{emery_revue_1d}.}
\begin{eqnarray} \nonumber
\langle n(x) n(0) \rangle & = & K_\rho/(\pi x)^2 + A_1 \cos (2k_{{\mathrm{F}}}x)
x^{-1-K_\rho} \ln^{-3/2}(x)  \\
\label{nn}
& & \mbox{} + A_2 \cos (4k_{{\mathrm{F}}}x) x^{-4K_\rho} + \ldots \virg \\
\label{ss}
\langle \S (x) \cdot \S (0) \rangle & = & 1/(\pi x)^2 + B_1 \cos
(2k_{{\mathrm{F}}}x)
x^{-1-K_\rho} \ln^{1/2}(x) + \ldots \virg
\end{eqnarray}
with model dependent constants $A_i,B_i$. The ellipses in (\ref{nn})
and (\ref{ss}) indicate higher harmonics of $\cos (2k_{{\mathrm{F}}} x)$
 which are
present but decay faster than the terms shown here.
Similarly, correlation functions for singlet (SS) and triplet (TS)
superconducting pairing are
\begin{eqnarray}
\nonumber
\langle O^\dagger_{\mathrm{SS}}(x)
 O_{\mathrm{SS}}(0) \rangle & = & C x^{-1-1/K_\rho}
\ln^{-3/2}(x) + \ldots \virg \\
\langle O^\dagger_{\mathrm{TS}_\alpha}(x)
 O_{\mathrm{TS}_\alpha}(0) \rangle & = & D
x^{-1-1/K_\rho} \ln^{1/2}(x) + \dots \point
\end{eqnarray}
The logarithmic corrections in these functions \cite{finkelstein_logs} have
been studied in detail recently
\cite{voit_logs,giamarchi_logs,affleck_log_corr,singh_logs}.
The corresponding susceptibilities (\ie \ the Fourier transforms of the
above correlation functions) behave at low temperatures as
\begin{eqnarray}
\chi_{\mathrm{CDW}}(T) \approx T^{K_\rho -1 } |\ln(T)|^{-3/2}  &\virg &
\chi_{\mathrm{SDW}}(T) \approx T^{K_\rho -1 } |\ln(T)|^{1/2}  \virg  \\
\chi_{\mathrm{SS}}(T) \approx T^{1/K_\rho -1 } |\ln(T)|^{-3/2} & \virg &
\chi_{\mathrm{TS}}(T) \approx T^{1/K_\rho -1 } |\ln(T)|^{1/2}  \virg 
\end{eqnarray}
\ie \ for $K_\rho < 1$ (spin or charge) density fluctuations at
$2 k_{{\mathrm{F}}}$ are enhanced and
diverge at low temperatures, whereas for $K_\rho >1$
pairing fluctuations dominate. 
The ``phase diagram'', showing in which part
of parameter space which type of correlation diverges for $T\rightarrow 0$
is shown in \fref{f4:4}.

These correlation functions with their power law variations actually
determine experimentally accessible quantities: the $2k_{{\mathrm{F}}}$ and
$4k_{{\mathrm{F}}}$ charge correlations lead to X--ray scattering
intensities $I_{2k_{{\mathrm{F}}}} \approx T^{K_\rho}$,
$I_{4k_{{\mathrm{F}}}} \approx T^{4K_\rho-1}$, and similarly the NMR
relaxation rate due to $2k_{{\mathrm{F}}}$ spin fluctuations varies as
$1/T_1 \approx T^{K_\rho}$. The remarkable fact in all the above results is
that there is only {\em one coefficient}, $K_\rho$, which determines all the
asymptotic power laws.

We here re--emphasize the two important properties of spin--1/2 interacting
fermions in one dimension: (i) correlation functions show power--law decay,
with interaction--dependent powers determined by one coefficient, $K_\rho$;
and (ii) ``spin--charge separation'': spin and charge degrees of freedom
propagate with different velocities. Both these properties are typical of
the Luttinger liquid and invalidate the Landau quasiparticle concept in one
dimension.

A nice experimental example of different spin--charge separation (different
velocities for the spin and charge modes) is provided by Raman scattering
experiments on single--channel quantum
wires.\cite{goni_gaas1d,schulz_wigner,sassetti98} On the other hand, the
situation in quasi--one--dimensional conductors \cite{jerome_revue_1d} is
less clear: in compounds of the TTF--TCNQ series, the observation of strong
$4k_F$ CDW fluctuations \cite{pouget_4kf} seems only to be explainable in
the Luttinger--liquid picture. On the other hand, in the much studies $\rm
(TMTSF)_2X$ family of compounds, the NMR data can be interpreted in the
Luttinger liquid framework \cite{wzietek_nmr}, but much of the
magnetotransport work of Chaikin and collaborators is explained rather
satisfactorily using a Fermi liquid like picture.

\subsubsection{Long--range interactions: Wigner crystallization}
The above calculations can be straightforwardly generalized to the case of
long--range interactions. Of interest is the case of unscreened Coulomb
interactions ($V(r) = e ^2/r$) which for example is of relevance for the
physics of an isolated quantum wire. The short--distance singularity has to
be cut off, and for example in a wire of diameter $d$ an approximate form
would be $V(r) = e^2/\sqrt{x^2+d^2}$, leading to a Fourier transform $V(q) =
2e^2 K_0(qd)$.  The long--range nature of the interaction is only of
importance for the forward--scattering processes, and these appear only in
the charge part of the Hamiltonian which is now given by
\begin{eqnarray}     \nonumber
 H_\rho &=& \frac{v_{{\mathrm{F}}}}{2 \pi} \int dx 
\left( \pippo \pi^2 (1 + \tilde{g}_1)
\Pi_\rho^2
       + (1-\tilde{g}_1) (\partial_x \phi_\rho )^2 \right) \\
\label{ham2}
& & + \frac{1}{\pi^2} \int dx dx'V(x-x') \partial_x \phi_\rho
\partial_{x'} \phi_\rho \virg
\end{eqnarray}
where $\tilde{g}_1 = g_1/(2\pi v_{{\mathrm{F}}})$.
The elementary excitations then are found to be charge oscillations (plasmons),
with energy--momentum relation
\begin{equation}           \label{pla}
\omega_\rho (q) = v_{{\mathrm{F}}} |q| \sqrt{(1 + \tilde{g}_1)(1 -
\tilde{g}_1+2\tilde{V}(q) \pippo)} \end{equation}
where $ \tilde{V}(q) = V(q) / (\pi v_{{\mathrm{F}}})$. 
The long--wavelength form ($q \to 0$),
$\omega_\rho (q) \approx |q^2 \ln q|^{1/2}$, agrees with RPA calculations
\cite{gold_1dplasmon,li_1dplasmon}, however, the effect of $g_1$, which is
a short--range exchange contribution, are usually neglected in those
calculations. The spin modes are still given by  $\omega_\sigma(q) = u_\sigma
|q|$, with $u_\sigma = v_{{\mathrm{F}}} \sqrt{1-\tilde{g}_1^2}$.

In the evaluation of correlation functions, the charge averages lead to
\begin{eqnarray}
\nonumber
\langle (\phi_\rho(x) - \phi_\rho(0))^2 \rangle & = & \int_0^\infty
dq \,
\frac{1 - \cos qx}{q}
\sqrt{\frac{1 + \tilde{g}_1}{1-\tilde{g}_1+2\tilde{V}(q)}} \\
    \label{av}
& \approx & c_2 \sqrt{\ln x} \;\;,
\end{eqnarray}
with $c_2 = \sqrt{(1+\tilde{g}_1) \pi v_{{\mathrm{F}}}/e^2}$.
One thus obtains for example
\begin{eqnarray}
\nonumber
\langle \rho(x) \rho(0) \rangle & = & A_1 \cos(2k_{{\mathrm{F}}}x)
\exp(- c_2 \sqrt{\ln x})/x \\
\label{corr}
& & + A_2 \cos(4k_{{\mathrm{F}}}x) \exp(- 4 c_2 \sqrt{\ln x}) + ... \;\;,
\end{eqnarray}
where $A_{1,2}$ are interaction dependent constants, and only the most
slowly decaying Fourier components are exhibited.  The most interesting
point here is the extremely slow decay (much slower than any power law!) of
the $4k_{{\mathrm{F}}}$ component, showing an incipient charge density wave
at wavevector $4k_{{\mathrm{F}}}$.  This slow decay should be compared with
the power--law decay found for short--range interactions (eq.(\ref{nn})).
The $4k_{{\mathrm{F}}}$ oscillation period is exactly the average
interparticle spacing, \ie \ the structure is that expected for a
one--dimensional {\em Wigner crystal}. Of course, because of the
one--dimensional nature of the model, there is no true long--range order,
however, the extremely slow decay of the $4k_{{\mathrm{F}}}$ oscillation
would produce strong quasi--Bragg peaks in a scattering experiment. It is
worthwhile to point out that this $4k_{{\mathrm{F}}}$ contribution arises
even if the Coulomb interaction is extremely weak and depends only on the
long--range character of the interaction. On the other hand, any
$2k_{{\mathrm{F}}}$ scattering is considerably weaker, due to the $1/x$
prefactor in (\ref{corr}) which has its origin in the contribution of spin
fluctuations. The $2\kf$ spin correlations equally contain a $1/x$ factor
and thus have the same asymptotic decay as the $2\kf$ charge correlations,
eq.(\ref{corr}). On the other hand, correlation functions that involve
operators changing the total number of particles (e.g.$\!$\ the single
particle Green function and pairing correlation functions) decay like
$\exp[- {\mathrm{cst.}} \times (\ln x)^{-3/2}]$,
\ie \ {\em faster} than any power law. This in particular means that the
momentum distribution function $n_k$ and all its derivatives are continuous
at $k_{{\mathrm{F}}}$, and there is only an essential singularity at 
$k_{{\mathrm{F}}}$.

It is instructive to compare the above result (\ref{corr}), obtained in the
limit of weak Coulomb interactions, with the case of strong repulsion (or,
equivalently, heavy particles). The configuration of minimum potential
energy is one of a chain of equidistant particles with lattice constant $a$,
and quantum effects are expected to lead only to small oscillations in the
distances between particles. The Hamiltonian then is
\begin{equation}
H = \sum_l \frac{p_l^2}{2m} + \frac14 \sum_{l \neq m} V''(ma)
(u_l - u_{l+m})^2 \;\;,
\end{equation}
where $u_l$ is the deviation of particle $l$ from its equilibrium position.
In the long--wavelength limit, the oscillation of this lattice have energy
$\omega(q) = \sqrt{2/(ma)} e q |\ln (qa)|^{1/2}$. The most slowly
decaying part of the density--density correlation function then is
\begin{equation}   \label{corr2}
    \langle \rho(x) \rho(0) \rangle \approx
\cos(2 \pi x/a) \exp \left( - \frac{4\pi}{(2 m e^2 a)^{1/2}} \sqrt{\ln
x}\right) .
\end{equation}
Noticing that $k_{{\mathrm{F}}} = \pi/(2 a)$, one observes that the results (\ref{corr})
and (\ref{corr2}) are (for $g_1=0$) identical as far as the
long--distance asymptotics are concerned,
{\em including the constants in the exponentials}.
Eq. (\ref{corr}) was obtained in the weak interaction
limit, whereas (\ref{corr2}) applies for strong Coulomb forces.
Similarly, the small--$q$ limit of the charge excitation energies is
identical. 
\subsection{Summary}
In this section we have developed the basic bosonization formalism for
one--dimensional interacting fermions and seen some elementary and direct
applications to the calculation of some physical properties, in particular
correlation functions. We have seen that the properties of the
one--dimensional interacting system, the {\em Luttinger liquid}, are
fundamentally different from two- or three--dimensional Fermi liquids. In
particular the elementary excitations are {\em not quasiparticles} but
rather collective oscillations of the charge and spin density, propagating
coherently, but in general at different velocities. This gives rise to the
interesting phenomenon of {\em spin--charge separation}. Finally, and again
contrary to the Fermi--liquid case, most correlation functions show {\em
non--universal powerlaws, with interaction--dependent exponents}. However,
all these exponents depend only on one parameter, $K_\rho$, the spin
analogue of which, $K_\sigma$, being fixed to unity by spin--rotation
invariance ($K_\sigma\neq 1$ is possible if spin rotation invariance is
broken). Beyond $K_\rho$, the only parameters that intervene in the
low--energy physics of a Luttinger liquid are the velocities of the spin and
charge modes, $u_{\rho,\sigma}$. In the spinless case only two parameters,
$K$ and $u$ are involved. Finally, we have seen that long--range (Coulomb)
interactions can profoundly modify these properties.

For lack of time and space, we have not touched here upon a number of
interesting uses and generalizations of the bosonization method, in
particular the Kondo effect \cite{emery_kivelson_kondo_review} and
applications to Fermi systems in more than one
dimension.\cite{houghton_bos_3d,fradkin_bos_3d} An extensive recent
review on Luttinger liquids and bosonization has been given by
Voit.\cite{voit95}

\sectio{Applications}
We will subsequently discuss some results from applications of the formalism
developed to interesting physical questions: transport and the effect of
disorder, and the physics of antiferromagnetic spin chains. Here, we will
come across the two cases where Luttinger liquid behavior is probably best
established experimentally: the physics of quantum Hall edge states, upon
which we will only touch briefly, and quantum spin systems, in particular the
spin-$1/2$ antiferromagnet, which we will discuss in some detail, with some
reference to experiment.
\subsection{Transport}
\subsubsection{Conductivity and conductance } \label{sec:cond}
In the previous section we were concerned with equilibrium properties
and correlation functions, in order to characterize the different phases
possible in a one--dimensional system of interacting fermions. Here, we
will investigate transport, in particular the {\sc dc} conductivity.
Finite--frequency effects have also been investigated, and the reader is
referred to the literature \cite{giamarchi_millis,giamarchi_rho}.

To clarify some of the basic notions, let us first consider a Luttinger
model in the presence of a weak space-- and time--dependent external
potential $\varphi$. The interaction of the fermions with $\varphi$ is
described by the extra term
\begin{equation}\label{eq:hext}
H_{\mathrm{ext}} = -e \int dx \hat{\rho}(x) \varphi(x,t)
\end{equation}
in the total Hamiltonian. We will assume that the external field is
slowly varying in space, so that in the particle--density operator
$\hat{\rho}$ only products of either two right-- or two left--going
fermion operators appear but no cross terms. Standard linear response
theory tells us that the current induced by the potential is given by
\begin{equation}\label{eq:lr1}
j(x,t) = - \frac{e^2}{\hbar} \int_{-\infty}^t dt' \int dx'
D_{j\rho}(x-x',t-t') \varphi(x',t') \virg
\end{equation}
where the {\em retarded current--density correlation function} is given
by
\begin{eqnarray} \nonumber
D_{j\rho}(x,t) & = & - \ii \heavi (t) \langle[j(x,t),\rho(0,0)]\rangle \\
& = & - \frac{u_\rho K_\rho}{\pi}\heavi (t) \left(
\pippo
\delta'(x-u_\rho t) +
\delta'(x+u_\rho t) \right)
\point
\label{eq:lr2}
\end{eqnarray}
The second line is the result for spin--$1/2$ electrons. For spinless
fermions one has to make the replacement $u_\rho K_\rho \rightarrow uK/2$.

Let us now first consider the situation where we adiabatically switch on
a potential of frequency $\omega$ and wavenumber $q$ along the whole
length of the system. From eq.(\ref{eq:lr2}) one then straightforwardly
obtains the $q$-- and $\omega$--dependent conductivity as
\begin{equation}\label{eq:sigqo}
\sigma(q,\omega) = \frac{4 e^2}{\hbar} u_\rho K_\rho
\frac{\ii(\omega+\ii 0^+)}{(\omega+\ii 0^+)^2-u_\rho^2 q^2} \point
\end{equation}
In particular, the real part of the conductivity for constant applied
field is
\begin{equation}\label{eq:sigom}
{\rm Re}\,\sigma(0,\omega) = \frac{2 e^2}{\hbar} u_\rho K_\rho \delta(\omega)
\virg
\end{equation}
in agreement with eq.(\ref{sig0}) (where units with $e^2 = \hbar =1$ were
used).

Applying on the other hand a static field over a finite part of the
sample, one obtains a current $j = 2e^2 K_\rho U/h$,
where $U$ is the applied tension. The conductance thus is
\begin{equation} \label{eq:conduct}
G = \frac{2e^2}{h} K_\rho \virg
\end{equation}
and depends on $K_\rho$ only, not on $u_\rho$. For the noninteracting case
$K_\rho=1$ this is Landauer's well--known result \cite{landauer}. Note that
interactions affect the value of the conductance. The conductance here is
independent on the length over which the field is applied. Noting that in
dimension $d$ the conductance is related to the {\sc dc} conductivity via $G
= L^{d-2}\sigma$, a length--independent conductance implies an infinite
conductivity in one dimension, in agreement with eq.(\ref{eq:lr2}). The fact
that $u_\rho$ does not appear in the expression for $G$ can be understood
noting that applying a static field over a finite (but large) part of the
sample, one is essentially studying the wavenumber--dependent conductivity
at strictly zero frequency, which from eq.(\ref{eq:lr}) is given by
$\sigma(q,0) =2e^2 K_\rho \delta(q)/\hbar$, indeed independent on
$u_\rho$. On the other hand, applying a field of finite frequency over a
finite length $\ell$, one can see that one measures the conductivity
$\sigma(q\rightarrow0,\omega)$ if $\omega > u_\rho/\ell$.

The situation of a finite static potential drop over only a finite part of a
wire is clearly difficult to realize experimentally. However, the result
(\ref{eq:conduct}), or more precisely its spinless analogue $G = Ke^2/h$,
applies to the chiral Luttinger liquid as realized on the edge of quantum
Hall systems.\cite{kane95,alekseev96,chamon97} On the other hand, for a
quantum wire connected to measuring contacts which impose a potential
difference over the whole length of the wire, one obtains
\cite{safi_trans,maslov_pure_wire,alekseev96,safi97}
\begin{equation} 
G = \frac{2e^2}{h}\virg
\label{eq:conduct2}
\end{equation}
independent of $K_\rho$, i.e. the momentum--conserving interactions play no
role in the dc conductance, as to be expected. On the other hand, the value
of $K_\rho$ plays an important role for the effects of random and contact
potentials.\cite{safi95,safi98} These results permit a consistent
explanation of some experiments on quantum wires \cite{tarucha95}, but
leave open questions in other cases.\cite{yacoby96} 

\subsubsection{Persistent current}
The Luttinger model description can be used straightforwardly to obtain the
current induced in a strictly one--dimensional ring threaded by a magnetic
flux $\Phi$ \cite{loss_ring}. The argument can in fact be made very
simply: in the one--dimensional geometry, the
vector field can be removed entirely from the Hamiltonian via a gauge
transformation, which then leads to the boundary condition $\psi(L) =
\exp(2\pi \ii \Phi/\Phi_0) \psi(0)$ for the fermion field operator. Here $L$
is the perimeter of the ring, and $\Phi_0 = hc/e$. For spinless fermions,
this is achieved by replacing
\begin{equation} \label{eq:Pi1}
\Pi(x) \rightarrow \Pi(x) + \frac{2\Phi}{L\Phi_0}
\end{equation}
in the bosonization formula, eq.(\ref{singlepsi0}). The total
$J$--dependent part of the Hamiltonian then becomes{}
\begin{equation}\label{eq:hj}
H_J = \frac{\pi u K}{2 L} (J+ 2\Phi/\Phi_0)^2 \virg
\end{equation}
giving rise to a number current
\begin{equation}\label{eq:jphi}
j = \frac{\Phi_0}{2 \pi} \frac{\partial E}{\partial \Phi} = \frac{uK}{L}
\left(J + \frac{2\Phi}{\Phi_0}\right) \point
\end{equation}
At equilibrium, $J$ is chosen so as to minimize the energy. Given that
at constant total particle number $J$ can only change by two units, one
easily sees that the equilibrium (persistent) current has periodicity
$\Phi_0$, and reaches is maximum value $uK/L$ at $\Phi=\Phi_0/2$,
giving rise to the familiar sawtooth curve for the current as a function
of flux.

For fermions with spin, as long as there is no spin gap ($g_1 >0$), the
above results can be taken over, with the simple replacement $uK
\rightarrow 2 u_\rho K_\rho$, the factor $2$ coming from the spin
degeneracy. Note in particular that the persistent current, an
equilibrium property, is given by the same combination of parameters as
the Drude weight in the conductivity. This is an illustration of Kohn's
result \cite{kohn_drude} relating the Drude weight to the effect of a
magnetic flux through a ring.

In the case of negative $g_1$, electrons can be transfered from the right to
the left--going branch only by pairs, Consequently, the periodicity of the
current and the ground state energy is doubled to $2\Phi_0$, and the maximum
current is equally doubled. This behavior has actually been found in
numerical calculations \cite{yu_hubbard_flux,sudbo_cuo}.

\subsubsection{Quantum Hall edge states} \label{qhes}
\begin{figure}[htb]
\centerline{\epsfxsize 6.5cm \epsffile{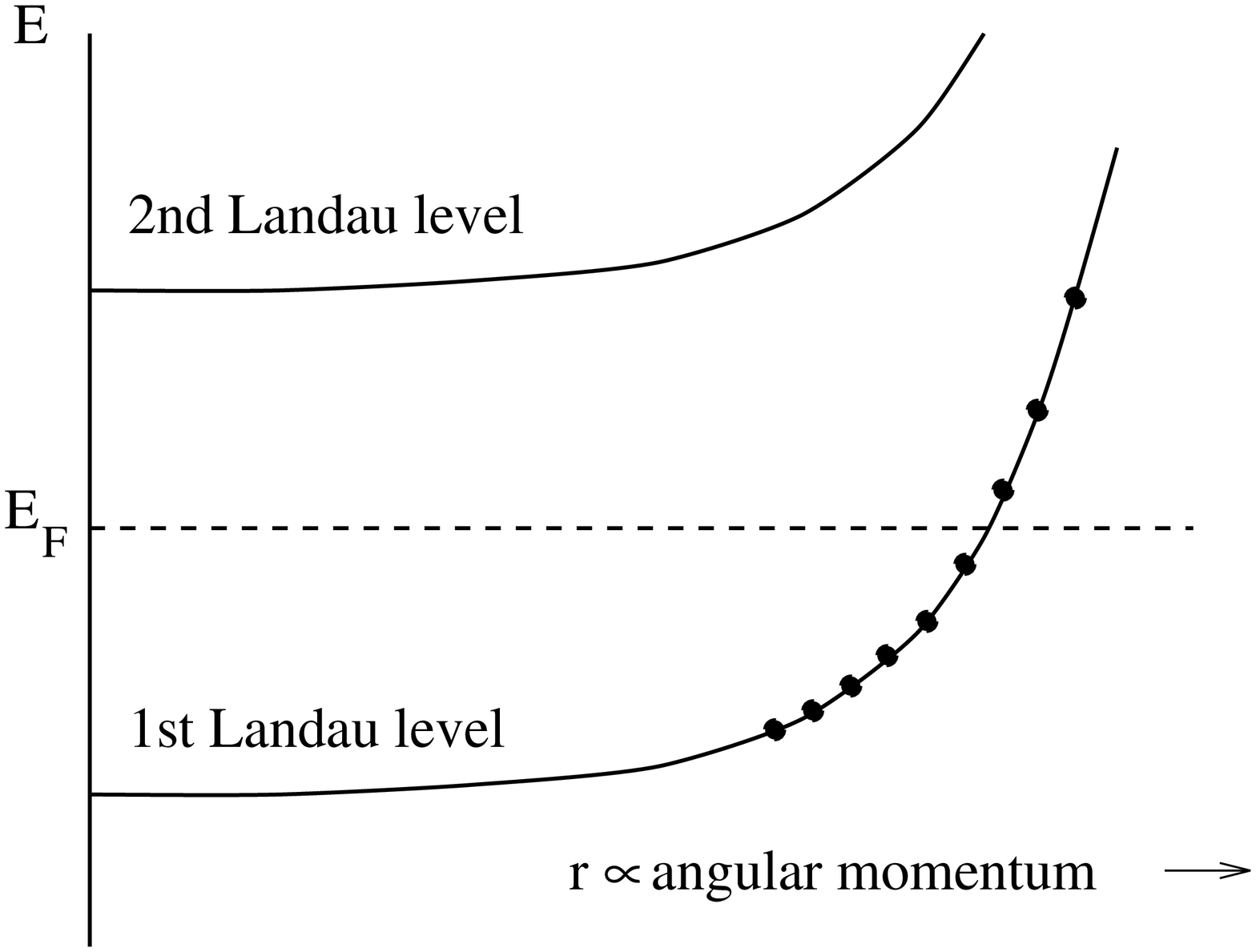}}
\SMALLCAP{b}{Quantum states in the lowest Landau level in the vicinity of
the edge of a quantum hall device. The spatial variation of the
confining potential is assumed to be slow on the scale of the magnetic
length, so that the energies of the different quantum states are
determined by the local value of the confining potential. For a circular
device the angular momentum of a state increases proportionally to its
distance $r$ from the origin.}
\label{f5:1}
\end{figure}

The Luttinger liquid picture has an interesting application to the physics
of the fractional quantum Hall effect, as discovered and discussed by Wen
\cite{wen_1,wen_2}. To see how this comes about, consider the states
available in the different Landau levels in the vicinity of the edge of the
quantum Hall device, as shown in figure~\ref{f5:1} \cite{macdonald95}.
It is clear that low--energy states only exist at the edge (the bulk
quantum Hall state is well-known to be characterized by a finite
excitation gap), and close to the Fermi energy (\ie \ the edge) the
states have a linear dispersion relation. This can be made particularly
clear if one assumes a disk--shaped sample: the states have a
wavefunction 
\begin{eqnarray}
\approx z^k \approx (\ee ^{\ii\theta})^k \virg
\end{eqnarray}
with $k$ increasing linearly with radial position. The angular momentum
quantum number $k$ thus plays a role very similar to linear momentum in
the linear geometry we have assumed up to now. One thus can linearize
the the dispersion in figure~\ref{f5:1} and obtains essentially the
spinless model discussed in section \ref{sec:bosonspl}, the only
difference being that here only right--going particles exist. This
difference is the origin of the term {\em chiral} Luttinger liquid (in
fact, the left going branch is to be found on the opposite edge of the
device). Because there are no left--going particles (or at least they
can be thought of as being at a macroscopic distance), there also is no
right--left interaction, and consequently one expects the noninteracting
value $K=1$. Moreover all the left--going components of the fields have
to be projected out, for example one has to replace $\phi \rightarrow
\phi_+=(\phi-\theta)/2$.

However, straightforward adoption of this scheme leads to trouble: from
the preceding subsection we know that $K=1$ leads to a conductance
(which in the present case is the Hall conductance) of $G=e^2/h$, {\em
different} from the well--known
\begin{equation}\label{eq:fraccond}
G = \nu \frac{e^2}{h}
\end{equation}
valid for a fractional quantum Hall state ($\nu = 1/m$ is the filling
factor). To repair this problem one makes the {\em hypothesis} that
instead of eqs.(\ref{eq:den}) and (\ref{eq:jx}) one has
\begin{equation}\label{eq:rhon}
\rho(x) = -\frac{\sqrt{\nu}}{\pi} \frac{\partial \phi_+}{\partial x}
\quad , \quad \quad j(x) = u \sqrt{\nu} \Pi_+(x) \virg
\end{equation}
where the subscripts indicate projection on right--going states. With
these definitions following the calculations of the previous subsection
one now straightforwardly reproduces the correct result,
eq.(\ref{eq:fraccond}). The appearance of the factors $\sqrt{\nu}$ in
eq.(\ref{eq:rhon}) indicates that the objects occupying the states in
figure~\ref{f5:1} are not free electrons but rather strongly affected
by the physics of the bulk of the samples. A more detailed derivation,
starting from a Chern--Simons field theory for the bulk physics, has
also been given by Wen \cite{wen_2}.

Beyond reproducing the correct value of the Hall conductance, the above
hypothesis leads to a number of interesting conclusions. Consider first
the creation operator for a real electron (charge $e$) on the edge.
Following the arguments of section \ref{sec:bosonspl}, because of
eq.(\ref{eq:rhon}), the bosonized version of the electron operator now
must create a jump of $\phi$ of height $\pi/\sqrt{\nu}$, rather than of
height $\pi$. This leads to
\begin{equation}\label{eq:spf}
\psi^{\phantom{\dagger}}_+(x) \approx \ee ^{- \ii \phi_+(x)/\sqrt{\nu}} \virg
\end{equation}
$x$ being the coordinate along the perimeter of the sample. Now, these
operators obey the relation
\begin{equation}\label{eq:spfcom}
\psi^{\phantom{\dagger}}_+(x') \psi^{\phantom{\dagger}}_+(x) 
= \ee ^{\pm \ii \pi/\nu} \psi^{\phantom{\dagger}}_+(x) 
\psi^{\phantom{\dagger}}_+(x') \point
\end{equation}
But the real electron is still a fermion, \ie \ $\psi^{\phantom{\dagger}}_+$
must obey anticommutation relations. Thus $m=1/\nu$ {\em has to be an odd
integer}. One thus reproduces one of the fundamental facts of the fractional
quantum Hall effect. From eq.(\ref{eq:spf}) one also finds a decay of the
single--electron Green function as
\begin{equation}\label{eq:spfg}
G(x,t) \propto \frac{1}{(x-ut)^{1/\nu}} \point
\end{equation}

Another fundamental property of the quantum Hall state appears when one
considers the fractionally charged elementary excitation of charge
$e\nu$ at the edge. A charge--$e\nu$ object is created by
\begin{equation}\label{eq:spff}
\psi^{\phantom{\dagger}}_{+\nu}(x) \approx \ee ^{- \ii \sqrt{\nu}\phi_+(x)} 
\virg
\end{equation}
leading to a slow decay of the corresponding Green function, with
exponent $\nu$, instead of $1/\nu$ in eq.(\ref{eq:spfg}). One now has
the relation
\begin{equation}\label{eq:spfcom2}
\psi^{\phantom{\dagger}}_{+\nu}(x') \psi^{\phantom{\dagger}}_{+\nu}(x) = 
\ee ^{\pm \ii \pi\nu} \psi^{\phantom{\dagger}}_{+\nu}(x)
\psi^{\phantom{\dagger}}_{+\nu}(x') ,
\end{equation}
\ie \ exchanging the fractionally charged objects one obtains nontrivial
($\neq\pm1$) phase factors and these quasiparticles thus also obey
fractional statistics.

A single hypothesis, the insertion of the factors $\sqrt{\nu}$ in
eq.(\ref{eq:rhon}), thus reproduces two of the fundamental facts about the
fractional quantum Hall effect! In addition one obtains a basis for the
treatment of transport phenomena mentioned at the end of sec.\ref{sec:iso}
and results for the asymptotics of Green functions.

\subsection{Disorder}
\subsubsection{Effects of isolated impurities} \label{sec:iso}
The infinite conductivity in the ideally pure systems considered up to here
is a natural but hardly realistic result: any realistic system will contain
some form of inhomogeneity. This in general leads to a finite conductivity,
and in one dimension one can anticipate even more dramatic effects: in a
noninteracting system any form of disorder leads to localization of the
single--particle eigenstates \cite{mott_loc,abrahams_loc}. How this
phenomenon occurs in interacting systems will be discussed in this and the
following section.

Following Kane and Fisher \cite{kane_fisher}, consider first the case of
a single inhomogeneity in an otherwise perfect one--dimensional system.
The extra term in the Hamiltonian introduced by a localized potential
$v(x)$ is (for spinless fermions)
\begin{equation}\label{eq:hb1}
H_{\mathrm{barrier}}  \propto \int d x \psi^{\dagger}(x) \psi(x) \point
\end{equation}
Decomposing the product of fermion operators into right-- and
left--going parts, one has
\begin{equation}\label{eq:lr}
\psi^{\dagger} \psi = \psi^{\dagger}_+ \psi^{\phantom{\dagger}}_+ 
+ \psi^{\dagger}_- \psi^{\phantom{\dagger}}_-
+ \psi^{\dagger}_+ \psi^{\phantom{\dagger}}_- + \psi^{\dagger}_- 
\psi^{\phantom{\dagger}}_+ \point
\end{equation}
In the bosonic representation, the first two terms are proportional to
$\partial_x \phi$ (see eq.(\ref{eq:den})), and therefore the
corresponding contribution
in eq.(\ref{eq:hb1}) can in fact be eliminated by a simple unitary
transformation of $\phi$. These terms represent scattering with momentum
transfer $q \ll 2k_{{\mathrm{F}}}$, \ie \ they do not transfer particles between
$k_{{\mathrm{F}}}$ and $-k_{{\mathrm{F}}}$ and therefore do not affect the conductance
in any noticeable way. On the other hand, the last two terms in
eq.(\ref{eq:lr}) represent scattering with $|q| \approx 2k_{{\mathrm{F}}}$,
 \ie \ from
the $+$ to the $-$ branch and {\it vice versa}. These terms certainly are
expected to affect the conductance, because they change the direction of
propagation of the particles. The bosonic representation of these terms is
\begin{equation}\label{eq:hb2}
H_{\mathrm{barrier}} = \frac{V(2k_{{\mathrm{F}}})}{\pi\alpha} 
\cos 2\phi(0) \virg
\end{equation}
where the potential $V(x)$ is assumed to be centered at $x=0$. For this
reason, only the value of the $\phi$ at $x=0$ intervenes in
eq.(\ref{eq:hb2}).

One now can integrate out all the degrees of freedom away from $x=0$, to
obtain an effective action implying only the time--dependence of
$\phi(0)$. Then a 
renormalization group equation for $V\equiv V(2k_{{\mathrm{F}}})$ can
be found as
\begin{equation} \label{eq:rgv5}
\frac{d V}{d \ell} = (1-K) V \virg
\end{equation}
where $E = E_0 \ee ^{-\ell}$, $E_0$ is the original cutoff, and $E$ is the
renormalized cutoff.

From eq.(\ref{eq:rgv5}) it is clear that there are three regimes:
\begin{enumerate}
\item For $K>1$ one has $V(\ell\rightarrow \infty) =0$, \ie \ as far as the
low--energy physics is concerned, the system behaves like one without the
barrier. In particular, the low--temperature conductance takes the ``pure''
value $G=e^2 K/h$, with corrections of order $T^{2(K-1)}$. We note that in
this case superconducting fluctuations dominate, and the prefect
transmission through the barrier can be taken as a manifestation of
superconductivity in the one--dimensional system.
\item For the noninteracting case $K=1$, $V$ is invariant, and one thus has
partial transmission and a non--universal conductance depending on $V$.
\item For $K<1$, $V(\ell)$ scales to infinity. Though the perturbative
calculation does not provide any direct way to treat this case, it is
physically clear that the transmission and therefore the conductance should
vanish.
\end{enumerate}
Note that the non--interacting case is marginal, separating the regions of
perfect and zero transmission. These results are very similar to earlier
ones by Mattis \cite{mat_bos} and by Luther and Peschel
\cite{luther_conductivite_disorder} who treat disorder in lowest--order
perturbation theory.

The case of $K<1$ can be further analyzed considering the case of two finite
Luttinger liquids coupled by a weak tunneling barrier, as would be
appropriate for a strong local potential. The barrier Hamiltonian then
is
\begin{equation} \label{eq:hb3}
H_{\mathrm{barrier}} = 
t \left( \psi^{\dagger}_1(0) \psi^{\phantom{\dagger}}_2(0)
+ \psi^{\dagger}_2(0) \psi^{\phantom{\dagger}}_1(0) \right) \approx
\frac{t}{\pi\alpha} \cos 2\theta(0) \point
\end{equation}
Here $\psi^{\phantom{\dagger}}_{1,2}$ are the field operators to the left
and to the right of the barrier. The operators have to satisfy the {\em
fixed boundary condition} $\psi^{\phantom{\dagger}}_i(x=0) =0$, different
from the periodic boundary conditions we have used so far. Noting that the
$\psi^{\phantom{\dagger}}_i$ can be decomposed into left-- and right--going
parts as $\psi^{\phantom{\dagger}}_i = \psi^{\phantom{\dagger}}_{i+} +
\psi^{\phantom{\dagger}}_{i-}$, 
this can be achieved, using eq.(\ref{singlepsi0}), by imposing the fixed
boundary condition $\phi_i(x=0) = \pi/2$ on the boson field
\cite{eggert_affleck,fabrizio_finiteLL}.

One can now proceed in complete analogy to the weak--$V$ case to obtain the
renormalization group equation
\begin{equation} \label{eq:rgt}
\frac{d t}{d \ell} = \left( 1-\frac{1}{K} \right) t \point
\end{equation}
Again, there are three different regimes: (i) for $K>1$ now
$t(\ell\rightarrow\infty) \rightarrow \infty$, \ie \ the tunneling amplitude
becomes very big. This can be interpreted as indicating perfect
transmission, e.g. $\!$ $G=e^2K/h$; (ii) the case $K=1$ remains marginal,
leading to a $t$--dependent conductance; (iii) for $K<1$ $t$ scales to zero,
there thus is no transmission, and $G=0$.
\begin{figure}[htb] 
\centerline{\epsfxsize 8.5cm {\epsffile{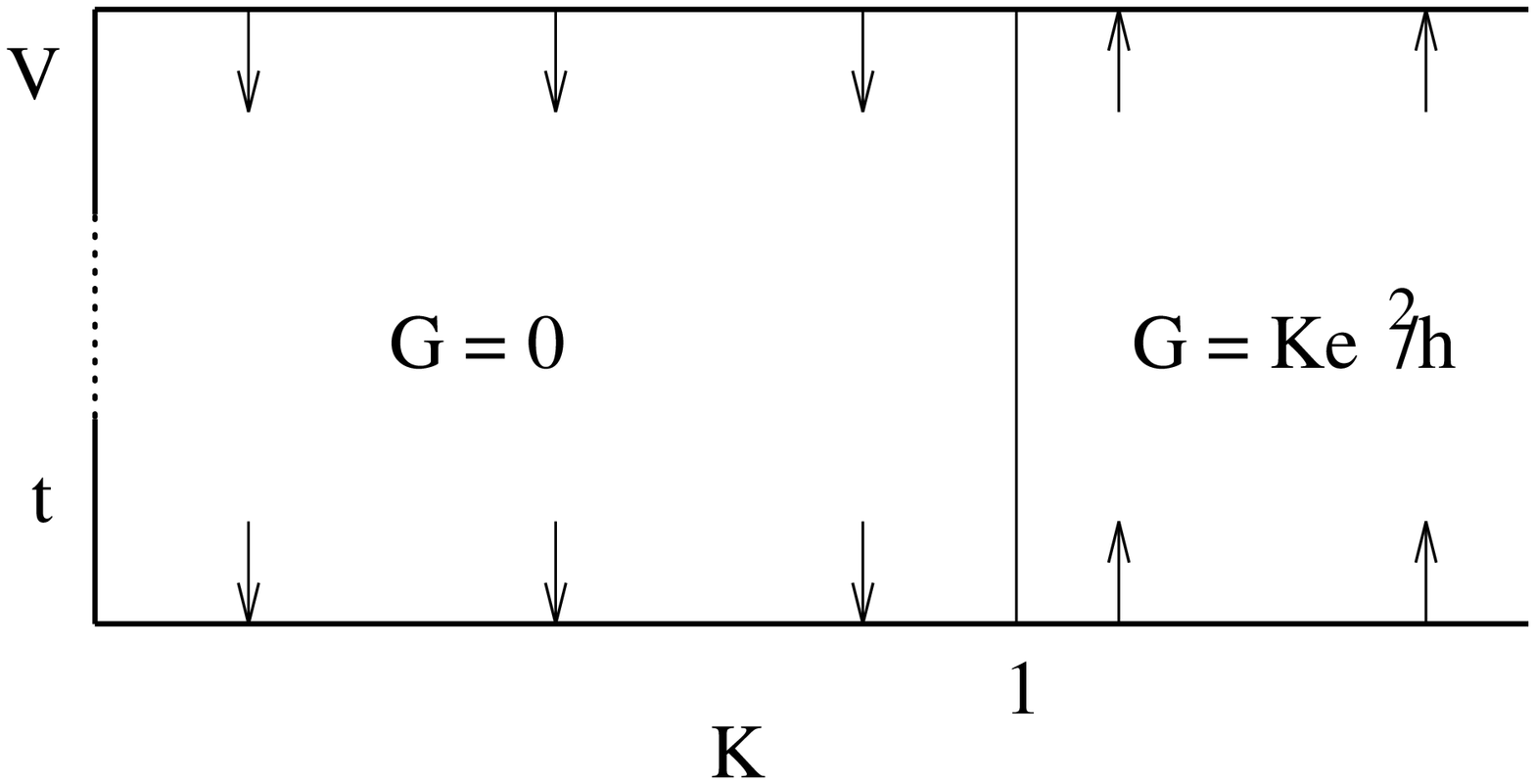}}} 
\SMALLCAP{}{``Phase diagram'' of a localized inhomogeneity in a 
spinless Luttinger liquid, characterized by an exponent parameter $K$, 
according to \cite{kane_fisher}. The scaling trajectories calculated for 
weak $V$ or $t$ are indicated by arrows. It is clearly plausible to
assume direct scaling from weak to strong coupling in the whole range
of $K$.}
\label{f5:2} 
\end{figure}
The results obtained in the two limiting cases of small $V$ (weak
scattering) and of small $t$ (weak tunneling) are clearly compatible:
e.g.$\!$  for $K<1$, $V$ becomes large, \ie \ at sufficiently low energies one
expects essentially a tunneling type behavior, and then from
eq.(\ref{eq:rgt}) the tunneling amplitude actually does scale to zero,
giving zero conductance in the low--energy (or low--temperature) limit. For
$K\ge 1$ a similar compatibility of the two limiting cases is found. The
global behavior can be represented by the ``phase diagram'' in 
figure~\ref{f5:2}.
For electrons with spin but spin--independent interactions, results are very
similar: the separation between zero and perfect transmission is at $K_\rho
= 1$, with $K_\rho=1$ again the marginal case. In the transmitting region
the conductance is $G=2K_\rho e^2/h$.

These considerations can be generalized to the case of two
barriers \cite{kane_fisher,furusaki_double_barriere}. In particular, assuming
that there are two identical, weakly scattering barrier at $\pm d$, the
effective scattering potential becomes $V_{\mathrm{eff}}(q) = 2v(q)
\cos(qd/2)$. Though in general this is non--zero when $V(q)$ is non--zero,
for particular values of $k_{{\mathrm{F}}}$, so 
that $\cos(k_{{\mathrm{F}}} d)=0$ this potential
vanishes, giving rise to perfect transmission even for $K<1$. This {\em
resonant scattering} condition corresponds to an average particle number
between the two barriers of the form $\nu+1/2$, with integer $\nu$, \ie \ the
``island'' between the two barriers is in a degenerate state. If
interactions between the electrons in the island are included, one can
recover the physics of the Coulomb
blockade \cite{kane_fisher,furusaki_double_barriere}.

For the chiral Luttinger liquid, discussed in section~\ref{qhes},
backscattering events a priori seem to be excluded because all the particles
are moving in the same direction. In that sense the chiral Luttinger liquid
can be considered as ``perfect''. However, if the quantum hall device has a
constriction that brings the two edges close to each other, scattering from
one edge to the other becomes possible and is the equivalent of
backscattering. Then similar considerations as made for the single--impurity
case are possible \cite{kane_fisher_fqhe}, and in particular the crossover
function describing the conductance through a resonance as a function of
temperature \cite{moon_fqhe,fendley95a} and the $I(V)$ characteristic
\cite{fendley95b} have been obtained.

\subsubsection{Anderson localization of one--dimensional
interacting fermions}
%local definitions
\def \cd{{\cal D}}

The discussion of the previous section was concerned with the effect of
at most two impurities, weak or strong. Clearly, in that case the effects of
coherent scattering from many impurities, which typically give rise to
Anderson localization, are absent. We now turn to this more complicated
case which had been studied in fact well before the single impurity
work \cite{chui_bray,suzumura_loc_scha,apel_loc,giamarchi_loc}.

In the absence of electron--electron interactions, localization effects
can be discussed in the framework of a scaling theory \cite{abrahams_loc}.
Under the assumption that at some short length scale one has elastic
scattering of electrons off impurities, this theory leads to the
following $\beta$--function for the variation of the conductance with
linear dimension $L$:
\begin{equation}
\beta (G) = \frac{d \ln (G)}{d \ln (L)} = d-2 - \frac{a}{G} + \dots
\virg
\end{equation}
where $a$ is a constant and $d$ the spatial dimensionality. In
particular in one dimension this leads to a conductance decaying
exponentially with the length of the system, exhibiting clearly the
localized character of the all single--electron states (a fact first
shown by Mott \cite{mott_loc} and studied in great detail since
\cite{erdos_loc1d,efetov_loc}).

In an interacting one--dimensional system (as described by the Luttinger
liquid picture of the previous section) now a number of questions
arise: what is the influence of disorder on the phase diagram obtained
previously? What are the transport properties?
Can one have true superconductivity in one dimension, \ie \ infinite
conductivity in a disordered system? To answer these question we
discuss below the generalizations necessary to include disorder in our
previous picture.

We start by the standard term in the Hamiltonian describing the coupling
of a random potential to the electron density
\begin{equation}
H_{\mathrm{imp}} = \sum_{i} \int dx \; V(x-R_i) \hat{\rho}(x) \virg
\end{equation}
where the $R_i$ are the random positions of impurity atoms, each acting
with a potential $V$ on the electrons. In one dimension one can
distinguish two types of processes:
(i) forward scattering, where the scattered particle remains in the
vicinity of its Fermi point. As in the single--impurity case, this leads
to a term proportional to $\partial \phi_\rho$ and can be absorbed by a
simple redefinition of the $\phi_\rho$ field. The physical effects are
minor, and in particular the {\sc dc} conductivity remains infinite.
(ii) backward scattering where
 an electron is scattered from $k_{{\mathrm{F}}}$ to
$-k_{{\mathrm{F}}}$ or {\it vice versa}. For small impurity density this can be
represented by a complex field
$\xi$ with Gaussian distribution of 
width $D_\xi = n_i V(q=2k_{{\mathrm{F}}})^2$:
\begin{equation}
H_b = \sum_\sigma \int dx \; \left( \xi(x) \psi^\dagger_{R\sigma}(x)
\psi^{\phantom{\dagger}}_{L\sigma}(x)  + \mathrm{h.c.} \right)
\end{equation}
This term has dramatic effects and in particular leads to Anderson
localization in the noninteracting case \cite{abrikosov_loc}.

From a perturbative expansion in the disorder one now obtains a set of
coupled renormalization group equations \cite{giamarchi_loc}:
\begin{eqnarray}
\nonumber
\frac{dK_\rho}{d\ell} & = & \lefteqn{- \frac{u_\rho}{2u_\sigma} K_\rho^2
{\cal D}} \\
\nonumber
\frac{dK_\sigma}{d\ell} & = & \lefteqn{-\frac12
({\cal D} + y^2) K^2_\sigma} \\
\nonumber
\frac{dy}{d\ell} & = & \lefteqn{2(1-K_\sigma)y -  {\cal D}} \\
\label{eq:dsca}
\frac{d{\cal D}}{d\ell} & = & \lefteqn{(3-K_\rho-K_\sigma-y){\cal D}}
\virg
\end{eqnarray}
where ${\cal D}= {2D_\xi\alpha}/(\pi u_\sigma^2)
\left({u_\sigma}/{u_\rho}\right)^{K_\rho}$ is the dimensionless
disorder, $y  =
{g_{1\perp}}/({\pi u_\sigma}) $ is the dimensionless backscattering
amplitude, and the $K_\nu$ are defined in eq.(\ref{uks}). These
equations are valid for {\em arbitrary}
$K_\nu$ (the usual strength of bosonization), but to lowest order in
${\cal D}$ and $y$.

As a first application of eqs.(\ref{eq:dsca}) one can determine the
effect of the random potential on the ``phase diagram'', as represented
in figure~\ref{f4:4}. In fact, there are three different regimes:
\begin{enumerate}
\item for $K_\rho > 2$ and  $g_{1\perp}$ sufficiently
positive the fixed point is ${\cal D}^*,y^* =0, K_\rho^* \geq
2$. Because the effective random potential vanishes this is a
delocalized region, characterized as in the pure case by the absence of
a gap in the spin excitations and dominant TS fluctuations.
\item For $K_\rho > 3$ and  $g_{1\perp}$ small or negative one has
${\cal D}^* =0$, $y \rightarrow -\infty $, $K_\rho^* \geq
3$. Again, this is a delocalized region, but now because $y \rightarrow
-\infty$ there is a  spin gap and one has predominant  SS fluctuations.
\item In all other cases one has ${\cal D} \rightarrow \infty,
y \rightarrow -\infty$. This corresponds to a localized regime. For small
$K_\rho$ the bosonized Hamiltonian in this regime is that of a charge
density wave in a weak random potential with small quantum fluctuations
parameterized by $K_\rho$. This region can therefore be identified as a
weakly pinned CDW, also called a ``charge density glass'' (CDG).  The
transition from the CDG to the SS region then can be seen as depinning of
the CDW by quantum fluctuations.
\end{enumerate}
One should notice that the CDG is a nonmagnetic spin singlet, representing
approximately a situation where localized single--particle states are doubly
occupied. Though this is acceptable for attractive or possibly weakly
repulsive interactions, for strong short--range repulsion single occupancy
of localized states seems to be more likely.  One then has a spin in each
localized state, giving probably rise to a localized antiferromagnet with
random exchange (RAF). 

\begin{figure}[htb]
\centerline{\epsfxsize 6.5cm
\epsffile{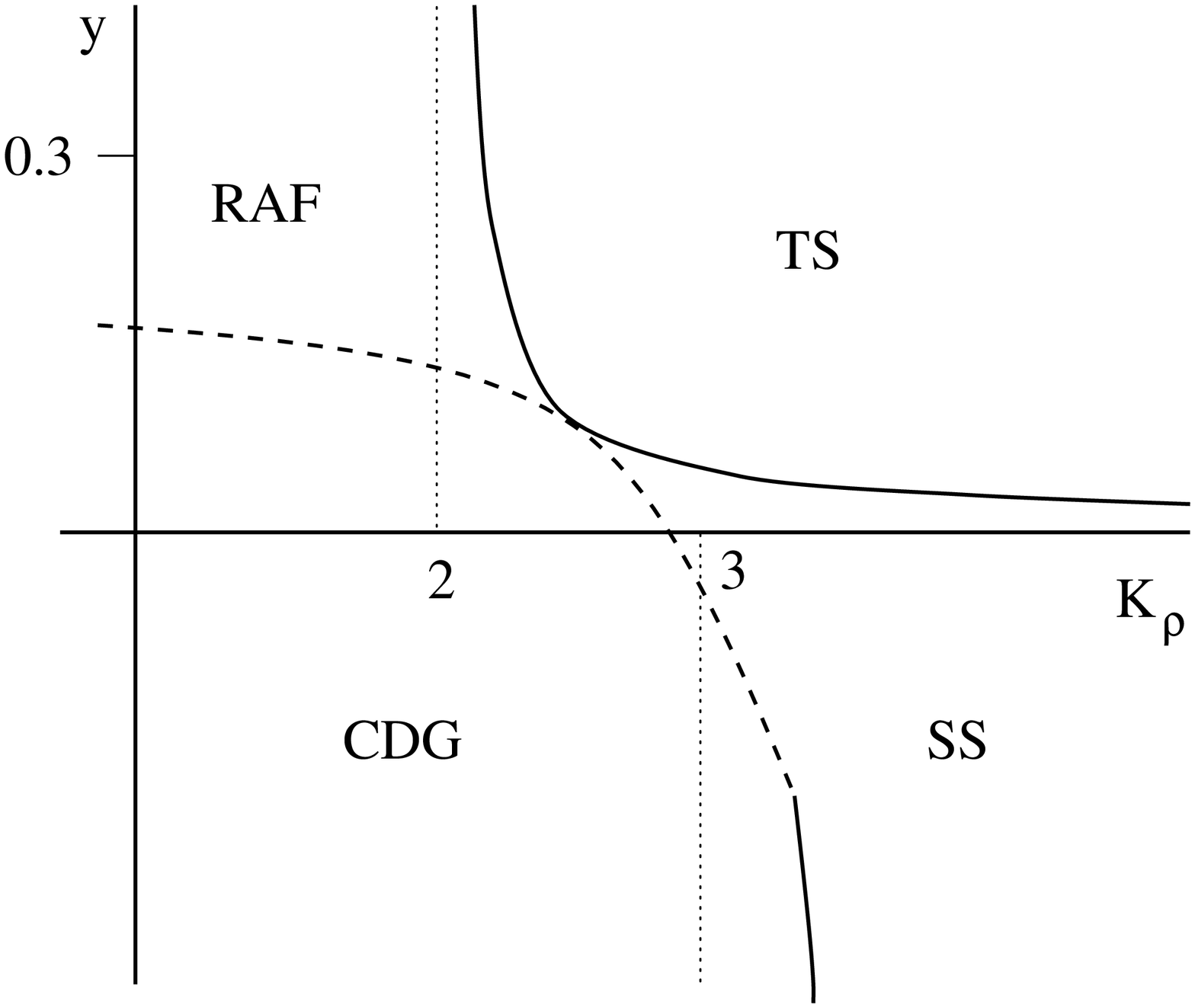}}
\SMALLCAP{t}{Phase diagram of a Luttinger liquid in the presence of a weak
random potential ($\cd=0.05$). The full lines represent results obtained
directly from the scaling equations (\ref{eq:dsca}), the dashed lines are
qualitative interpolations. The dotted lines are the phase boundaries in the
limit $\cd \rightarrow 0$.}
\label{f5:3}
\end{figure}

A detailed theory of the relative stability of the
two states is currently missing and would certainly at least require
higher-order perturbative treatment.  The boundaries of the different
regimes can be determined in many cases from eqs.(\ref{eq:dsca}), and the
resulting phase diagram is shown in figure~\ref{f5:3}.

The localization length for small disorder can be obtained from standard
scaling arguments:
suppose that a system with some fixed disorder $\cd_0$ has a
localization length $\xi_0$. Then in the general case one has
$\xi ( {\cal D}) = \xi_0 \ee ^{\ell (\cd_0,\cd)}$, where $\ell(\cd_0,\cd)$
is the time it takes for the ``bare'' disorder $\cd$ to scale up to
$\cd_0$. From this reasoning one finds, for the case without a spin gap
($g_1>0$) and weak disorder
\begin{equation}
\label{eq:xi1}
\xi (\cd )  \propto  (1/\cd )^{1/(2-K_\rho)}\point
\end{equation}
Note that for $K_\rho>1$, \ie \ superconducting fluctuations
predominating in the pure case $\xi$ is greater than the mean free path
$\lambda \propto 1/\cd$, there is a kind of diffusive regime,
contrary to the noninteracting case. On the other hand, for $K_\rho<1$
one has $\xi<\lambda$. In the vicinity of the TS--RAF boundary one has
\begin{equation}
\label{eq:xi2}
\xi (\cd )  \propto  \exp \left( \frac{K_\rho -2}{\cd - y (K_\rho -2)}
\right) \point
\end{equation}

The analogous results for the case with a spin gap ($g_1<0$) are
\begin{eqnarray}
\label{eq:xi3}
\xi (\cd ) & \propto & (1/\cd )^{1/(3-K_\rho)} \\
\label{eq:xi4}
\xi (\cd ) & \propto & \exp \left( \frac{2 \pi}{\sqrt{9 \cd - (K_\rho
-3)^2}} \right)
\end{eqnarray}
There are two points to be noted about this result: (i) for $K_\rho=0$ one
has $\xi \propto \cd^{-1/3}$, which is the same result as that found for the
pinning length of a classical CDW \cite{fukuyama_pinning}. (ii) the results
(\ref{eq:xi1}, \ref{eq:xi2}) and (\ref{eq:xi3}, \ref{eq:xi4}) are
qualitatively different, both in the vicinity of the phase boundaries and in
the localized states. The transitions are thus in different universality
classes, and this strongly supports the idea that the localized phases
reached through the transition are themselves different (RAF or CDG).

The temperature dependence of the {\sc dc} conductivity can be obtained noting
that at finite temperature there are no coherent effects on length scales
larger than $v_{{\mathrm{F}}}/T$. One
 therefore stops renormalization at $\ee^{\ell^*} =
v_{{\mathrm{F}}}/(\alpha T)$. 
As long as $\cd$ remains weak one can still use the Born
approximation to obtain
\begin{equation}
\sigma(\cd)  = \sigma_0 \frac{\ee^\ell \cd}{\cd(\ell)} \virg
\end{equation}
where $\sigma_0 = e^2v_{{\mathrm{F}}}^2/2\pi\hbar D_\xi$ is the lowest order
conductivity.  In the delocalized phases one then finds a conductivity
diverging as $ \sigma(T) \sim T^{-1-\gamma} $ where $\gamma = K_\rho^*-2$ in
the TS case and $\gamma = K_\rho^*-3$ in the SS case. On the phase
boundaries one has universally $\sigma \sim 1/T$.  In the localized region
$\cd$ diverges at low temperatures, and a perturbative calculation thus
becomes meaningless. However, the conducting--localized crossover can still
be studied at not too low temperatures \cite{giamarchi_loc}. In particular,
the high--temperature conductivity is found to vary as $\sigma \sim
T^{1-K_\rho}$. This is the perturbative result first found by Mattis
\cite{mat_bos} and by Luther and
Peschel \cite{luther_conductivite_disorder} and also reproduced by the
single--impurity calculations \cite{kane_fisher}. The high--temperature
behavior thus can be understood in terms of scattering off the individual
impurities. On the other hand, at lower temperatures one necessarily comes
into the region where $\cd$ increases sharply. This has its origin in
coherent scattering from many impurities and ultimately gives rise to
localization.

One can finally notice the effects of different types of interactions on
localization. Roughly speaking, for forward scattering  repulsion ($g_2>0$)
enhances localization whereas attraction weakens it. In particular, strong
attraction leads to vanishing effective random potential. The delocalized
state then can be considered to be a true superconductor in the sense that
there is infinite conductivity even in an impure system. The effect of backward
scattering interactions is opposite to that of forward interactions.

\subsection{The spin--1/2 chain as a Luttinger liquid}
\label{sec:s12lutt}
One of the fundamental models of solid state physics is the Heisenberg model
of insulating magnets. In the one--dimensional case (``spin chains'') its
Hamiltonian takes the form
\begin{eqnarray}
\nonumber
H & = &  \sum_{l=1}^L (S ^x_l  S ^x_{l+1}
                 +  S ^y_l  S ^y_{l+1}
                 + \Delta \, S ^z_l  S ^z_{l+1}) \\
\label{eq:hh}
& = & \sum_{l=1}^L \left( \frac12(S ^+_l  S ^-_{l+1}
                 +  S ^-_l  S ^+_{l+1})
                 + \Delta \, S ^z_l  S ^z_{l+1} \right) \point
\end{eqnarray}
Here ${{\mathbf{S}}}_l=(S_l^x,S_l^y,S_l^z)$ is a spin operator (we will first
concentrate on the case of spin $1/2$ so that
${\mathbf{S}}_l\cdot{\mathbf{S}}_l=3/4$) acting on site $l$, $\Delta$ is an
anisotropy parameter that allows one to treat the antiferromagnetic
($\Delta=1$), the ferromagnetic ($\Delta = -1$), and general anisotropic
cases, and periodic boundary conditions imply ${\mathbf{S}}_{L+1} =
{\mathbf{S}}_1$. We notice that the Hamiltonian conserves the $z$--component of
the total spin (for $|\Delta|=1$ total spin is also conserved).

The spin model can be transformed into a model of spinless fermions, noting
that $S_l^+$ and $S_l^-$ anticommute. The {\em Jordan--Wigner
transformation} \cite{jor_tran} then relates spin to fermion operators
($a^{\phantom{\dagger}}_l, a^\dagger_l$) via
\begin{equation} \label{eq:jw}
S_l^+ = a_l^\dagger \exp \left( \ii \pi\sum_{j=1}^{l-1} a^\dagger_j
a^{\phantom{\dagger}}_j \right) \virg ~~~~ S_l^z = a^\dagger_l
a^{\phantom{\dagger}}_l - \frac12 \point
\end{equation}
Presence or absence of a fermion now represent an up or down spin, and the
exponential factor insures that spin operators on different sites commute,
whereas fermionic operators of course anticommute. The transformation can
now be used to rewrite the spin Hamiltonian (\ref{eq:hh}) in terms of
fermions as
\begin{equation} \label{eq:hhfer}
H  =  \sum_{l=1}^L \left( \frac12 \left(a^\dagger_l a^{\phantom{\dagger}}_{l+1}
+ a^\dagger_{l+1} a^{\phantom{\dagger}}_l\right) + \Delta \, 
\left(a^\dagger_la^{\phantom{\dagger}}_l -\frac12 \right)\left(a^\dagger_{l+1}
a^{\phantom{\dagger}}_{l+1} -\frac12 \right) \right) \point
\end{equation}
The ``spin--flip'' terms thus give rise to motion of the fermions, whereas
the $S^z$--$S^z$ interaction leads to a fermion--fermion interaction between
adjacent sites.

It is instructive to see in detail how one can pass to the continuum limit
and then to a bosonic model starting from eq.(\ref{eq:hhfer}). We start by
passing to momentum space in the standard way:
\begin{equation} 
a_l = \frac1{\sqrt{L}} \sum_k a_k \ee ^{ \ii kl} \virg
\label{eq:alk}
\end{equation}
with the momentum sum restricted to the first Brillouin zone: $-\pi < k \leq
\pi$. Insertion into eq.(\ref{eq:hhfer}) then straightforwardly leads to
\begin{eqnarray}
H & = & \sum_k ( -\cos k - \Delta) a^\dagger_k a^{\phantom{\dagger}}_k +
  \frac{\Delta}{L} \sum_{k_i} \delta(k_1+k_2-k_3-k_4) \ee ^{ \ii (k_1-k_4)}
  a^\dagger_{k_1} a^\dagger_{k_2} a^{\phantom{\dagger}}_{k_3}
  a^{\phantom{\dagger}}_{k_4}
\label{eq:hhk} \\
& = & \sum_k ( -\cos k - \Delta) a^\dagger_k a^{\phantom{\dagger}}_k +
\frac{\Delta}{2L} \sum_{k_1k_2q} \left( \cos q - \cos(k_1 - k_2 +
q)\right)
a^\dagger_{k_1+q} a^\dagger_{k_2-q} a^{\phantom{\dagger}}_{k_2}
a^{\phantom{\dagger}}_{k_1} \virg
\label{eq:hhkb}
\end{eqnarray}
where the $\delta$--symbol insures momentum conservation modulo $2\pi$:
$\delta(x)=1$ if $x = 0$~mod~$2\pi$, and $\delta(x)=0$ otherwise. In order to
be close to the standard case of a band minimum at $k=0$ we have shifted the
origin of $k$--space by $\pi$. This amounts to a rotation of every other
spin in (\ref{eq:hh}) by $\pi$ around the $z$ axis. In eq.(\ref{eq:hhkb})
the form factor appearing in the interaction has been properly
symmetrized. This will be useful subsequently.

At least for weak interaction (small $\Delta$) only interactions involving
states close to the Fermi energy are important, and one therefore can then
map eq.(\ref{eq:hhk}) onto the spinless Luttinger model discussed in section
\ref{wcsec}. For the $g_2$ interaction, processes with either 
$k_1 \approx k_4$ or $k_1 \approx k_3$ contribute (the $\approx$ sign is
meant to indicate that both momenta are close to the same Fermi point). To
cast the second type of processes into the form of (\ref{hint}) one has to
commute two fermion operators, giving rise to an extra minus sign. The
coupling constant at small momentum transfer then is
\begin{equation} 
g_2 = 2\Delta(1-\cos2k_{{\mathrm{F}}})\point
\label{eq:g2h}
\end{equation}
Note that this vanishes when the spin chain is nearly fully spin--polarized
($k_{{\mathrm{F}}}\rightarrow 0,\pi$). 

The determination of the appropriate parameter $g_4$ is slightly less
straightforward. Naively, one would expect that processes where all four
states are in the vicinity of the same Fermi point contribute, giving $g_4 =
2
\Delta$. However, as pointed out by Fowler \cite{fowler_80}, this is not
correct: in fact, in the lattice model there are corrections to the bare
fermion energy coming from the exchange part of the first--order
Hartree--Fock selfenergy, given by
\begin{equation} 
\Sigma_{\mathrm{HF}}(k) = - \frac{2\Delta}{L} \sum_p 
\langle a^\dagger_p a^{\phantom{\dagger}}_p \rangle
= - \frac{2 \Delta}{\pi} \sin k_{{\mathrm{F}}} \cos k \point
\label{eq:shf}
\end{equation}
This leads to a renormalization of the Fermi velocity
\begin{equation} 
v_{{\mathrm{F}}} = 
\sin k_{{\mathrm{F}}} \rightarrow \sin k_{{\mathrm{F}}} + 
\frac{2 \Delta}{\pi} \sin^2k_{{\mathrm{F}}} \point
\label{eq:rvf}
\end{equation}
On the other hand, in the continuum model of section~\ref{wcsec}, such
Hartree--Fock selfenergy corrections must not be considered (formally by
appropriately normal ordering the interaction terms), otherwise they
would be infinite due to sums over infinitely many occupied states. The
proper way to account for the finite velocity renormalization,
eq.(\ref{eq:rvf}), then is to include this via a properly chosen $g_4$. In
the present case, in order to reproduce (\ref{eq:rvf}) one then has to set
\begin{equation} 
g_4 = g_2 =  2\Delta(1-\cos2k_{{\mathrm{F}}})\point
\label{eq:g4h}
\end{equation}
Note that at half--filling this is twice the naive expectation.

Following the steps of section~\ref{wcsec}, the low--energy excitations of the
spin chain model then are described by the Hamiltonian \cite{luther_chaine_xxz}
\begin{equation} 
H = \int dx \left( \frac{\pi u K}{2} \Pi(x)^2
     +\pippo \frac{u}{2\pi K} (\partial_x \phi )^2 \right) \point
\label{eq:hbos0p}
\end{equation}
This is of course our standard Hamiltonian, \ie \ {\em spin chains are
Luttinger liquids} \cite{haldane_bosonisation}. Specializing to the case of
zero applied field, so that the total magnetization vanishes and one has
$k_{{\mathrm{F}}}=\pi/2$, the parameters are given perturbatively by
\begin{equation} 
u=\sqrt{1+\frac{4\Delta}{\pi}} \quad , 
\qquad K = \frac{1}{\sqrt{1+\frac{4\Delta}{\pi}}}
\label{eq:perpar}
\end{equation}
These results agree to first order in $\Delta$ with the exact ones, to be
discussed in section~\ref{sec:3b} below, and are close to them over much of the
parameter range.

For an exactly half--filled band extra umklapp scattering processes are
possible, with $k_{1,2} \approx \pi/2$, $k_{3,4} \approx -\pi/2$, and {\it vice
versa} \cite{haldane_xxzchain,bla_equ,nijs_equivalence}. In a continuum
representation, these operators become
\begin{equation} 
H_{\mathrm{u}} = \Delta \int dx \left 
( \left(\psi^\dagger_R \partial_x \psi^\dagger_R\right)
\left(\psi^{\phantom{\dagger}}_L \partial_x 
\psi^{\phantom{\dagger}}_L\right) + {\textrm{h.c.}}
\right) =
\frac{\Delta }{2(\pi\alpha)^2} \int \cos 4\phi(x)
\label{eq:humkl}
\end{equation}
This operator has scaling dimension $4K$ and therefore is strongly
irrelevant for small $\Delta$. However, it becomes relevant for $K<1/2$ and
then in particular is responsible for the creation of a gap in the
excitation spectrum and long--range antiferromagnetic order in the case of
an Ising type anisotropy ($\Delta>1$).
		\subsubsection{Physical properties of the spin 1/2 chain
		(small $\Delta$)}
\label{sec:3b}
From the bosonized form of the Hamiltonian one immediately obtains the
specific heat of the spin chain as
\begin{equation} 
C(T) = -\frac{T}{L} \frac{\partial^2 F}{\partial T^2} = 
\frac{\pi}{6} \frac{k_B^2 T}{u} \point
\label{eq:cspc}
\end{equation}
The susceptibility for a field applied along the $z$--direction can also be
obtained, noting that $\partial_x \phi$ is proportional to the fermion
density, \ie \ the $z$--component of the magnetization:
\begin{equation} 
\chi = \frac{4\mu_B^2}{\pi} \frac{K}{u} \point
\label{eq:susKu}
\end{equation}

Beyond these thermodynamic properties the fermionic analogy allows one to
study correlation functions of the spin chain using the bosonization
formalism developed in section \ref{sec:bosonspl}. We start by rewriting the
fermion operators in a continuum form:
\begin{equation} 
a_l = \psi_R(x) + \psi_L(x) \virg
\label{eq:psicont}
\end{equation}
where $x = la$ and in most of what follows the lattice constant $a$ is set
to unity. In order to calculate spin--spin correlation functions we further
need a representation of the exponential factor in eq.(\ref{eq:jw}). This is
given by
\begin{equation} 
\exp \left( \ii\pi\sum_{j=1}^{l-1} a^\dagger_j
a^{\phantom{\dagger}}_j \right) = 
\ee ^{\ii k_{{\mathrm{F}}}x} \ee ^{- \ii \phi(x)} \virg
\label{eq:steph}
\end{equation}
where the first exponential factor comes from the mean value of the density
and the second one represents the effect of fluctuations about that mean,
see eq.(\ref{eq:den}).

We can now use the continuum representations (\ref{eq:psicont}) and
(\ref{eq:steph}) together with the fermionic representation of the
spin--$1/2$ operators, eq.(\ref{eq:jw}), to give a bosonic representation of
the spins:
\begin{eqnarray}
\nonumber 
S_l^z & = & \psi^+_R \psi_R(x) + \psi^+_L \psi_L(x)
 + \psi_L^+ \psi_R(x)  + \psi_R^+ \psi_L(x)  =
-\frac1{\pi} \partial_x \phi + \frac1{\pi\alpha}
 \cos(2\phi(x)-2k_{{\mathrm{F}}}x) \\
S_l^+ &  = & \frac1{\sqrt{2\pi\alpha}}
\left( (-1)^n \ee ^{- \ii \theta} + \ee ^{ \ii (2k_{{\mathrm{F}}}-\pi)x} 
\ee ^{-2i\phi - \ii \theta}
\right) \point
\label{eq:sbos}
\end{eqnarray}
Here we have omitted the Klein factors which will be unimportant in the
following, and have kept a finite $\alpha$ as a short--distance cutoff, to
be taken of the order of a lattice constant. Further, in $S_l^+$ we have
restored the factors $(-1)^l$ that were lost in going from
eq.(\ref{eq:hhfer}) to eq.(\ref{eq:hhkb}). Using these expressions we can
now follow the calculations of section~\ref{sec:bosonspl} to obtain the
spin--spin correlation functions. In particular, at zero temperature
\cite{luther_chaine_xxz}
\begin{eqnarray}
\nonumber 
\langle T_\tau S_l^z(\tau) S_0^z(0) \rangle &
  = & \frac{K}{\pi^2} \frac{x^2-u^2\tau^2}{(x^2+u^2\tau^2)^2} + 
    \frac{A\cos(2k_{{\mathrm{F}}}x)}{(x^2+u^2\tau^2)^K} \\
\langle T_\tau S_l^+(\tau) S_0^-(0) \rangle & = &
 B_1 \cos((2k_{{\mathrm{F}}}-\pi)x)
 \frac{x^2-u^2\tau^2}{(x^2+u^2\tau^2)^{K+1/4K+1}}
+ \frac{B_2 \cos \pi x}{(x^2+u^2\tau^2)^{1/4K}} \point
\label{eq:scfct}
\end{eqnarray}
Here the constants $A,B_{1}$ cannot be reliably determined by the present
methods, however in the nonoscillating part of the $z$--$z$ correlation
function there is no undetermined parameter, and very recently $B_2$ has
been determined.\cite{lukyanov98} In most cases the dominant contributions
in eq.(\ref{eq:scfct}) come from the second, quickly oscillating terms.  The
alternation indicates the expected tendency towards antiferromagnetic order,
but correlations do decay (with a rather slow power law), and there is thus
no long--range order, as to be expected in one dimension.

Fourier transforming eq.(\ref{eq:scfct}) gives, after analytic continuation
to real frequencies and generalizing to nonzero temperature \cite{schu_spins}
\begin{eqnarray} 
\nonumber
\chi_\parallel(q,\omega) & = &
- \frac{A \sin(\pi K)}{u} \left( \frac{2\pi T}{u} \right)^{2K-2}
 B\left(\frac{K}{2} - \ii \frac{\omega+uq'}{4\pi T},1-K\right)
 B\left(\frac{K}{2} - \ii \frac{\omega-uq'}{4\pi T},1-K\right)\\
& & \nonumber \quad + \frac{\pi A}{u(1-K)} \\
& = & - \sin(\pi K) \Gamma^2(1-K) \frac{A}{u^{2K+1}} [u^2 q'^2 -
(\omega+\ii\delta)^2]^{K-1} \quad \textrm{at $T=0$} \point
\label{eq:spara}
\end{eqnarray}
Here $B(x,y) = \Gamma(x) \Gamma(y)/\Gamma(x+y)$ is Euler's beta function,
$q'=q-2k_{{\mathrm{F}}}$, and the last term has been introduced ``by hand'' 
in order to
reproduce the known logarithmic result for $K=1$. For the transverse ($+-$)
correlations, an analogous result holds, with $K\rightarrow 1/(4K)$ and $q'
= q -\pi$. One should notice that eq.(\ref{eq:spara}) is valid uniformly for
arbitrary ratios between the energies $T$, $\omega$, and $u|q'|$, as long as
all of them are small compared to the exchange energy (which here play the
role of the bandwidth). At zero temperature, a finite imaginary part of
$\chi_\parallel$ and $\chi_\perp$ exists in the V--shaped region $|\omega|
\geq u|q'|$. This represents precisely the regions where the Bethe ansatz
solution produces the two--spinon continuum.\cite{faddeev_spin1/2} The
magnetic scattering cross section obtained from eq.(\ref{eq:spara}) is in
excellent agreement with
experiment.\cite{tennant_kcuf,tennant_kcuf_1d,dender97}  

\begin{figure}
\centering
\mbox{	\subfigure[]{\epsfxsize 6cm \rotate[l]{\epsffile{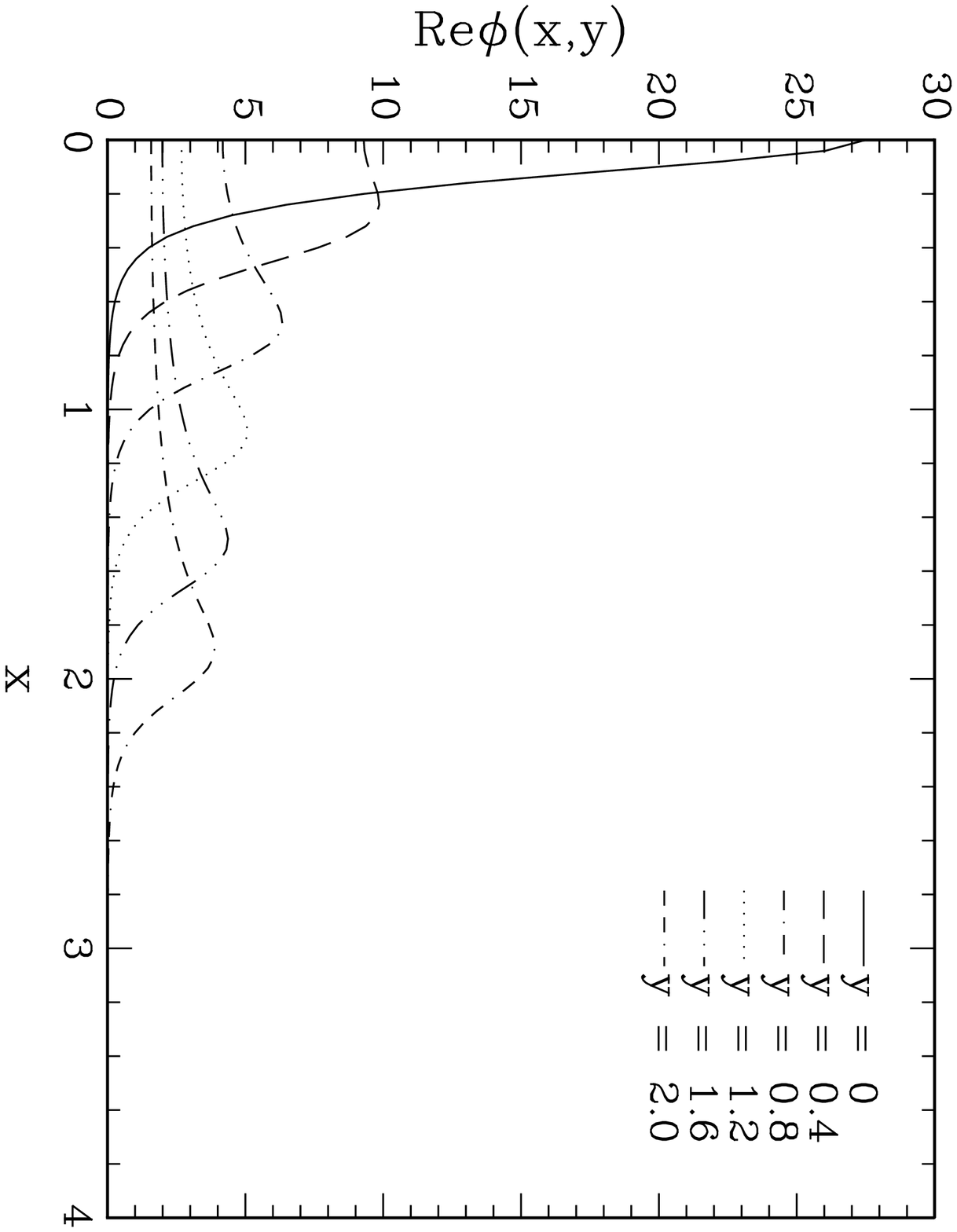}}}
        \subfigure[]{\epsfxsize 6cm \rotate[l]{\epsffile{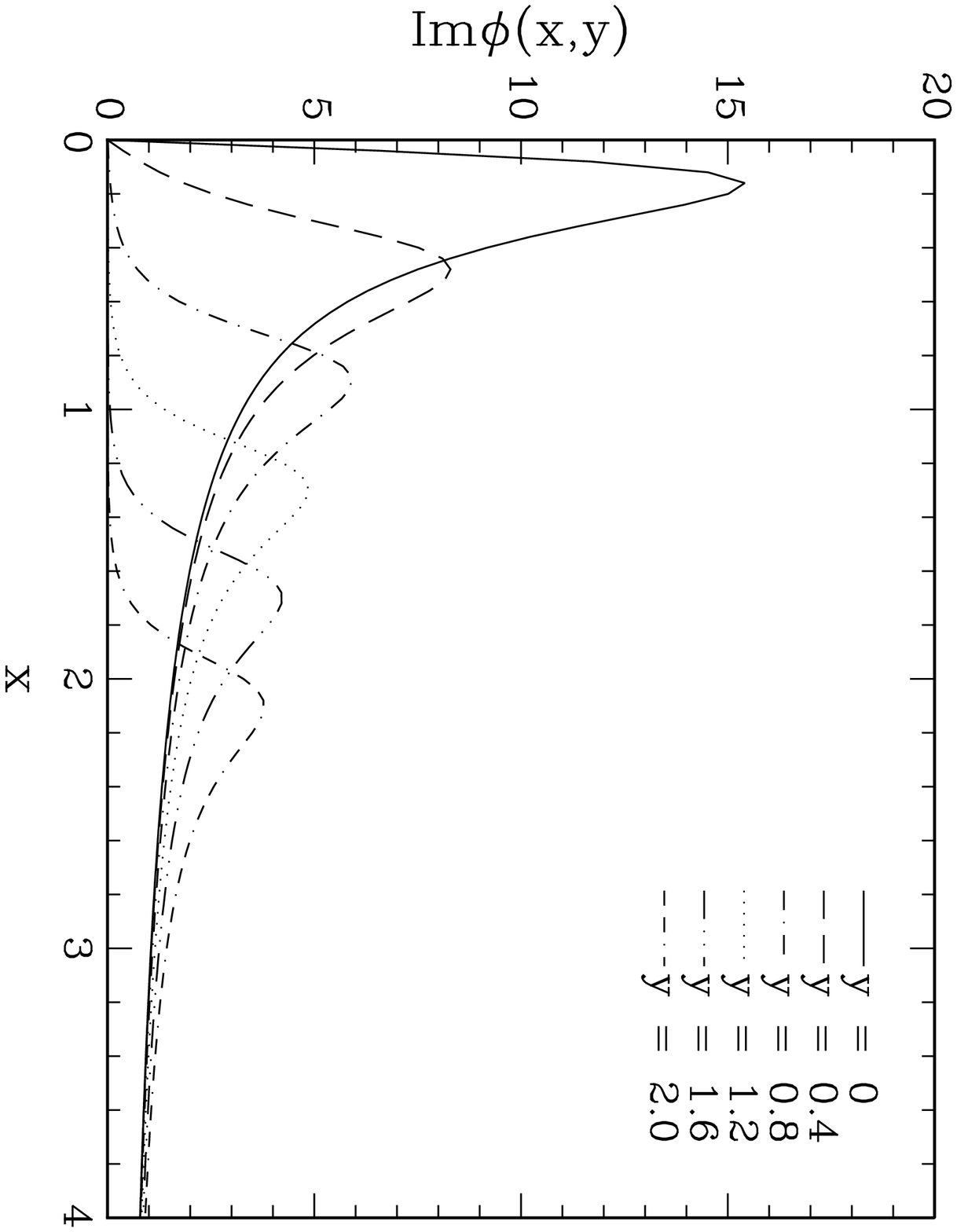}}}
	}
\SMALLCAP{t}{Real (a) and imaginary (b) part of the quantum critical scaling 
function $\phi(x,y)$ for the isotropic Heisenberg antiferromagnet ($K=1/2$).}  
\label{f:3b1}
\end{figure}

At finite temperature, the susceptibilities take the general quantum
critical scaling form \cite{chubukov94}
\begin{equation} 
\chi_{\parallel,\perp}(q,\omega) = \frac{A_{\parallel,\perp}}{T^{2-2K}}
\phi_{\parallel,\perp}\left(\frac{uq'}{4\pi T},\frac{\omega}{4\pi T}\right)
\virg
\label{eq:chiscal}
\end{equation}
with the scaling functions $\phi_{\parallel,\perp}$ given by the products
of beta functions in eq.(\ref{eq:spara}). Figure~\ref{f:3b1} shows plots of
this function
for the particularly relevant case of the isotropic antiferromagnet,
$K=1/2$. Note that these function has maxima for $x \approx y$. For $x
\approx y \rightarrow \infty$ these maxima develop into the
(zero--temperature) square root singularity of eq.(\ref{eq:spara}).

Generalizing the arguments leading to eq.(\ref{eq:sbos}) it has been
shown \cite{cro_spinpei} that the nearest--neighbor interaction contains
an oscillating part of the form
\begin{equation} 
{\mathbf{S}}_l \cdot {\mathbf{S}}_{l+1} \approx \frac1{\pi\alpha}
\sin(2\phi(x)-2k_{{\mathrm{F}}}x)
\point
\label{eq:dim}
\end{equation}
Correlation functions of this operator, which describes the tendency toward
dimerization (or spin--Peierls order), then decay with the same power law as
those of the $z$ component of the spin.

An important consequence of the above discussion is that the low temperature
correlations are only determined by two parameters: one, $K$, determines the
power laws of the decay of correlations, the other one, $u$, is the velocity
of the excitations. In the previous section, we have given explicit formulae
for these constants, eq.(\ref{eq:perpar}), however these can only be
expected to be valid for small $\Delta$ where the linearization of the
fermion spectrum is expected to be {\em quantitatively} reliable. It is
clearly interesting to determine these parameters outside the perturbative
regime. The exact Bethe wavefunction is, at least up to now, too complicated
as to allow calculation of correlation functions (however, for some recent
progress see \cite{bougourzi96,karbach97}). One point where a
non--perturbative result can be obtained is the isotropic Heisenberg
antiferromagnet $\Delta=1$ in zero field, so that $k_{{\mathrm{F}}} =
\pi/2$.  Then spin rotation invariance requires the longitudinal and
transverse correlation functions to be equal, and thus $K=1/2$ is needed in
eq.(\ref{eq:scfct}).  Note that this implies that the dimerization
correlations described by eq.(\ref{eq:dim}) also decay with a $1/r$ power
law.

For general $\Delta$ a more indirect approach to the determination can be
used \cite{haldane_xxzchain}. One notices that the bosonized Hamiltonian
leads to a variation of the ground state energy with the number of right--
and left--going particles as 
\begin{equation} 
E(N,J) = \frac{\pi}{2L} \left( u K J^2 + \frac{u}{K} N^2 \right) \virg
\label{eq:enj}
\end{equation}
where $N$ and $J$ are the sum and difference of the number of added right--
and left--going particles, and eqs.(\ref{phi0}) to (\ref{eq:NJ}) have been
used. Haldane introduces the charge and current velocities $v_N = u/K$ and
$v_J = u K$. On the other hand, the variation of the ground state energy with
$N$ and $J$ can also be obtained from the Bethe solution (or even from a
numerical solution on a finite lattice, if an exact solution is not available).
In particular, for the spin--$1/2$ chain in zero field \cite{yang_xxz}
\begin{equation} 
E(N,0) = \frac{(\pi-\arccos \Delta)\sqrt{1-\Delta^2}}{2\arccos \Delta}
\frac{N^2}{L} \point
\label{eq:en0}
\end{equation}
This fixes $u/K$. Note that the coefficient of $N^2$ in eq.(\ref{eq:enj}) is
proportional to the inverse susceptibility. The variation with $J$ can also
be obtained, noting that $J \propto \partial_x \theta$ can be related to
the stiffness constant of the $x$ and $y$ components of the spin (compare
eq.(\ref{eq:sbos})).  However, it is simpler to use directly the known
result for the velocity of the elementary excitations \cite{descloizeaux66}:
\begin{equation} 
 u = \frac{\pi \sqrt{1-\Delta^2}}{2\arccos\Delta} \point
\label{eq:uofd}
\end{equation}
Alternatively, the velocity can also be obtained from the low--temperature
specific heat \cite{takahashi73}, leading to the same result.  This then
fixes the correlation exponent as
\begin{equation} 
K = \frac{\pi}{2(\pi-\arccos\Delta)} \virg
\label{eq:kofd}
\end{equation}
first obtained in \cite{luther_chaine_xxz} by a scaling argument using the
exact solution of the fully anisotropic XYZ model \cite{bax_xyz}. The exact
results (\ref{eq:uofd}) and (\ref{eq:kofd}) agree to linear order in
$\Delta$ with eq.(\ref{eq:perpar}), as expected, and even beyond first order
the approximate results are quite close to the exact ones. One should also
notice that to first order in the interaction $\Delta$ the coefficients in
eq.(\ref{eq:enj}) can be calculated using the zeroth--order wavefunction,
\ie \ the noninteracting Slater determinant. This of course reproduces
eq.(\ref{eq:g4h}) and probably represents the ``cheapest'' way to calculate
the anomalous exponent $K$. The exact expressions (\ref{eq:uofd}) and
(\ref{eq:kofd}) are valid for vanishing magnetization, \ie \ no applied
field. Numerical results for a finite magnetization along $z$ (\ie \ a
non--half--filled fermion band) have been given by
Haldane \cite{haldane_xxzchain}.

\subsubsection{The isotropic antiferromagnet ($\Delta=1$)}
Up to now we have ignored the effect of the umklapp operator,
eq.(\ref{eq:humkl}). Indeed, as long as it is irrelevant, \ie \ in the region
of planar anisotropy $\Delta <1$, umklapp interactions only lead to
subleading corrections to correlation functions \cite{giamarchi_logs}.
However, at the isotropic antiferromagnetic point $\Delta=1$, the umklapp
operator is only marginally irrelevant. Then the corresponding coupling
constant is renormalized
as \cite{cardy86,affleck_log_corr,giamarchi_logs,singh_logs} 
\begin{equation} 
\frac{dg}{d\ell} = -\pi bg^2 \virg
\label{eq:rgg}
\end{equation}
with solution
\begin{equation} 
\pi bg(\ell) = \frac{\pi bg}{1 + \pi b g \ln L} = \frac1{\ln{L/L_0}} \point
\label{eq:rgg2}
\end{equation}
Here $b= 4/\sqrt{3}$ is a normalization constant fixed by the requirement
that the correlations of the marginal operator decay as $r^{-4}$ with unit
coefficient, $L = \ee ^\ell$ is the rescaled short distance cutoff (lattice
constant), and $L_0 = \exp(-1/(\pi b g))$. The determination of the bare
coupling constant $g$ is not entirely trivial. Naively from
eq.(\ref{eq:humkl}) one would set it of order $\Delta$. However, for the
isotropic antiferromagnet $\Delta =1$, giving a coupling of order unity
which would be expected to be outside the perturbative domain of validity of
eq.(\ref{eq:rgg}). A precise determination can be achieved noting that the
marginal operator affects the low--lying excited states in a finite system
in a well--understood way involving only the combination $\pi b
g(\ell)$ \cite{cardy86,affleck_log_corr}. One finds in particular that the
predicted linear variation of $\pi b g(\ell)$ with $\ln(L)$ is satisfied to
within a few percent, and an order of magnitude better agreement is found if
two--loop corrections are included \cite{nomura93}. From the two--loop
calculation the bare coupling constant (at $L=1$) can be estimated as $g
\approx 0.24$. A priori, this would be expected to be an effective coupling,
reproducing correctly the long--distance behavior, however in fact the
spectra of all but the shortest chains are rather well fitted by this
value. On the other hand, there are considerable uncertainties associated
with this estimate: fitting the same data to the one-loop function one finds
$g \approx 0.11$.

The umklapp operator lead to logarithmic corrections in the temperature
dependence of the spin susceptibility, of the form \cite{eggert94}
\begin{equation} 
\chi(T) = \frac{4\mu_B^2}{\pi^2} \left( 1 + \frac{1}{2 \ln (T_0/T)} \right)
\virg 
\label{eq:chit}
\end{equation}
with $T_0 \approx 7.7$, obtained from fitting eq.(\ref{eq:chit}) to exact
finite--temperature Bethe ansatz results. Experimental results on the nearly
perfectly one--dimensional compound $\rm Sr_2CuO_3$ \cite{motoyama_srcuo}
are in excellent agreement with the logarithmic law (\ref{eq:chit}) and with
Bethe ansatz results
\cite{eggert94} over a wide temperature regime. Similar logarithmic corrections
also exist in the zero temperature magnetization curve $M(H)$
\cite{yang_xxz}. On the other hand, there are no such corrections in the
low--temperature specific heat.

The marginal operator further produces multiplicative logarithmic
corrections in various correlation functions,
e.g.$\!$ \cite{affleck_log_corr,giamarchi_logs,singh_logs}
\begin{equation} 
\langle T_\tau \S_x(\tau) \cdot \S_0(0) \rangle  = \frac{3}{(2\pi)^{3/2}}
 \frac{(-1)^x}{x^2+u^2\tau^2} \ln^{1/2} 
\left(\frac{\sqrt{x^2+u^2\tau^2}}{L_0} \right)
\virg
\label{eq:scfctl}
\end{equation}
where the prefactor has only been determined very recently.\cite{affleck98}
An identical logarithmic correction is found in correlations of the
dimerization operator, eq.(\ref{eq:dim}), however with exponent $-3/2$
instead of $1/2$, \ie \ the dimerization fluctuations are logarithmically
suppressed compared to the antiferromagnetic ones. Also the numerical
prefactor in those correlations is not currently known. Similarly, the
staggered susceptibility has a logarithmic correction factor:
\begin{equation} 
\chi(\pi,0;T) = \frac{A}{T} \ln^{1/2} (T_0/T) \point
\label{eq:cstl}
\end{equation}
Numerical investigations of correlation functions initially shed doubt on
the existence of these logarithmic corrections \cite{kubo88}, however more
recent numerical \cite{hallberg95,koma96,eggert96,starykh97a,starykh97b} and
analytic \cite{bougourzi96,karbach97} work provides ample evidence for their
existence. In particular, the constants in eq.(\ref{eq:cstl}) have been
determined as $A=0.32 \pm 0.01$, $T_0 = 5.9 \pm 0.2$ \cite{starykh97b}. The
difference of $T_0$ with the value obtained from the susceptibility is
possibly due to the fact that the susceptibility data were fitted in the
asymptotic low--temperature region ($T\le10^{-2}$), whereas the staggered
susceptibility was calculated at somewhat higher temperatures.

Apart from being directly accessible in neutron scattering
experiments,\cite{tennant_kcuf,tennant_kcuf_1d,dender97} the above staggered
spin correlation function leads to characteristic temperature dependences in
the longitudinal and transverse NMR relaxation times \cite{starykh97a}
(under the assumption of a hyperfine--dominated relaxation)
\begin{equation} 
\frac{1}{T_1} \propto \left(\ln \frac{T_0}{T} \right)^{1/2} \quad ,
\quad  \frac{1}{T_{2G}} \propto \left(\frac1{T} \ln \frac{T_0}{T} \right)^{1/2}
\virg
\label{eq:nmrrel}
\end{equation}
in good agreement with experiments on $\rm Sr_2CuO_3$.\cite{takigawa97}
In the absence of the marginal operator the logarithmic factors would be
absent \cite{sachdev94a}, leading in particular to a
temperature--independent $T_1$.

We finally mention that for $\Delta >1$ the spins are preferentially aligned
along the $z$--direction, and one then has a long--range ordered ground
state of the Ising type. There is thus a phase transition exactly at the
isotropic point $\Delta=1$. In the fermionic language, this corresponds to a
metal--insulator transition.\cite{shankar_spinless}

We can summarize this section by noting that quantum spin chains provide one
of the experimentally best established cases of Luttinger liquid
behavior. This largely due to two facts: (i) the relative ease
with which one can define a microscopic Hamiltonian for a given experimental
system: one rarely has to go beyond a slightly modified Heisenberg model,
with very few interaction constants to be determined; this is to be compared
with the difficulties one encounters in conducting systems: long--range
interactions, electron--phonon interactions, etc.; (ii) the availability of
a large number of experimental techniques which give results directly
comparable with experiment and, concomitantly, the possibility of using
either very well--controlled (often exact) theoretical methods or numerical
approaches which for spin systems are much more reliable than for itinerant
fermions.

\sectio{Spin ladders and coupled Luttinger liquids}
\subsection{Coupled spin chains}
\label{sec:spins}
Investigating models of coupled parallel chains is of interest for a number
of reasons: (i) {\em quasi}--one--dimensional antiferromagnets always have
some form of interchain coupling, usually leading to three--dimensional
ordering at sufficiently low
temperatures;\cite{satija_kcuf,tennant_kcuf_q1d,keren_srcuo} (ii) there is a
number of ``spin--ladder'' compounds containing a small number of coupled
chains;\cite{azuma_srcuo,dagotto_ladders} (iii) coupled spin--1/2 chain models
can be used to describe higher spin quantum numbers.\cite{luther_spin1}

Consider $N$ coupled spin--1/2 chains with spin degrees of freedom
$\bbox{S}_{j}$, $j=1..N$, described by the Hamiltonian
\begin{equation} \label{eq:twoc}
H = \sum_{j=1}^N H(\bbox{S}_j) + 
\sum_{j<k} \lambda_{jk} H_c(\bbox{S}_j,\bbox{S}_k) \virg \quad
H_c(\bbox{S}_j,\bbox{S}_k) = \sum_i \bbox{S}_{j,i} \cdot 
\bbox{S}_{k,i} \virg
\end{equation}
where $i$ labels sites along the chains, $j$,$k$ label the chains, and
$H(\bbox{S}_j)$ is of the form (\ref{eq:hh}). For $\lambda_{jk}\equiv -1$
the ground state of this model is exactly that of the spin--$N/2$ chain, each
site being in a state of total spin $N/2$,\cite{luther_spin1} but the model
can be considered for general $\lambda$, both ferromagnetic ($\lambda<0$)
and antiferromagnetic ($\lambda>0$). Performing now the Jordan--Wigner
transformation for the $\bbox{S}_j$ separately and going to the boson
representation the Hamiltonian becomes\cite{schulz_spins}
\begin{eqnarray} 
\nonumber
H & = & \int dx \left[ \frac{\pi u K}{2} \bbox{\Pi}(x)^2
       + \frac{u}{2\pi K} (\partial_x \bbox{\phi} )^2 
       + \frac{u}{2\pi \bar{K}} (\partial_x \bar{\phi} )^2 \right] \\
\label{eq:hams1}
+&  \lefteqn{%    
 \frac{1}{(\pi\alpha)^2}  \sum_{j<k}
\int dx \{ \lambda_{1,jk} \cos(2( \phi_j+\phi_k)) 
+ \lambda_{2,jk} \cos(2( \phi_j-\phi_k))
+ \lambda_{3,jk} \cos( \theta_j-\theta_k)    \} \virg} &
\end{eqnarray}
Here $\bbox{\phi} = (\phi_1,\phi_2,..,\phi_N)$, $\bar{\phi} = \sum_j
\phi_j/\sqrt{N}$, the constants $u,K,\bar{K}$ all depend on the different
constants in the original Hamiltonian, and the $\lambda_{1,2,3;jk}$ are all
proportional to $\lambda_{jk}$. 

Elementary power counting, using the result eq.(\ref{eq:scfctl}) for the
spin correlations, shows that the coupling term $H_c$ always is a strongly
relevant perturbation. Consequently, an explicit renormalization group
calculation \cite{schulz_spins} shows that either $\lambda_2$ or $\lambda_3$
always scale to strong coupling, i.e. the ``relative'' degrees of freedom
$\phi_j - \phi_k$ all acquire a gap.  In particular, for not too strong
anisotropy, the $\lambda_3$ operator dominates, giving rise to long--range
order in the $\theta_j-\theta_k$ and correspondingly exponential decay of
the $\phi_j-\phi_k$ correlations. Integrating out these massive degrees of
freedom an effective Hamiltonian for the ``global'' $\bar{\phi}$ mode is
found:
\begin{equation}
\label{eq:heff}
H = \frac{u}{2} \int \left\{\pi K \bar{\Pi}(x)^2 + \frac{1}{\pi K}
(\partial_x \bar{\phi})^2 + g \cos( \mu \sqrt{N} \bar{\phi})
- \frac{\sqrt{N}}{\pi} h \partial_x \bar{\phi} \right\} \virg
\end{equation}
where $\mu=2$ for even $N$ and $\mu=4$ for odd $N$, the coefficients $u,K,g$
are renormalized, and $h$ is an external magnetic field applied along the
$z$ direction. Similarly, the leading contribution to spin correlations
comes from the operators
\begin{equation}
\label{eq:ops}
S^+(x) \propto e^{i\pi x} e^{-i \bar{\theta}/\sqrt{N}} \virg \quad
S^z(x) \propto e^{i\pi x} \cos(2\sqrt{N} \bar{\phi}) \virg
\end{equation}
where the second equation applies to odd $N$ only. 

From this a number of important conclusions can be drawn.\cite{schulz_spins}
We first notice that massless excitations and the corresponding slow
algebraic decay of correlation functions are only possible if the $\cos$
term in eq.(\ref{eq:heff}) is irrelevant. Moreover from eq.(\ref{eq:ops}) it
follows that spin correlation functions are isotropic only if $K=1/(2N)$,
implying a decay as $1/x$, as in the $S=1/2$ case. For the case of odd $N$
(equivalently, for ferromagnetic $\lambda$, for half--odd--integer S) this
is indeed the correct behavior: for $\mu=4$ the $\cos$ term is marginally
irrelevant. Thus both antiferromagnetic spin chains with half--odd--integer
S and for odd numbers of coupled $S=1/2$ chains massless behavior is
predicted, with correlations asymptotically decaying like those of a
spin--1/2 chain.\cite{schulz_spins} There is both numerical
\cite{ziman_spin3/2,hallberg_s3/2,white_nchains,frischmuth_nchains} and
experimental \cite{azuma_srcuo} evidence that this is correct.

On the other hand, for even $N$ (equivalently, integer $S$) the $\cos$ term
in eq.(\ref{eq:heff}) is strongly relevant and therefore generates a gap
$\Delta_s$ in the spin excitations. For the integer--$S$ spin chains this is
the well--known and verified Haldane prediction \cite{haldane_gap}, but
there also is a gap for any even number of coupled
chains.\cite{schulz_spins,strong_millis} Analogous conclusions concerning
qualitative differences between even and odd numbers of coupled chains have
also been reached based on the nonlinear
$\sigma$--model.\cite{sierra_nchains} The gap implies exponential decay of
spin correlations. Numerical \cite{white_nchains,frischmuth_nchains} and, at
least for $N=2$, experimental work\cite{azuma_srcuo} confirms this picture.
Another prediction, again valid both for integer--$S$ antiferromagnets and
even numbers of coupled chains, concerns the effect of an applied magnetic
field:\cite{schulz_spins,affleck_field,tsvelik_field} as long as the field
is smaller than a critical field $h_c \propto \Delta_s$, the ground state is
unchanged and has zero magnetization. However, beyond $h_c$ the
magnetization is expected to increase as $M\propto
\sqrt{h-h_c}$. Experiments on $S=1$ antiferromagnetic chains confirm this
prediction.\cite{katsumata_nenp_field} 

A natural question left open by the above considerations concerns
quasi--one--dimensional antiferromagnets like $\rm KCuF_3$
\cite{satija_kcuf,tennant_kcuf_q1d} or $\rm
Sr_2CuO_3$,\cite{keren_srcuo,motoyama_srcuo} which can be considered as the
$N \rightarrow \infty$ limit of the above model. One clearly expects (and
observes) true antiferromagnetic order at sufficiently low temperatures, at
first sight in contradiction both with the exponential decay of spin
correlation predicted for even $N$ and the universal $1/x$ law for odd
$N$. However, one should note that on the one hand the correlation length of
the even--$N$ systems is expected to increase quickly with increasing $N$,
and that on the other hand the $1/x$ correlation law of the odd--$N$ systems
also is expected to be only valid beyond a correlation length $\xi(N)$ which
increase with $N$. In the thermodynamic limit $N\rightarrow \infty$ this
then is perfectly consistent with the existence of long--range
order. Theoretical treatments of magnetic order in quasi--one--dimensional
antiferromagnets can be based on a mean-field treatment of the interchain
interaction \cite{scalapino_q1d,schulz_q1daf} which gives quantitative
predictions for systems like $\rm KCuF_3$ or $\rm Sr_2CuO_3$.\cite{kojima97}

\subsection{Two coupled Luttinger liquids}
\begin{figure}[htb]
\centerline{\epsfysize 6cm
{\epsffile{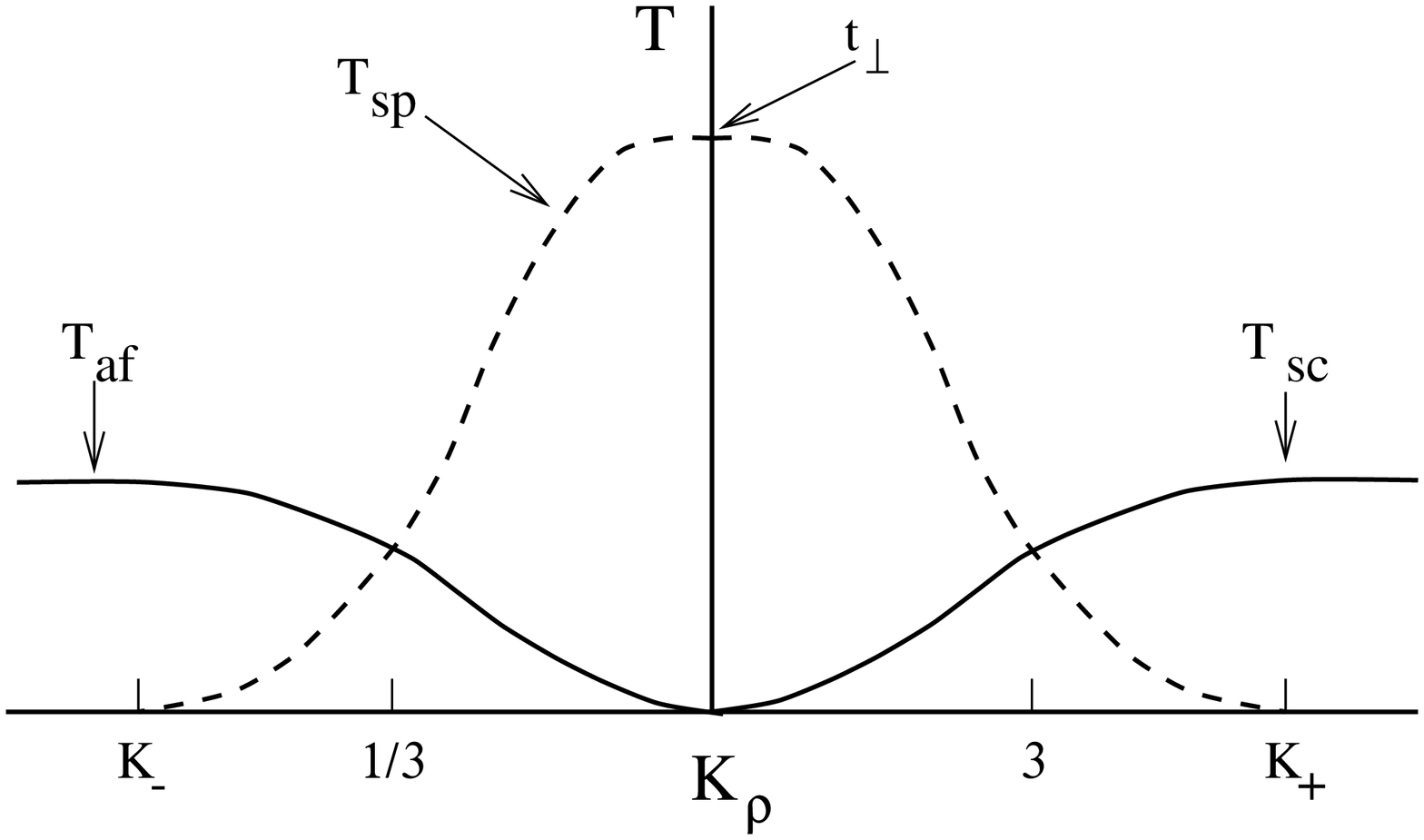}}
}
\caption[to]{Qualitative phase diagram in the temperature--$K_\rho$ plane
for Luttinger liquids coupled by interchain hopping. $K_\pm = 3 \pm \sqrt8$.}
\label{f:1}
\end{figure}

It is clearly of interest to see what happens to the peculiar
one--dimensional behavior when one puts chains in parallel. This question is
of relevance for the understanding of quasi--one--dimensional
conductors,\cite{jerome_revue_1d} doped spin ladders,\cite{uehara96}
few--channel quantum wires,\cite{yacoby96} and generally for the
understanding of possible non--Fermi--liquid behavior and
correlation--induced superconductivity in higher--dimensional solids. Of
particular interest is the effect of an interchain single--particle
tunneling term of the form
\begin{equation} \label{eq:tun}
H_{ij} = -t_\perp \int dx (\psi^\dagger_{rs i}
\psi^{\phantom{\dagger}}_{rs j}+h.c.)  \virg
\end{equation}
where $\psi^{\phantom{\dagger}}_{rs j}$ is the fermion field operator for
right ($r=+$) or left ($r=-$) going particles of spin $s$ on chain
$i$. Simple scaling arguments \cite{schulz_trieste} lead to the ``phase
diagram'' shown in fig.\ref{f:1}. The dashed line represents the crossover
below which single--particle tunneling becomes strongly relevant and below
which one thus expects Fermi liquid like behavior (an alternative
interpretation is due to Anderson\cite{anderson_q1d,clarke_couplage}).  The
full lines indicate where two--particle or particle--hole hopping becomes
relevant. The most plausible interpretation is that this is the temperature
where three--dimensional long--range order of some type sets in.  A more
detailed discussion of this in quasi--one--dimensional systems (a
thermodynamically large number of chains) has been given
elsewhere.\cite{bourbon_couplage,boies_moriond} It is also worth noting that
a different approach to the crossover to higher dimensions, working with
continuously varying spatial dimension, comes to similar
conclusions.\cite{castellani_continuousD}

As a model for doped spin--ladder systems, as well as a first step
towards a many--chain system, one can study the two--chain
case.\cite{fabrizio_2chain,finkelstein_2chain,schulz_2chain} The tunneling
term, eq.(\ref{eq:tun}), then leads to a splitting of the single--particle
bands into symmetric and antisymmetric combinations which we label by
transverse wavenumbers $k_\perp = 0,\pi$. Now each $k_\perp$ mode can be
bosonized separately. Introducing the linear combinations $\phi_{\nu\pm} =
(\phi_{\nu 0}\pm\phi_{\nu\pi})/\sqrt2$ ($\nu=\rho,\sigma$) the Hamiltonian
(including $t_\perp$) then takes the form
\begin{eqnarray} \nonumber
H &=& H_0 + H_{int,1} + H_{int,2}\virg \quad 
H_0 = \frac{\pi v_F}{2} \sum_{{\nu=\rho,\sigma } \atop {\alpha=\pm}}
\int dx
\left[\Pi^2_{\nu\alpha}
+\frac{1}{\pi^2}(\partial_x\phi_{\nu\alpha})^2\right] \\
\nonumber
H_{int,1} &=& -  \frac{g_1}{4} \int dx \left[ \frac{1}{\pi^2} \{(\partial_x
\phi_{\rho+})^2 + (\partial_x \phi_{\sigma+})^2 \} -\Pi_{\rho+}^2
-\Pi_{\sigma+}^2  \right] \\
\nonumber
 & + & \frac{g_1}{2(\pi\alpha)^2}
\int dx \{
\cos2\phi_{\sigma+}(\cos 2\theta_{\rho-} + \cos 2\phi_{\sigma-} -
\cos 2\theta_{\sigma-}) - \cos 2\theta_{\rho-} \cos 2\theta_{\sigma-} \} \\
\nonumber 
H_{int,2} & = & \frac14 \int dx \sum_{\gamma=\pm} g^{(2)}_\gamma
\left[ \frac{1}{\pi^2} (\partial_x \phi_{\rho\gamma})^2 -
\Pi^2_{\rho\gamma} \right] \\
\label{eq:h2c}
&&+  \frac{g^{(2)}_{00\pi\pi}}{2(\pi\alpha)^2} \int dx \cos
2 \theta_{\rho-} (\cos 2\phi_{\sigma-}+\cos 2 \theta_{\sigma-}) \point
\end{eqnarray}
Here $g^{(2)}_\gamma= g^{(2)}_{0000} + \gamma g^{(2)}_{0\pi\pi 0}$, and
$g^{(i)}_{abcd}$ is the coupling constant for an interaction scattering two
particles from $k_\perp$--states $(a,b)$ into $(d,c)$. The signs of the
different interaction terms in eq.(\ref{eq:h2c}) have been determined
following the reasoning explained in the appendix. Here we consider a
case where there is only intrachain interaction, implying that all {\em
bare} coupling constants are independent of $k_\perp$. 

For the pure forward scattering model ($g_1 \equiv 0$) the only nonlinear
interaction ($g^{(2)}_{00\pi\pi}$) scales to infinity, leading to a gap in
the $(\rho-)$ modes and in {\em half} of the $(\sigma-)$ modes. The
remaining $(\sigma-)$ modes are protected by the duality symmetry under
$\phi_{\sigma-} \leftrightarrow \theta_{\sigma-}$, the $(\sigma-)$ sector is
in fact a critical point of the Ising type for $g_1 \equiv
0$.\cite{schulz_2chain} For repulsive interactions here the dominant
fluctuations are of CDW type, with decay proportional to
$r^{-(3+2K_{\rho})/4}$, and $K_{\rho}^2=(\pi v_F - g_2 + g_1/2)/(\pi v_F +
g_2 - g_1/2) $. This state can be labeled by the number of massless modes as
$C1S1\frac12$, where quite generally $CnSm$ denotes a state with $n$
massless charge and $m$ massless spin modes.\cite{balents_2chain}

\begin{figure}[htb]
\centerline{\epsfysize 5cm
{\epsffile{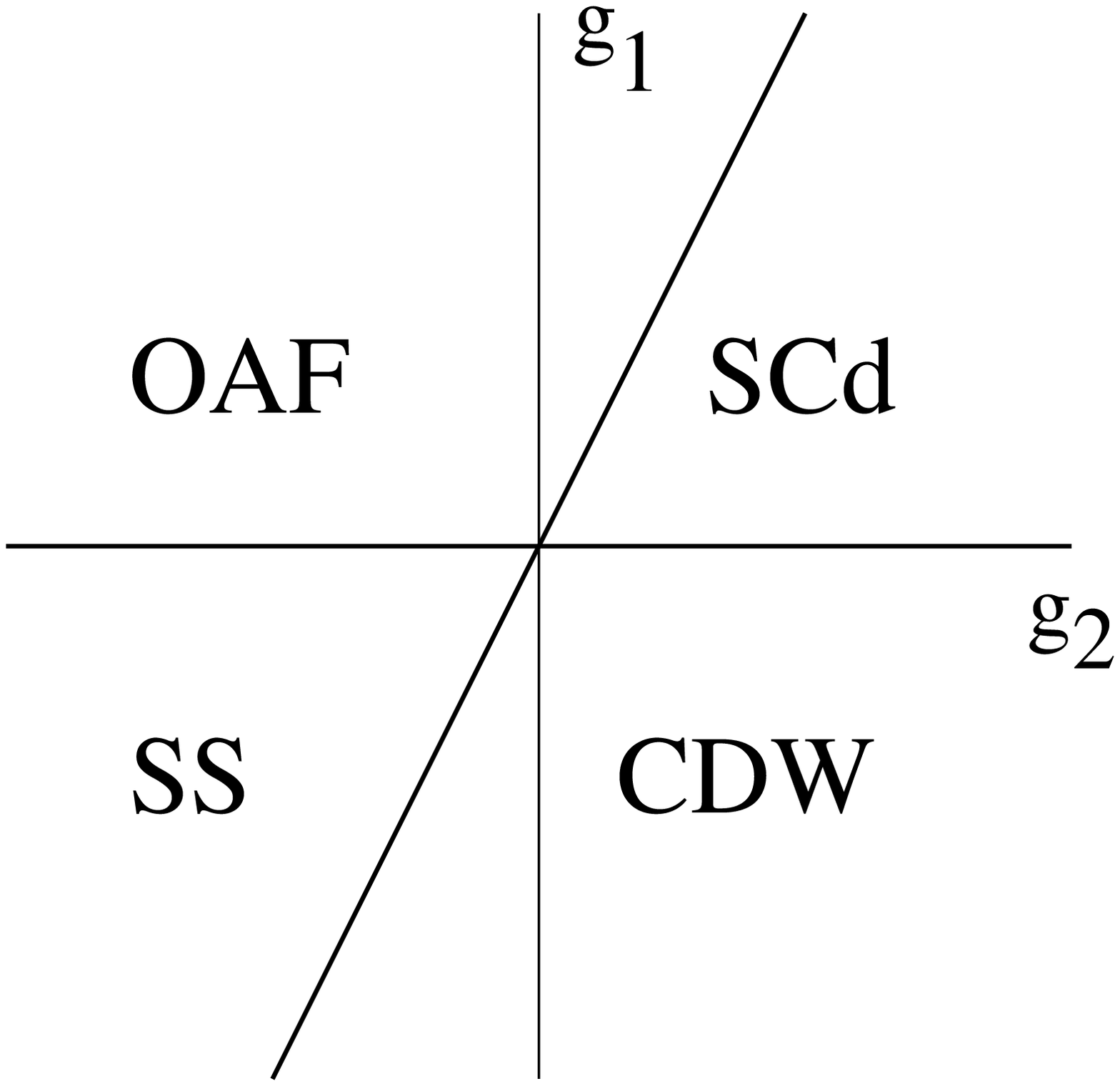}}
}
\caption[to]{Phase diagram of the two--chain model. The different dominating
fluctuations are: SCd: ``d--type'' pairing; OAF: orbital antiferromagnetism;
SS: ``s--type'' pairing; CDW: charge density wave. The critical lines
$g_1=0$ and $g_1=2g_2$ are of Ising type.}
\label{f:2}
\end{figure}

For nonzero $g_1$ all interactions scale to strong coupling, only the total
charge mode remains massless ($C1S0$), reflecting the translational invariance
of the system, and all spin excitations have a gap. The physics in this
regime can be determined looking for the semiclassical minima of the
different $\cos$ terms in eq.(\ref{eq:h2c}). One then finds that the CDW
correlations now decay exponentially, and for the interesting case $g_1>0,
g_2> g_1/2$, corresponding to purely repulsive interaction, the strongest
fluctuations are now of ``d--type'' superconducting
pairing,\cite{fabrizio_2chain,schulz_2chain} with decay as
$r^{-1/(2K_{\rho})}$. Labeling this state as ``d--type'' seems appropriate
because the pairing amplitudes at $k_\perp=0$ and $\pi$ intervene with
opposite sign. In real space, this corresponds to pairs formed by two
fermions on the two different chains. Note that even for weak interactions
where $K_{\rho} \rightarrow 1$ this decay is very slow.  The $4k_F$
component of the density correlations also has a power law decay, however
with an exponent $2K_{\rho}$, much bigger than the SCd
exponent.\cite{schulz_2chain,balents_2chain} The full phase diagram in the
$g_1$--$g_2$ plane is shown in fig.\ref{f:2}. For $g_1<0$ the diagram is
identical to the
single--chain case, however, for $g_1>0$ the behavior is changed
dramatically, and in particular superconductivity is predicted for repulsive
interactions, for example for the Hubbard model which would be represented
as $g_1=g_2$ in the present language.

Remarkably, results basically identical to this weak coupling analysis can
be obtained assuming strong repulsive interactions in the individual chains,
so that one is for example in the regime where interchain electron--hole
pair hopping is more relevant than single particle tunneling, $K_\rho<1/3$
in fig.\ref{f:1}). It is then more appropriate to bosonize the degrees of
freedom of individual chains, rather than working in $k_\perp$--space. Under
renormalization one then generates an interchain interaction of the form
\cite{schulz_trieste,khvesh_2chain}
\begin{eqnarray} 
\nonumber
H_{jk} &=&\frac{1}{(2\pi\alpha)^2} \cos(\sqrt2(\phi_{\rho j}-\phi_{\rho
k}))\\
\label{eq:hij} \times & \!\!\!\!\! \lefteqn{%
\left\{J_\perp[\cos(\sqrt2(\theta_{\sigma j}-\theta_{\sigma k}))
+ \cos(\sqrt2\phi_{\sigma j}) \cos(\sqrt2\phi_{\sigma k})]
+ V\sin(\sqrt2\phi_{\sigma j}) \sin(\sqrt2\phi_{\sigma k})] \right\}\virg}&
\end{eqnarray}
where $j$, $k$ now are chain indices.  Remarkably, this term leads to
properties identical to those found in weak coupling.\cite{schulz_2chain}
First, for $g_1=0$ one has $V=J_\perp$. Then $H_{jk}$ is invariant under the
duality $\phi_{\sigma j} \leftrightarrow \theta_{\sigma j}$, and one has an
Ising critical theory. Secondly, for $g_1 >0$ one finds $V \neq
J_\perp$. Then the fields $\phi_{\rho 1} - \phi_{\rho 2}$, $\phi_{\sigma 1}
+ \phi_{\sigma 2}$, and $\theta_{\sigma 1} - \theta_{\sigma 2}$ become
long--range ordered and one is in a $C1S0$ state. Power law correlations again
exist for SCd and $4k_F$ density fluctuations, with the same scaling
relation between the two exponents as in weak coupling. However, because now
$K_{\rho}<1/3$ the $4k_F$ density fluctuations actually dominate. The
equivalent results in the weak and strong coupling regime very strongly
suggest that {\em the two--chain model is in the same phase for weak and
strong repulsion}. This point is further supported by considering the ``t--J
ladder model'' for strong interchain exchange.\cite{troyer96}

Numerical work on the two--chain model is in agreement with the existence of
d--type
pairing,\cite{dagotto_tjladder,noack_2chain,hayward_2chain,hayward_2chain_2,kuroki96}
the evidence for the $4k_F$ density fluctuations is however
inconclusive.\cite{noack_2chain_2} Concerning experimental observation, one
should notice that the SCd state becomes localized by weak
disorder.\cite{orignac_2ch_disorder}

\subsection{Summary}
In this section we have discussed a number of results, mainly analytical, on
the effect of different forms of interchain coupling on the Luttinger liquid
behavior of strictly one--dimensional systems. As far as spin chains are
concerned, the most spectacular result is the ``oscillation'' between even
and odd numbers of chains, reminiscent of (and formally related to) the
Haldane phenomenon in spin--S antiferromagnetic chains. There is both
experimental and numerical evidence for this behavior, as discussed above.

For conducting chains, we have only discussed the two--chain case. The most
interesting conclusion here was that for the Hubbard model (and a rather
wide class of its generalizations) with purely repulsive interactions a
d--wave superconducting state is predicted from weak up to rather strong
repulsion. This seems to be one of the first cases where there is a reliable
theoretical argument in favor of superconductivity in the (repulsive)
Hubbard model. Only for very strong repulsion does a $4k_F$ CDW predominate.
These results in principle apply directly to doped spin
ladders,\cite{uehara96} and to few--channel quantum wires. Concerning doped
spin ladders, one of the most interesting experimental questions certainly
is whether the superconducting state is indeed of d type as predicted.

A number of results exist for larger numbers of
chains.\cite{arrigoni_3chain,schulz_moriond,kimura97,lin97} In the
perturbative region for weak repulsion generally again a d--wave
superconducting state is found, however for stronger coupling different
phases are found.\cite{schulz_moriond}

\section*{Acknowledgments} We wish to thank the organizers of the Chia
Laguna school, in particular G. Morandi, for arranging an interesting and
stimulating school in a very pleasant environment. One of us (H.J.S.) thanks
his colleagues in Orsay, in particular T. Giamarchi, D. J\'erome,
P. Lederer, G. Montambaux, and J. P. Pouget, for many stimulating
discussions and shared insights over the years.

\appendix\sectio{When to bosonize in peace}
\renewcommand{\theequation}{\Alph{section}\arabic{equation}}
The fundamental ingredient for the following is the expression
eq.(\ref{singlepsi0}) for the
single--fermion field operators 
\begin{equation} \label{eq:singlepsi1}
\psi_{\pm,\sigma}(x) = \frac{1}{\sqrt{L}} \sum_k a_{\pm,\sigma,k} e^{ikx} =
\lim_{\alpha \rightarrow 0} \frac{\eta_{\pm,\sigma}}{\sqrt{2\pi
\alpha}} \exp \left[
\pm i k_F x \mp i \phi_\sigma(x) + i \theta_\sigma(x) \right]
\virg
\end{equation}
where $\sigma$ can be the spin or some other internal degree of freedom of
the fermions and we have replaced the $U$--operators by the Majorana
(``real'') fermion operators ${\eta_{\pm,\sigma}}$ introduced to guarantee
proper anticommutation between the $\psi$'s.\cite{banks_sun} They satisfy
the anticommutation relation
\begin{equation} \label{eq:ecom}
[\eta_r,\eta_s]_+ = 2\delta_{r,s}  \virg
\end{equation}
where $r$ and $s$ are compound indices containing both the chirality $\pm$
and the internal degree of freedom $\sigma$. Eq.(\ref{eq:ecom}) implies
in particular $(\eta_r)^2=1$. The Majorana fermions can be represented by
standard (Dirac) fermion operators $c_r$ as $\eta_r = c^\dagger_r +
c_r$. Note that there is just one isolated fermionic degree of freedom per
branch, and that these degrees of freedom will not appear in the bosonized
Hamiltonian if properly handled.

Using eq.(\ref{eq:singlepsi1}) and its generalization to cases with spin and
other ``internal'' degrees of freedom like perpendicular momentum indices in
coupled chain problems, a typical fermion interaction term becomes
\begin{equation} \label{eq:}
\psi^\dagger_\alpha \psi^\dagger_\beta \psi^{\phantom{\dagger}}_\gamma \psi^{\phantom{\dagger}}_\delta = \eta_\alpha
\eta_\beta \eta_\gamma \eta_\delta \times (\mbox{boson operators})
= h_{\alpha\beta\gamma\delta} \times (\mbox{boson operators}) \virg
\end{equation}
where the second equality defines $h_{\alpha\beta\gamma\delta}$. This
operator, responsible for taking into account fermion anticommutation
properly, is nevertheless carefully passed under the rug in the vast
majority of the literature, thus leaving a purely bosonic Hamiltonian to be
considered, as implied in the term ``bosonization''. I will here investigate
under which conditions this is allowed. 

The relevant situation is that all the indices $\alpha,\beta,\gamma\,\delta$
in $h_{\alpha\beta\gamma\delta}$ are different from each other. Otherwise
the anticommutation rule
\begin{equation} \label{eq:ecom2}
[\eta_r,\eta_s]_+ = 2\delta_{r,s} \Rightarrow \eta_r^2=1 \virg
\end{equation}
allows to simplify $h_{\alpha\beta\gamma\delta}$. I will therefore only
consider the general case. First note that
\begin{equation} \label{eq:h2}
h_{\alpha\beta\gamma\delta}^2 =1 \virg
\end{equation}
$h_{\alpha\beta\gamma\delta}$ thus has eigenvalues $\pm 1$. Secondly,
\begin{equation} \label{eq:com3}
[h_{\alpha\beta\gamma\delta},h_{\kappa\lambda\mu\nu}]_\pm =0 \virg
\end{equation}
where according to whether an even or odd number of pairs of indices taken
from the two sets $(\alpha,\beta,\gamma\,\delta)$ and
$(\kappa,\lambda,\mu,\nu)$ are equal the commutator (even case) or
anticommutator (odd case) is to be used. Finally, permutation of indices
leads to sign changes:
\begin{equation} \label{eq:sig}
h_{\alpha\beta\gamma\delta}=-h_{\beta\alpha\gamma\delta}
= - h_{\alpha\gamma\beta\delta} = - h_{\alpha\beta\delta\gamma}
\end{equation}

It is now clear that if all the $h$'s occurring in a given Hamiltonian
commute, they can be simultaneously diagonalized, which means that it will
be possible to replace each of the $h$'s by $\pm1$, leading to a purely
bosonic Hamiltonian. This clearly is the case if all the $h$'s occurring have
an even number of indices in common. In the opposite case some of the $h$'s
do not commute, therefore can not be simultaneously diagonalized and not be
eliminated from the Hamiltonian. Bosonization then is not possible.

As a simple example consider the single--chain Luttinger model with spin,
sec.\ref{spinhalfsec}. the four allowed values of the discrete indices are
$1\equiv (+,\uparrow),2\equiv (+,\downarrow),3\equiv (-,\uparrow),4\equiv
(-,\downarrow)$. Consequently, only $h_{1234}$ can occur, and according to
the eigenvalue chosen the backward scattering interaction takes the form
$\pm g_1 \cos( \sqrt{8}\phi_\sigma)$. The choice of eigenvalue of $h_{1234}$
affects however the expressions for correlation functions: for example
$h_{1234} = \pm1$ implies $\eta_1\eta_3=\pm\eta_2\eta_4$, and consequently
the $2k_F$ charge density operator contains either a factor
$\cos(\sqrt{2}\phi_\sigma)$ (plus sign) or $\sin(\sqrt{2}\phi_\sigma)$. A
similar discrete ``gauge covariance'' exists of course for all correlation
functions.

In more complicated cases like the two--chain problem, more then one
$h$--operator occurs. Even if they all commute, as is the case for the
two--chain problem, additional constraints  on the permissible eigenvalues
of the $h$'s exist due to the existence of relations of the type
\begin{equation} \label{eq:rel}
h_{\alpha\beta\gamma\delta} h_{\kappa\lambda\mu\nu} h_{\pi\rho\sigma\tau}=
\pm1\virg
\end{equation}
and similar relations involving more than three $h$'s. However, a discrete
gauge freedom of the type mentioned above often remains. For the particular
case of fermions with an internal SU(N) symmetry,\cite{banks_sun}
bosonization can be performed without problem and all the $h$--operators can
be given eigenvalue $+1$, a fact not noticed in the original work.

\bibliographystyle{prsty}
\bibliography{revues,1dtheory,1dexp,hubbard,2dtheory,la,kondo,semic,fermi,remo,loc}
%\bibliography{totphys}

\end{document}